\documentclass[footinbib,a4paper,aps,prx,superscriptaddress,reprint,twocolumn,preprintnumbers,amsmath,amssymb,nobalancelastpage,10pt]{revtex4-1}
\usepackage{amsmath}
\usepackage{verbatim}
\usepackage{graphicx}
\usepackage{color}
\usepackage{tikz}
\usepackage{subfigure}
\usepackage{float}

\usepackage{mathptmx}
\usepackage{amssymb}
\usepackage{amsmath}
\usepackage{amsfonts}
\usepackage{array}

\usepackage{bm}
\usepackage{xcolor}
\graphicspath{{figures/}}

\usepackage{braket}

\definecolor{mygrey}{gray}{0.35}
\definecolor{myblue}{rgb}{0.2,0.2,0.8}
\definecolor{myzard}{cmyk}{0,0,0.05,0}
\definecolor{mywhite}{rgb}{1,1,1}
\definecolor{myred}{rgb}{1,0.,0.3}

\usepackage[colorlinks=true,citecolor=myblue,linkcolor=myred]{hyperref}

 \def\ee{\mathord{\rm e}}
 
 \def\ii{\mathord{\rm i}}

\def\half{\textstyle\frac{1}{2}}

\def\fourth{\textstyle\frac{1}{4}}

\renewcommand{\ii}{{\rm i}}
\renewcommand{\ee}{{\rm e}}
\def\beq{\begin{equation}}
\def\eeq{\end{equation}}
\def\barray{\begin{eqnarray}}
\def\earray{\end{eqnarray}}

\begin{document}

\title{Topological chiral currents in  the   Gross-Neveu  model extension}


\author{E. Tirrito}
\affiliation{International School for Advanced Studies (SISSA), via Bonomea 265, 34136 Trieste, Italy}
%
\author{M. Lewenstein}
\affiliation{ICFO - Institut de Ci\`encies Fot\`oniques, The Barcelona Institute of Science and Technology, Av. Carl Friedrich Gauss 3, 08860 Castelldefels (Barcelona), Spain} 
\affiliation{ICREA, Lluis Companys 23, 08010 Barcelona, Spain}
\author{A. Bermudez}
\affiliation{Departamento de F\'{i}sica Te\'{o}rica, Universidad Complutense, 28040 Madrid, Spain}
\affiliation{Instituto de F\'{i}sica Te\'{o}rica, UAM-CSIC, Universidad Aut\'onoma de Madrid, Cantoblanco, 28049 Madrid, Spain.}

\begin{abstract}
We unveil  an interesting connection of  Lorentz-violating quantum field theories, studied in the context of the standard model extension, and  Hubbard-type models of   topological crystalline phases. These models can be interpreted as a  regularisation of the former and, as hereby discussed,    explored with current quantum simulators based on ultra-cold atoms in optical Raman lattices. 
In particular, we present a complete  analysis  of the  Creutz-Hubbard ladder under  a generic magnetic  flux, which regularises a Gross-Neveu model extension, and presents a characteristic  circulating chiral current whose non-zero value arises from  a specific violation of Lorentz invariance. We present a complete phase diagram with trivial insulators,  ferromagnetic and anti-ferromagnetic phases, and  current-carrying topological crystalline phases.
These predictions are benchmarked using tools from condensed matter and quantum-information science,  showing that  self-consistent Hartree-Fock  and  strong-coupling Dzyaloshinskii-Moriya  methods capture the essence of the phase diagram in different regimes,  which is further explored  using extensive numerical simulations based on   matrix-product states.
\end{abstract}

\maketitle

\setcounter{tocdepth}{2}
\begingroup
\hypersetup{linkcolor=black}
\tableofcontents
\endgroup

  \section{\bf Introduction}
The standard model (SM) of particle physics is one of the big triumphs of  theoretical physics, as it provides simultaneously an accurate and economic description of nature~\cite{Peskin:1995ev}.   
The SM introduces a reduced set of  quantum fields and, building upon elegant and simple symmetry principles, it  fixes the form of their interactions and precisely accounts for all   fundamental particles observed to date. However, existing difficulties in the incorporation of gravity to the SM suggest that  this model could  be  
a  low-energy limit of a more fundamental  theory that describes  physics at a much higher  scale, the Planck scale. In these theories, fundamental symmetries such as Lorentz invariance may be broken, as occurs in string theory~\cite{PhysRevD.39.683} and non-commutative field theories~\cite{PhysRevLett.87.141601}. The remnants of this {\it Lorentz violation}  can be included in the so-called standard model extension, an effective quantum field theory (QFT) in a larger parameter space that includes all possible  Lorentz-violating terms  with new bare couplings~\cite{PhysRevD.58.116002,PhysRevD.65.056006,COLLADAY2001209}. A wide variety of experiments~\cite{Mattingly2005}, including recent quantum-information-enhanced interferometry~\cite{Pruttivarasin2015}, establish  tight bounds on  these  couplings and, so far, have shown no evidence of the violation of Lorentz symmetry.

\begin{figure*}[t]
\centering
  \includegraphics[width=0.99\linewidth]{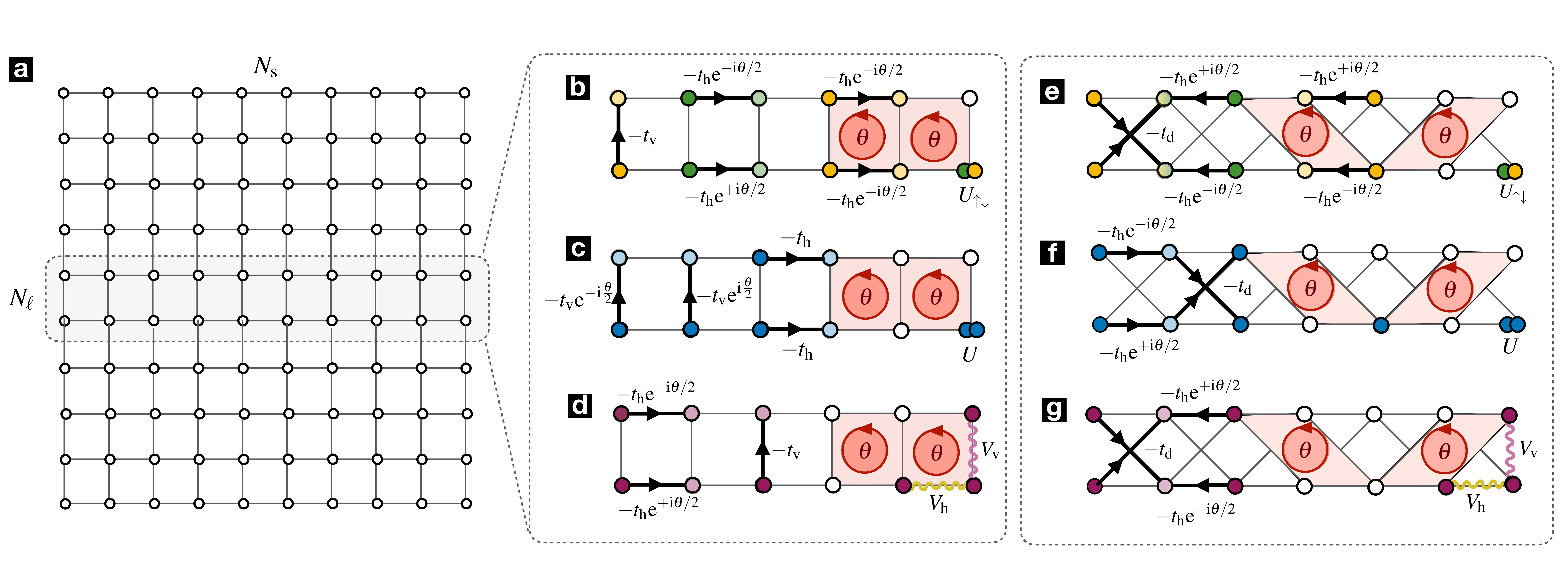}
\caption{\label{fig:scheme_ladders} {\textbf{Hubbard ladders pierced by a magnetic field:}} {\bf (a)} Diagram of $N_\ell$-leg ladder with open boundary conditions. Particles  move across the ladder hopping along horizontal and vertical links that join the different lattice sites. 
{\bf (b)} Rectangular 2-leg ladder for spin-full fermions subjected to an external magnetic flux $\theta$. The fermions
can hop along the legs and rungs  with strength  $t_{\rm h} e^{\pm i \theta /2}$ and $t_{\rm v}$, respectively, and interact through a repulsive local Hubbard interaction $U_{\uparrow \downarrow}$ depending on the spin. {\bf (c)} Rectangular ladder for spin-less bosons subjected to  an external homogeneous magnetic field. The tunneling strengths are arranged similarly to {\bf (b)},  and $U$ represents the  interaction strength when two bosons meet at the same lattice site.  Exploiting the gauge freedom, one can  arrange the Peierls' phases in the tunnelings along the vertical directions maintaining the same  flux, such that the bosons pick an Aharonov-Bohm phase proportional to the flux $\theta$  when tunneling across a plaquette.
{\bf (d)} Rectangular ladder for spin-less fermions subjected to  an external magnetic flux $\theta$, here depicted through an anti-clockwise directed circle. The one-site Hubbard interaction is forbidden by Pauli exclusion, and  only  nearest-neighbor ones $V_{\rm v}$ and $V_{\rm h}$ can contribute.  {\bf (e)-(g)} Same as {\bf (b)-(d)} but for a cross-linked, so called, Creutz-Hubbard ladder. Here, the tunneling between neighbouring chains occurs diagonally instead of vertically $t_{\rm v}\to t_{\rm d}$, forming a crossed-linked pattern. In this case, particles hopping along two possible trapezoidal plaquettes gain an Aharonov-Bohm phase proportional to the magnetic flux $\theta$. }
\end{figure*}

There is, however, an alternative arena  where Lorentz invariance is not an exact  symmetry  but, instead, appears  in the infrared long-wavelength limit~\cite{CHADHA1983125}. In condensed-matter physics, many microscopic theories are formulated in terms of non-relativistic quantum field theories~\cite{fradkin_2013} and, yet, Lorentz invariance and relativistic effects can appear in effective descriptions of  low-energy phenomena. The standard scenario where this happens  is that of phase transitions~\cite{landau}, where the physics around some critical points can be  described by Lorentz-invariant QFTs, and the renormalization group allows to understand why  Lorentz-breaking microscopic corrections become irrelevant at low energies~\cite{WILSON197475,RevModPhys.66.129}. Indeed, the dialogue between condensed matter and high-energy physics has proved to be extremely fruitful in this context, see e.g.~\cite{PhysRev.106.162} and~\cite{PhysRev.122.345,PhysRev.124.246}, or~\cite{PhysRev.130.439} and~\cite{PhysRevLett.13.321,PhysRevLett.13.508}.  An alternative scenario where  the emergence of Lorentz invariance   has turned out to be relevant is in graphene~\cite{RevModPhys.81.109}, Weyl semi-metals~\cite{RevModPhys.90.015001} and topological insulators and superconductors~\cite{RevModPhys.83.1057}. 
  
In recent years, this fertile dialogue has broadened its scope,   as Lorentz invariance has also become manifest as an effective symmetry in systems of atomic, molecular, and optical (AMO) physics~\cite{Bloch2012,Blatt2012}. Building upon Feynman's idea of a quantum simulator~\cite{Feynman_1982}, there has been a number of experiments targeting Lorentz- and gauge-invariant QFTs using ultracold atoms~\cite{Tarruell2012,Duca288,Schweizer2019,Mil1128,1902.09551,2003.08945} and trapped ions~\cite{Gerritsma2010,PhysRevLett.106.060503,Martinez2016} (see e.g.~\cite{Zohar_2015, doi:10.1080/00107514.2016.1151199, 1911.00003,aidelsburger2021cold,klco2021standard} for  recent reviews). To the best of our knowledge, however, quantum simulations of Lorentz-violating extensions of the SM of particle physics    have not been considered  yet. Although  Lorentz-breaking terms should, in fact, be the rule rather than the exception, since one starts from a non-relativistic model that accurately describes these AMO systems, the difficulty lies in tailoring  those systems so that the Lorentz-violating terms that survive in the long-wavelength limit correspond precisely to those considered in the standard model extension (SME)~\cite{PhysRevD.58.116002}. This may change the role of the SME, which has routinely  provided a framework for some of the most precise null experiments to date~\cite{RevModPhys.83.11}. Instead, if successful, AMO quantum simulators have the potential to  turn the SME into a framework of experimentally-testable  analogue physics beyond the standard model. In this work, we follow this route for a particular case,  unveiling interesting connections between Lorentz-violating quantum field theories and lattice models of correlated topological phases of matter that display a persistent chiral current. We show that the latter provides a lattice regularisation of a  specific matter sector  of the SME  including, in addition, four-Fermi Gross-Neveu-type interactions. Moreover, we also discuss how these models  can be implemented  in   experiments of ultra-cold atoms with synthetic spin-orbit coupling~\cite{Galitski2013,liu_zhang_materials_liu}.

This article is organised as follows: 
In Sec. \ref{sec:CHL_Arbitrary_Flux}, we start by   reviewing  the physics of Hubbard ladders, which are the minimal fermionic lattice models  that allow for the effect of external (background) gauge fluxes, and may support a  circulating, so-called  chiral, current. This sets the stage to introduce a particular lattice regularization of low-dimensional Lorentz-violating QFTs,  the {\it cross-link 
Creutz-Hubbard ladder} for {generic magnetic flux}  $\theta$. We argue that this model can host a current-carrying topological crystalline  phase,  which is simultaneously described by a non-zero topological invariant and a circulating chiral current. A short summary of the main results of this work is then presented.
In Sec. \ref{sec:CHL_Continuum_Limit}, we explicitly construct the 
continuum limit of this ladder,
and  show that it corresponds to a {\it  Gross-Neveu model extension}
with a particular { Lorentz violation} that has been explored in the SME. This  gives a neat perspective on  the appearance of the persistent chiral current, which coexists with a non-zero topological invariant,  and can be used to chart the phase diagram via the associated susceptibility, as discussed in Sec.~\ref{sec:Chiral_Current}.  
In Sec. \ref{section:mf_mps}, using a self-consistent  mean-field approach analogue to a large-$N$ limit of the Gross-Neveu model,  we study how this chiral current can be used to characterise the robustness  of  
the topological crystalline phase as the strength of the four-Fermi interactions are increased and  discuss the full phase diagram of the model. 
We benchmark the mean-field predictions using tools developed by  the condensed-matter and quantum-information communities, i.e. tensor-network variational techniques based on matrix-product states. As discussed in the text, these quasi-exact numerics confirm qualitatively the mean-field prediction of the phase diagram and correct the typical shortcomings of mean-field treatments. Finally, in Sec.~\ref{sec:cold_atoms}, we present a proposal for a potential experimental realization with spin-orbit-coupled  ultra-cold atoms. In this way, models of high-energy physics considered in the context of the Standard Model extension could be accessed with table-top  experiments of non-relativistic neutral atoms at ultra-low temperatures.

\section{\bf The Creutz-Hubbard ladder with arbitrary flux} \label{sec:CHL_Arbitrary_Flux}

In this section, we describe the model under study: the cross-linked Creutz-Hubbard ladder under an arbitrary magnetic flux. Let us start, however, by  reviewing  briefly the context of strongly-correlated behaviour in ladder compounds, which has a long and fruitful history in condensed matter. 

\subsection{Previous studies on  Hubbard ladders}
In ladder models, the particles are arranged on a lattice composed  of $N_{\ell}$   chains with $N_{\rm s}$ sites each. Typically, these particles have nearest-neighbour couplings along the vertical and horizontal directions resembling a rectangular ladder structure of $N_{\ell }$ legs (see Fig.~\ref{fig:scheme_ladders}{\bf (a)}). These systems interpolate between 1D and 2D as one increases the numbers of legs $N_{\ell}\to N_{\rm s}$ and, sometimes, can host unexpected phenomena. For instance,  in the context of  Heisenberg ladders and their connection to the high-$T_{\rm c}$ cuprates~\cite{Dagotto_1999}, the parity of the number of legs determines the gapped/gapless nature of the model~\cite{Dagotto618,Sierra_1996}, drawing a neat connection to the Haldane conjecture of spin chains and the interplay of topology and constrained QFTs~\cite{HALDANE1983464,AFFLECK1985397}. Through an analogue of the  super-exchange mechanism~\cite{PhysRev.79.350}, these Heisenberg ladders can be seen as the strong-coupling limit of a half-filled Hubbard ladder, where electrons interact via a strong local density-density coupling, the so-called Hubbard interaction~\cite{PRSLSA_276_238}. Being quasi-1D, ladders yield a neat playground to generalise analytical methods~\cite{PhysRevB.48.15838,PhysRevB.53.12133,PhysRevB.58.1794,PhysRevB.56.6569} developed for one-dimensional systems~\cite{Giamarchi:743140}. Moreover, Hubbard and Heisenberg ladders can be explored numerically~\cite{NOACK1996281,PhysRevLett.73.886} using  efficient  and very accurate numerical schemes~\cite{PhysRevLett.69.2863,RevModPhys.77.259}.

Getting closer to the subject of the present work, we note that  the rectangular ladders are the minimal lattice structures that can be pierced by an external magnetic flux and have a well-defined thermodynamic limit (see Figs.~\ref{fig:scheme_ladders}{\bf (b)}-({\bf d})). By increasing the number of legs, these ladders yield a clear route towards the  integer~\cite{PhysRevLett.45.494} and fractional~\cite{PhysRevLett.48.1559} quantum Hall effects. For the two-leg ladder, the presence of a magnetic flux modifies the fermion tunneling in a  Hubbard ladder via the so-called Peierls' substitution~\cite{Peierls1933} depicted in Fig.~\ref{fig:scheme_ladders}{\bf (b)}. This in turn leads to a rich phase diagram that contains Luttinger and Luther-Emery phases~\cite{PhysRevB.76.195105}, depending on the magnetic flux and the specific filling factor. In the case of spinless fermions, on-site Hubbard interactions must be exchanged for nearest-neighbour ones (see Fig.~\ref{fig:scheme_ladders}{\bf (d)}), the interplay of which with the magnetic flux can also give rise to various strongly-correlated effects. While in the absence of the magnetic field, these finite-range interactions typically stabilise charge- and bond-density wave patterns depending on the  particular filling, the situation becomes much richer as the magnetic field is switched on. For instance,  one may find staggered flux phases with a pattern of local currents along the vertical and horizontal links, encompassing alternating circulations (i.e vortices)  in the  plaquettes~\cite{PhysRevB.71.161101,PhysRevB.73.195114}.  Moreover, one can also find kinks/anti-kinks  that interpolate between the two possible symmetry-broken patterns of vortices, each of which  hosts fractionally-charged excitations in analogy  to  fractionalisation phenomena in  QFTs~\cite{PhysRevD.13.3398} and polymers~\cite{PhysRevLett.42.1698,JACKIW1981253}.

The remarkable  progress of ultra-cold atoms in optical  lattices has stimulated a renewed interest in the physics of the bosonic counterparts of these ladders (see Fig.~\ref{fig:scheme_ladders}{\bf (c)}). In particular,  recent experiments~\cite{Atala2014,Tai2017} have realised a Bose-Hubbard model~\cite{PhysRevB.40.546,PhysRevLett.81.3108} in a two-leg ladder under an external magnetic flux. As discussed in previous theoretical works~\cite{PhysRevB.64.144515}, this bosonic model can host liquid phases analogous to the  Meissner and vortex phases of superconductors. These states of matter can be distinguished by a {\it circulating chiral current}, which has actually been measured in   experiments~\cite{Atala2014}. This current quantifies a boson flow with  a natural connection to the edge states and skipping orbits of quantum Hall samples as the number of legs is increased~\cite{PhysRevA.89.023619}. However, as the microscopic particles are bosons, the nature of the phases differs, and one can find novel Meissner- and vortex-type groundstates. In the Meissner phase, the bosonic chiral current plays the role of the screening current in superconductors (i.e. it flows in a different direction so as to screen  the external magnetic flux). Alternatively, in the vortex phase,  currents run along the ladder rungs   leading to vortices where the magnetic flux is not fully screened, which draws a clear analogy to type-II superconductors. More recently, this distinction  has also been identified as one increases the Hubbard interactions, where one finds Meissner- and vortex-type Mott insulators~\cite{PhysRevLett.111.150601,PhysRevB.91.140406}, as well as the so-called vortex lattices~\cite{PhysRevB.87.174501,PhysRevLett.115.190402}. Let us also remark that the  chiral current can also be used to unveil a quantitative connection of these quasi-1D  ladders, both bosonic and fermionic, with the topologically-ordered phases~\cite{PhysRevLett.50.1395} of the fractional quantum Hall effect~\cite{PhysRevB.92.115446,PhysRevX.7.021033}.

Once the relevant literature on fermionic and bosonic Hubbard models on rectangular ladders has been reviewed, let us describe the literature on {\it cross-linked Hubbard ladders}. In this case, the particles can  tunnel between neighbouring chains forming a crossed-linked pattern, as  depicted in  Fig.~\ref{fig:scheme_ladders}{\bf (e)-(g)}. To the best of our knowledge, this ladder geometry was first introduced by M. Creutz~\cite{PhysRevLett.83.2636,RevModPhys.73.119}  in the context of domain-wall fermions~\cite{KAPLAN1992342,JANSEN1992374,GOLTERMAN1993219,Kaplan:2009yg} in lattice gauge theories~\cite{RevModPhys.51.659,RevModPhys.55.775}. The presence of the external magnetic field changes considerably the band structure, and can even lead to completely flat bands  in the $\pi$-flux regime~\cite{flat_band_review}, yielding an instance of Aharonov-Bohm cages~\cite{PhysRevLett.81.5888} with fermions locked to the  plaquettes  due to destructive interference~\cite{PhysRevLett.83.2636}. In fact, this so-called Creutz ladder is an archetype of flat-band physics~\cite{flat_band_review}. In addition to these localised bulk states, there are additional edge states exponentially localised to the left- and right-most boundaries of the ladder. These states can be understood as lower-dimensional versions~\cite{PhysRevLett.83.2636} of  domain-wall fermions~\cite{KAPLAN1992342}, and correspond to  the  edge states of a topological insulator~\cite{RevModPhys.83.1057} in the symmetry class $\mathsf{AIII}$/$\mathsf{BDI}$ ~\cite{Mazza_2012,1907.11460}. The persistence of these topological features at finite temperatures~\cite{PhysRevB.86.155140,PhysRevLett.112.130401}, and adiabatic~\cite{PhysRevLett.102.135702} or sudden~\cite{PhysRevB.99.054302} dynamical quenches, has also been discussed in the literature. Moreover, the Creutz ladder also yields a neat scenario to explore the phenomenon of topological charge pumping~\cite{PhysRevB.96.035139}. 
  
Let us now briefly review some recent works exploring the strongly-correlated physics of this cross-linked ladder in the presence of Hubbard-type interactions. As  occurred for the rectangular ladder, there has  been a recent interest in studying the phases of a Bose-Hubbard model in this cross-linked geometry (see  Fig.~\ref{fig:scheme_ladders}{\bf (f)}). Here, a flat-band-induced frustration can change the standard condensation of bosons~\cite{PhysRevB.82.184502}, and lead to new phases of matter even for weak interactions. Depending on the filling per site, the contact interactions can lead to  valence bond crystals,  quasi-superfluids of paired bosons, or supersolids that arise from bound pairs of domain walls that interpolate between two   valence-bond orderings~\cite{PhysRevA.88.063613,PhysRevB.88.220510}. We note that, in addition to the cold-atom quantum simulators, the bosonic Creutz-Hubbard ladder has also been discussed in the context of microwave photons in superconducting circuits~\cite{PhysRevA.99.053834}, photonic waveguide arrays~\cite{doi:10.1002/qute.201900105}, and  superradiant tight-binding lattices in momentum space~\cite{PhysRevLett.126.103601}. Moving on to the spinful fermionic case of  Fig.~\ref{fig:scheme_ladders}{\bf (e)}, this model yields a neat playground to understand correlation effects in topological phases of matter~\cite{Rachel_2018}. For instance, for attractive interactions and in the context of flat-band superconductivity~\cite{PhysRevB.83.220503,Peotta2015}, one can find  an exact Bardeen-Cooper-Schrieffer(BCS) groundstate and emerging $SU(2)$ symmetries~\cite{PhysRevB.94.245149}.  In this case, a crossover between such BCS superconductor and a superfluid of tightly-bound pairs for attractive Hubbard interactions has also been found~\cite{PhysRevB.98.134513,PhysRevB.98.155142}. The competition of   BCS pairing and repulsive Hubbard interactions was discussed in~\cite{PhysRevB.89.115430} and shown to yield an interesting phase diagram with topological superconductors characterised by Majorana edge states. The interplay of flat-bands interactions and disorder has also been explored for the fermionic Creutz-Hubbard ladder~\cite{Kuno_2020,PhysRevB.102.041116} in connection to many-body localisation~\cite{doi:10.1146/annurev-conmatphys-031214-014726}.
  
The inclusion of various hyperfine states in cold-atom gases in optical lattices opens a new perspective for quantum simulations, as one may use the internal states as a synthetic dimension~\cite{PhysRevLett.108.133001,PhysRevLett.112.043001,Ozawa2019}, as recently demonstrated in  quantum simulators of quantum Hall ladders~\cite{Mancini1510,Stuhl1514,PhysRevLett.122.065303}, or  of spin-orbit coupled interacting wires~\cite{PhysRevLett.117.220401,Kolkowitz2017,Bromley2018}. In this context, the $s$-wave scattering  that  gives rise to $SU(N)$ Hubbard models~\cite{Gorshkov2010,Pagano2014}, can be understood as non-local interactions along the synthetic dimension of a rectangular ladder (see  Fig.~\ref{fig:scheme_ladders}{\bf (b)-(d)}), giving rise to a new family of models with a rich interplay between  interaction-induced strong correlations and  flux-induced kinetic frustration~\cite{PhysRevX.7.021033,Barbarino_2016,PhysRevA.94.023630,PhysRevA.95.043613,PhysRevLett.118.230402,PhysRevA.97.013634,PhysRevB.99.245101}. As discussed in~\cite{PhysRevX.7.031057}, the synthetic ladders are not restricted to rectangular geometries, as one can exploit Floquet-engineering techniques~\cite{RevModPhys.89.011004} to induce cross-link tunnelings, giving rise to the subject of this work: the {\it synthetic Creutz-Hubbard ladder}, which we now describe. 
  
\subsection{The synthetic Creutz-Hubbard  ladder} 
  \label{sec:ch_previous_results}
We consider  a system of spinless fermions that can be created (annihilated) at the sites of  a two-leg ladder by  the operators $c^\dagger_{i,\ell}$ ($c^{\phantom{\dagger}}_{i,\ell}$) , which are labelled by  $i\in\{1,\cdots,N_{\rm s}\}$ and $\ell\in\{\rm u,d\}$. The cross-link geometry of the ladder, as displayed in Fig.~\ref{fig:scheme_ladders}{\bf (g)}, is determined by the tunnelling terms of the lattice Hamiltonian
\beq
\label{eq:H_C}
H_{\rm C}\!=-\!\!\sum_{i,\ell}\!\!\left(\!t_{\rm h}\ee^{-\ii \frac{s_\ell\theta}{2}}c^\dagger_{i+1,\ell}c^{\phantom{\dagger}}_{i,\ell}+t_{\rm d}c^\dagger_{i+1,\ell}c_{i,\overline{\ell}}- \!\frac{s_\ell\Delta\epsilon}{2} c^\dagger_{i,\ell}c^{\phantom{\dagger}}_{i,\ell}+{\rm H.c.}\!\!\!\right)\!\!,
\eeq
where we have introduced the horizontal (diagonal) tunnelling strength $t_{\rm h}$ ($t_{\rm d}$), the energy imbalance $\Delta\epsilon$, and the notation $s_\ell=\{+1,-1\}$ and $\overline{\ell}=\{\rm  d, u\}$  for $\ell=\{\rm u,d\}$, respectively. Finally, we note that the complex Peierls phases of the tunnellings  lead to a non-zero Aharonov-Bohm phase $\theta$  when the fermions tunnel across the trapezoidal plaquettes depicted in Fig.~\ref{fig:scheme_ladders}{\bf (g)}. Accordingly, the parameter $\theta$ can be understood as the magnetic flux $\Phi_{\rm B}$    of a background magnetic field $\bm{B}_{\rm bg}=\bm{\nabla} \bm{\times} \bm{A}_{\rm bg}$ that is directed perpendicularly to the plaquette $S$, 
$\theta=\frac{e}{\hbar}\oint_\gamma{\rm d}\bm{l}\cdot\bm{A}_{\rm bg}=2\pi\frac{e}{h}\int_{S}{\rm d}\bm{S}\cdot\bm{B}_{\rm bg}=2\pi{\Phi_{\rm B}/\Phi_0}$, expressed in  units of the flux quantum $\Phi_0=e/h$.
  
As advanced in the previous section, the  $\theta=\pi$-flux limit is obtained when half a flux quantum pierces each  plaquette ${\Phi_{\rm B}=\Phi_0}/2$, the  tunnelings have equal strengths $t_{\rm h}=t_{\rm d}$, and the imbalance vanishes $\Delta\epsilon=0$. This leads to an Aharonov-Bohm destructive interference and a pair of  flat bands $\epsilon_{\pm}(k)=\pm 2t_{\rm h}$, such that  particle or hole excitations do not propagate $v_{\rm g}=\partial_k\epsilon_{\pm }(k)=0$, where $k\in{\rm BZ}=[-\pi/a,\pi/a)$ is the quasi-momentum of the reciprocal lattice for  lattice spacing $a$. Moreover, these flat bands are topological, as they can be  characterised by a non-zero topological invariant  $\gamma_{\pm}=\mp\pi$~\cite{PhysRevX.7.031057}, the so-called Zak's phase $\gamma_{\pm}=\int_{\rm  BZ}{\rm d}k\mathcal{A}_\pm(k)$~\cite{PhysRevLett.62.2747}, where we have introduced the Berry connection $\mathcal{A}_\pm(k)=\bra{\epsilon_\pm(k)}\ii\partial_k\ket{\epsilon_\pm(k)}$~\cite{doi:10.1098/rspa.1984.0023}. As discussed in~\cite{PhysRevX.7.031057}, the Berry connection in this $\pi$-flux limit is actually homogeneous $\mathcal{A}_\pm(k)=\mp1/2$ which, in higher dimensions, would lead to a vanishing Berry curvature and one would speak of a topological flat-band structure. Let us also note that, by switching the energy imbalance $\Delta\epsilon\neq 0$, the bands (connection) will no longer be flat (homogeneous), but the topological invariant remains quantised for $|\Delta\epsilon|\leq 4t_{\rm h}$. According to the underlying symmetries~\cite{PhysRevB.78.195125,doi:10.1063/1.3149495}, this imbalanced Creutz ladder  hosts an $\mathsf{AIII}$ topological insulator. 
   
The interplay of topology and interactions can lead to exotic phases of matter in topological flat-band systems~\cite{PARAMESWARAN2013816}. In the context of  synthetic dimensions in the cold-atom scheme~\cite{PhysRevX.7.031057}, as discussed in more detail in Sec.~\ref{sec:cold_atoms}, the upper and lower leg of the ladder  actually correspond to two different hyperfine states of the groundstate manifold. Therefore, a contact Hubbard interaction due to the $s$-wave scattering of the ultra-cold  atoms can be interpreted as a nearest-neighbour interaction along the vertical direction  connecting the  legs of the synthetic ladder (see Fig.~\ref{fig:scheme_ladders}{\bf (g)}). This leads to the following quartic term
\beq
 \label{eq:H_CH}
H_{\rm CH}=  H_{\rm C}+\frac{V_{\rm v}}{2}\sum_{i,\ell}c^{{\dagger}}_{i,\overline{\ell}\phantom{\overline{\ell}}\!\!}c^\dagger_{i,\ell\phantom{\overline{\ell}}\!\!\!}c^{\phantom{\dagger}}_{i,\ell\phantom{\overline{\ell}}\!\!\!}c^{\phantom{\dagger}}_{i,\overline{\ell}},
\eeq
where $V_{\rm v}>0$ represents a repulsive interaction strength. As shown in~\cite{PhysRevX.7.031057} using various analytical and numerical techniques, the non-interacting $\mathsf{AIII}$ topological insulator is adiabatically connected to a {\it correlated topological insulator } in a wide lobe of parameter space $(V_{\rm v},\Delta\epsilon)$, the tip of which corresponds to $(V_{\rm v},\Delta\epsilon)=(8t_{\rm h},0)$. This symmetry-protected topological phase is surrounded by a correlated trivial band insulator and an orbital ferromagnet. This symmetry-broken phase is characterised by the magnetic order parameter  $M_y=\langle T_i^y\rangle\neq0$, where $T_i^y=\ii (c^{{\dagger}}_{i,{\rm d}}c^{\phantom{\dagger}}_{i,{\rm u}}-c^{{\dagger}}_{i,{\rm u}}c^{\phantom{\dagger}}_{i,{\rm d}})/2$ can be understood as a spin-1/2 operator when the upper and lower leg components are interpreted as the two spin projections. 
  
Interestingly, this $\pi$-flux regime has a direct connection to a Wilson-type discretization~\cite{Wilson1977} of the Gross-Neveu (GN) model~\cite{PhysRevD.10.3235}, a  QFT of self-interacting fermions in (1+1) dimensions that shares some features with higher-dimensional non-Abelian gauge theories~\cite{RevModPhys.55.775}. The GN model belongs to the family of four-Fermi field theories, originally introduced in the context of nuclear interactions by Enrico Fermi~\cite{Fermi1934,doi:10.1119/1.1974382}. Specifically, it can be understood as a low-dimensional version of  Nambu-Jona-Lasinio (NJL)  models~\cite{PhysRev.122.345,PhysRev.124.246}, and  allows to explore chiral symmetry breaking  by dynamical mass generation   and asymptotic freedom in a renormalizable framework~\cite{PhysRevD.10.3235}.
As discussed in~\cite{BERMUDEZ2018149,PhysRevB.99.125106}, the $\pi$-flux synthetic Creutz-Hubbard ladder can be unitarily mapped onto a Wilson-type discretization of the   Gross-Neveu  model, which is described by the following lattice Hamiltonian 
\begin{widetext}
\beq
\label{eq:H_GNW}
H_{\rm GNW}=a\sum_{x\in\Lambda_{\rm s}}\left[\left(-\overline{\Psi}(x)\left(\frac{\ii\gamma^1}{2a}+\frac{1}{2a}\right)\Psi(x+a)+\overline{\Psi}(x)\left(\frac{m}{2}+\frac{1}{2a}\right)\Psi(x)+{\rm H.c.}\right)-\frac{g^2}{2N}\left(\overline{\Psi}(x){\Psi}(x)\right)^2\right],
\eeq
\end{widetext}
where $\Psi(x)=(\psi_1(x),\cdots,\psi_N(x))^{\rm t}$ contains $N$ flavours of the two-component Dirac spinors $\psi_n(x)$, $\overline{\Psi}(x)=\Psi(x)^\dagger\gamma^0$ is the  adjoint, and $\gamma^0,\gamma^1$ are the gamma matrices in a  (1+1)-dimensional  spacetime $\{\gamma^\mu,\gamma^\nu\}=2g^{\mu\nu}$, where   $g^{\mu\nu}={\rm diag}(1,-1)$ is the flat Minkowski metric. For the synthetic Creutz-Hubbard ladder~\eqref{eq:H_C}-\eqref{eq:H_CH}, one needs to  set $N=1$, $\psi_1(x)=\frac{1}{\sqrt{a}}\sin(\pi j/2)\left(c_{j,u},c_{j,d}\right)^{\rm t}$ with $x=ja\in\Lambda_{\rm s}$, and chose the gamma matrices $\gamma^0=\sigma^z$, and $\gamma^1=\ii\sigma^y$. The correspondence between the microscopic parameters is 
\beq
\label{eq:parameters}
\theta=\pi,\hspace{2ex} t_{\rm d}=t_{\rm h}, \hspace{2ex}ma=\frac{\Delta\epsilon}{4t_{\rm h}}-1,\hspace{2ex}\frac{g^2}{2N}=\frac{V_{\rm v}}{4t_{\rm h}},
\eeq
and the topological insulator lies in the $\mathsf{BDI}$ class. We note that digital quantum simulations  of the GN model, including other discretizations  such as the staggered-fermion approach,  have also been discussed recently~\cite{PhysRevA.98.012332,li2020simulation,czajka2021quantum}.
  
This connection yields  interesting insights. On the one hand, it shows that the  pseudo-scalar condensate $\Pi_0=\langle \overline{\Psi}(x)\ii\gamma^5{\Psi}(x)\rangle$, where $\gamma^5=\gamma^0\gamma^1=\sigma^x$, corresponds exactly to the order parameter of the aforementioned orbital ferromagnet $ \Pi_0\leftrightarrow M_y$.  This condensate  acquires a non-zero value in the parity-breaking Aoki phase~\cite{PhysRevD.30.2653}, and also plays a role in lattice discretization of quantum chromodynamics~\cite{PhysRevD.58.074501,kaplan2012chiral}.  Accordingly, the numerical results of~\cite{BERMUDEZ2018149} show that the existence of the Aoki phase, typically predicted in the basis of  a large-$N$ expansion~\cite{PhysRevD.30.2653}, 
actually  survives down to the ultimate quantum limit of $N=1$. Moreover, from the perspective of the  phase diagram of the  Creutz-Hubbard  ladder, this Aoki phase is not an artefact of the lattice scaffolding of the continuum fields, but rather a physical phase that delimits the correlated $\mathsf{BDI}$ topological insulator. On the other hand, the connection between these two models also shows that the standard continuum limit, where one recovers the continuum GN field tehory~\cite{PhysRevD.10.3235}, $a\to0$ and $g^2\to 0$, is actually the critical region separating the  trivial and correlated topological insulators, and that it can be used to understand the strongly-coupled features of the topological phase diagram~\cite{PhysRevB.99.125106}.
 
As can be seen in Eq.~\eqref{eq:parameters}, however, the connection to the GN field theory and the interplay of topological phases and lattice discretizations~\eqref{eq:H_GNW}   only apply to the $\pi$-flux regime of the synthetic Creutz-Hubbard ladder.  It thus remains open to {\it (i)} explore the full phase diagram of the model~\eqref{eq:H_C}-\eqref{eq:H_CH} for arbitrary magnetic flux $\theta$, {\it (ii)} find experimentally-accessible observables to characterise it,  and {\it (iii)} unveil its connection to phenomena discussed in a high-energy context. In this work, we pursue this three-fold goal and find the results  summarised in the following subsection.
 
\subsection{Summary of the results}
 \label{sec:summary}
We have found that the Creutz-Hubbard ladder for arbitrary magnetic flux $\theta$~\eqref{eq:H_C}-\eqref{eq:H_CH} has interesting connections between correlated topological phases of matter and  Lorentz-violating QFTs in the SME. Interestingly, both the topological invariant that characterises these phases, and the  coupling strength of  the Lorentz-violating terms, can be controlled by modifying the value of the magnetic flux that pierces the ladder. 
As discussed in Sec.~\ref{sec:CHL_Continuum_Limit}, in the absence of interactions and for a particular set of bare microscopic couplings, the band structure of the model  describes a semi-metal with a single Fermi point at
momentum $k_{+}= \pi/2a$ or, otherwise, at $k_{-}= -\pi/2a$. 
The dispersion relation has a different propagation speed for right- and left-moving excitations, which clearly breaks the emergent Lorentz invariance that appeared in the $\pi$-flux model~\cite{PhysRevX.7.031057}. Remarkably, we find that this Lorentz-breaking mechanism  corresponds to a particular Lorentz violation of the SME, which also breaks parity and time-reversal symmetry~\cite{PhysRevD.58.116002}. In combination with  Hubbard interactions, we show that the continuum limit is described by a Lorentz-violating GN  model with four-Fermi terms that couple the Dirac spinors around each of these two Fermi points. We refer to this continuum limit as a Gross-Neveu model extension (GNME).

In Sec.~\ref{sec:Chiral_Current}, we discuss how the Lorentz violation in the GNME  brings in a new feature. Whereas the $\pi$-flux Creutz-Hubbard ladder cannot support persistent  currents in the groundstate, the breaking of Lorentz invariance due to the asymmetry in the speed of left- and right-moving particles can be responsible for a net current. In particular, in the non-interacting limit, we show that the  groundstates can simultaneously display a non-zero topological invariant and a  large  chiral current circulating around the ladder, both of which are consistent with an inversion symmetry. Using the  susceptibility associated to this chiral current,   we chart the phase diagram of the model, and identify wide regions of parameter space with a  current-carrying topological crystalline groundstate. In the limit of very strong Hubbard interactions, the half-filled ladder is described by an effective spin model that corresponds to the XY model with Dzyaloshinskii-Moriya super-exchange, and  subjected to a transverse  field. Depending on the value of the magnetic flux, one can either find  ferromagnetic or anti-ferromagnetic phases, and  second-order phase transitions that separate them from a disorder paramagnet that is preferred for sufficiently-large transverse fields.

 \begin{figure}[t]
  \centering
  \includegraphics[width=1.05\linewidth]{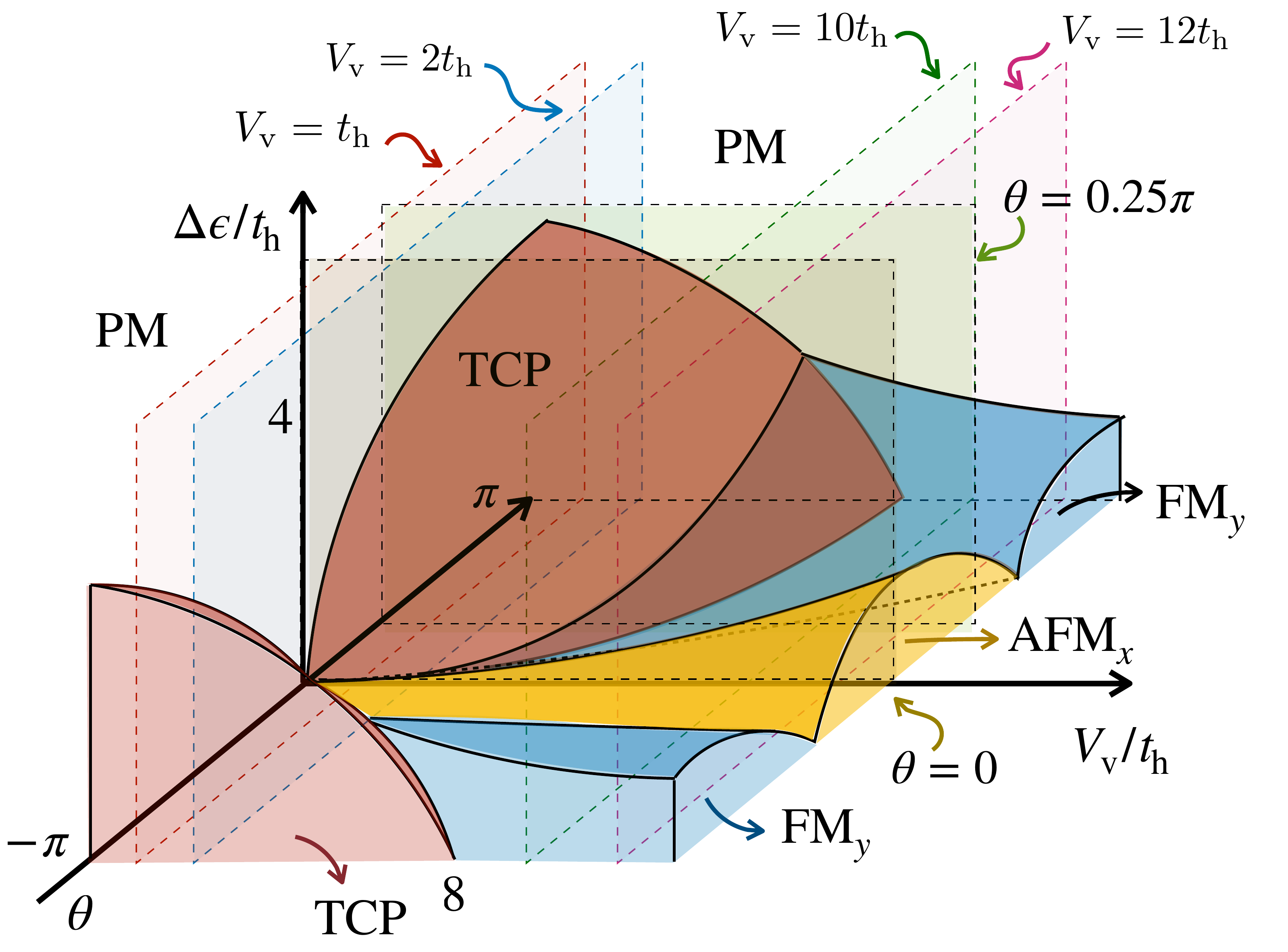}
\caption{\label{fig:scheme_phase_diagram} {\textbf{Schematic phase diagram for  arbitrary fluxes:} Qualitative phase diagram of the Creutz-Hubbard ladder as a function of the Hubbard  interactions $V_{\rm v}$, the energy imbalance $\Delta \epsilon$,  and the magnetic flux $\theta$. At $|\theta|=\pi$, the topological  phase in red is protected by a sublattice symmetry and belongs to the $\mathsf{AIII}$ class of topological insulators, which is surrounded by a topologically-trivial paramagnet (PM) in white, and a symmetry-breaking ferromagnet (FM$_y$) in blue,  and delimited by second-order critical lines. For $|\theta|\lesssim\pi$ }, the sublattice symmetry is broken, and the topological phase is instead protected by inversion symmetry, representing a topological crystalline phase (TCP) which, in contrast to the $\pi$-flux limit, supports a large circulating chiral current. As the magnetic flux is decreased further towards $\theta\approx 0$, the TCP phase is also surrounded by a new symmetry-broken phase, an antiferromagnet (AFM$_x$) in yellow. We also depict shaded planes in parameter space where detailed numerical results are presented in the main text. }
\end{figure}

In Sec.~\ref{section:mf_mps}, we show that the  topological crystalline phases with a non-zero Zak's phase  disappear in favour of these magnetic phases for sufficiently strong interactions.  As a visual aid for the following discussions, we present  a qualitative sketch of the phase diagram containing the various results in Fig.~\ref{fig:scheme_phase_diagram}. In the parameter regime $\Delta \epsilon / 4 t_{\rm h} \in (-1,1)$ and $\theta \in (-\pi,\pi)$, we find that the  Creutz-Hubbard ladder  displays four distinct phases: a paramagnet (PM), a ferromagnet (FM$_y$), an anti-ferromagnet (AFM$_x$), and a topological crystalline phase (TCP).
By using a mean-field reasoning in the effective GNME of the continuum limit, we understand how the effect of interactions contributes to various mass terms for the effective Dirac fermions, which provides a qualitative understanding of the fate of the topological phases when the four-Fermi interactions are increased. We make this discussion quantitative by finding the mean-field parameters self-consistently, and exploring the full phase diagram in detail. Although the specific critical lines are not accurate, we show that the structure of the phase diagram predicted by this self-consistent mean-field agrees with that obtained from a quasi-exact DMRG method.  For weak interactions $V_{\rm v}\ll t_{\rm h} $, we show that the system has
two topological  lobes in  parameter space $(V_{\rm v},\theta,\Delta \epsilon)$. This  topological phase is protected by inversion symmetry, and can be thus identified with a correlated topological crystalline  phase that is surrounded by a trivial 
band insulator, or   by a symmetry broken phase with ferromagnetic
(anti-ferromagnetic) long range order for large (small) fluxes $|\theta|\approx \pi$ ($\theta\approx0$). 
Within these two lobes, the chiral current has a  large absolute value and, despite the Lorentz-invariance breaking, the topological invariant remains to be non-zero  $\gamma=\pm \pi$. The topological nature of this correlated phase is further confirmed by showing a 2-fold degeneracy of the entanglement spectrum, as calculated with our DMRG numerics.
As one increases the interactions, the topological  lobes become smaller, and there is a SSB process where the crystalline topological
phase disappears in favour of the magnetic phases.
Finally, for $V_{\rm v} \gg t_{\rm h}$, we show that the DMRG results agree very well with those obtained from the effective spin model with  anisotropic XY couplings and a Dzyaloshinskii-Moriya  super-exchange.

An important aspect of the current work is that this interplay of topological phenomena, persistent chiral currents, and correlations  in regularised Lorentz-violating QFTs can be realised in experiments of ultra-cold atoms in optical lattices. We show that by modifying recent schemes of   synthetic spin-orbit coupling using Raman optical lattices, one can realise the Creutz-Hubbard ladder with an arbitrary flux, being the latter tunable by adjusting the propagation direction and the frequency of the Raman beams. This scheme avoids the use of state-dependent optical lattices or Floquet-assisted tunnelings in interacting Fermi gases, minimising in this way the effective heating due to residual photon scattering from the auxiliary excited states. In light of the recent experimental progress with Raman optical lattices~\cite{PhysRevLett.121.150401,Songeaao4748,Wu83,liang2021realization}, we believe that the current scheme opens a new direction in the realisation of strongly-coupled QFTs that incorporate Lorentz-violating terms of the SME, such as the GNME studiend in this work. Additionally, the particular lattice regularization explored in this work would allow to explore the interplay of these relativistic models with current-carrying topological phases.


\section{\bf The Gross-Neveu model extension in the  continuum limit} \label{sec:CHL_Continuum_Limit}
    
\begin{figure*}[t]
  \centering
  \includegraphics[width=0.8\linewidth]{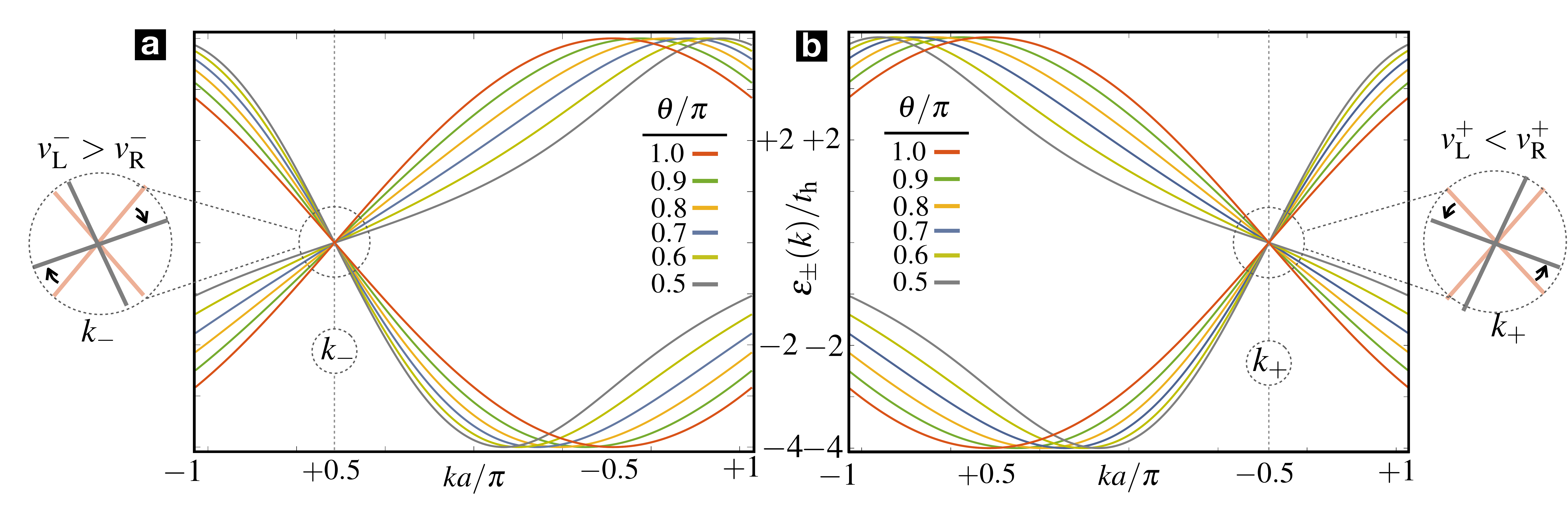}
\caption{\label{fig:band_structure} {\textbf{Band structure and Lorentz-violating Fermi points:} Dispersion relations $\epsilon_{\pm}(k)$~\eqref{eq:bands} for the two bands of the imbalanced Creutz Ladder for different fluxes $\theta\in[\pi/2,\pi]$ (see the coloring labels of the insets). 
{\bf (a)} The flux is set such that $\Delta \epsilon=-4t_{\rm h} \sin (\theta/2)$, and one can see that the low-energy properties are controlled by a massless Dirac fermion at $k_{-}=-\pi/2a$. {\bf (b)} The flux is set such that $\Delta \epsilon=4t_{\rm h} \sin (\theta/2)$, and one can see that the low-energy properties are controlled by the massless Dirac fermion at $k_{+}=\pi/2a$. The left- and right-most insets show that the corresponding Dirac cones have different propagation speeds for right- and left- moving excitations, which read $v^{\pm}_{\rm R}$ and $v^{\pm}_{\rm L}$.}}
\end{figure*}

We start by discussing the continuum limit of the   Creutz-Hubbard ladder~\eqref{eq:H_C}-\eqref{eq:H_CH}  for  arbitrary flux $\theta\in(-\pi,\pi]$.  In the rest of this work, we will set $t_{\rm d}=t_{\rm h}$ and consider half-filling conditions. One can diagonalise the non-interacting part~\eqref{eq:H_C} by a Fourier transform $c_{j,\ell}=\sum_{k\in{\rm BZ}}\ee^{ika j}c_{k,\ell}/\sqrt{N_{\rm s}}$, which yields
\beq
\label{eq:creutz_generic_theta_BZ}
H_{\rm C}=\!\!\sum_{k\in{\rm BZ}}\sum_{\ell,\ell'}c^\dagger_{k,\ell}h_{\ell\ell'}^{\phantom{\dagger}}(k) c^{\phantom{\dagger}}_{k,\ell'}, \hspace{1ex} h_{\ell\ell'}^{\phantom{\dagger}}(k)=\epsilon_0(k)\mathbb{I}_2+\boldsymbol{d}(k)\cdot\boldsymbol{\sigma}^{\phantom{\dagger}},
\eeq
where we have introduced the scalar and vector functions
\beq
\label{eq:non_int_parameters}
\begin{split}
\epsilon_0(k)&=-2t_{\rm h}\cos\left(\frac{\theta}{2}\right)\cos(ka),\\
\boldsymbol{d}(k)&=2t_{\rm d}\cos(ka)\textbf{ e}_x+\!\left(\!\frac{\Delta\epsilon}{2}-2t_{\rm h}\sin\left(\frac{\theta}{2}\right)\!\sin(ka)\!\right)\!\!\!\textbf{ e}_z,
\end{split}
\eeq  
and $\boldsymbol{\sigma}=\sigma^x\textbf{  e}_x+\sigma^y\textbf{  e}_y+\sigma^z\textbf{  e}_z$. Accordingly, the band structure is determined by diagonalizing the corresponding matrix, which yields the following two bands
\beq
\label{eq:bands}
\epsilon_{\pm}(k)=\epsilon_0(k)\pm|\boldsymbol{d}(k)|.
 \eeq
In the  $\theta=\pi$-flux regime, and setting $\Delta\epsilon=4t_{\rm h}$ ($\Delta\epsilon=-4t_{\rm h}$), the band structure describes a semi-metal with a  single Fermi point at $k_{+}=+\pi/2a$ ($k_{-}=-\pi/2a$). As one modifies the flux $\theta\neq\pi$ and, simultaneously, adjusts the imbalance to $\Delta\epsilon=\pm4t_{\rm h}\sin(\theta/2)$, the band structure still contains the single  Fermi points at the same points of the Brillouin zone(see Figs.~\ref{fig:band_structure}  {\bf (a)} and {\bf (b)}). Since the low-energy properties are controlled by particle/hole excitations around those Fermi points, the continuum limit is   obtained by  a long-wavelength approximation around them, where we  define a spinor field
\beq
\label{eq:slowly_varying}
\psi(x)=\frac{1}{\sqrt{a}}(c_{j,u},c_{j,d})^{\rm t}\approx\ee^{\ii k_+a}\Psi_+(x)+\ee^{-\ii k_-a}\Psi_-(x),
\eeq
and separate the rapidly-oscillating parts $\ee^{\ii k_\eta a}$ from the slowly-varying operators $\Psi_\eta(x)$ for each of these Fermi points $\eta\in\{+,-\}$. These operators satisfy the desired  anti-commutation relations in the continuum limit $a\to0$, where $\{\Psi_{\eta,\ell}(x),\Psi^\dagger_{\eta',\ell'}(x')\}=\delta_{\eta,\eta'}\delta_{\ell,\ell'}\delta_{j,j'}/a\to \delta_{\eta,\eta'}\delta_{\ell,\ell'}\delta(x-x')$.

To obtain the QFT that governs this continuum limit, we use Eq.~\eqref{eq:slowly_varying} and perform a gradient expansion for the fields $\Psi_\eta(x\pm a)=\Psi_\eta(x)\pm a\partial_x\Psi_\eta(x)+\mathcal{O}(a^2)$ to find the following lowest-order expressions
\beq
\label{eq:cont_limit_bilinears}
\begin{split}
\psi^\dagger(x)\hat{O}\big(\psi(x+a)+\psi(x-a)\big)&\approx2\ii a\sum_\eta\eta\Psi^\dagger_\eta(x)\hat{O}\partial_x\Psi_\eta(x),\\
\psi^\dagger(x)\hat{O}\big(\psi(x+a)-\psi(x-a)\big)&\approx2\ii\sum_\eta\eta\Psi^\dagger_\eta(x)\hat{O}\Psi_\eta(x),\\
\end{split}
\eeq
for any operator $\hat{O}$ acting on the spinor degrees of freedom. Using these expressions, together with Eq.~\eqref{eq:slowly_varying} and introducing the Hubbard interactions, we find that the continuum limit of the  Creutz-Hubbard ladder~\eqref{eq:H_C}-\eqref{eq:H_CH} can be expressed as
\beq
\begin{split}
\label{eq:cont_limit}
H_{\rm CH}&=\!\int\!{\rm d}x\left(\sum_\eta\overline{\Psi}_\eta(x)\left(-\ii (c\gamma^1_\eta+\tilde{c}_\eta\gamma^0_\eta)\partial_x+m_\eta c^2\right){\Psi}_\eta(x)\!\!\right)\\
&-\!\int\!{\rm d}x\frac{g^2}{2}\left(\sum_\eta\overline{\Psi}_\eta(x){\Psi}_\eta(x)+(-1)^{\frac{x}{a}}\overline{\Psi}_\eta(x){\Psi}_{\overline{\eta}}(x)\right)^{\!\!\!2}.
\end{split}
\eeq  
The first line of this equation can be interpreted  as the Hamiltonian QFT of a massive Dirac field for each of the species $\eta\in\{+,-\}$. These correspond to the so-called fermion doublers in the context of lattice gauge theories, which always yield an even number of species in the continuum limit~\cite{NIELSEN198120,NIELSEN1981173}. The corresponding gamma matrices are
\beq
\gamma^0_\pm=\sigma^z,\hspace{1ex}\gamma^1_\pm=\pm\ii\sigma^y, \hspace{1ex} \gamma^5_\pm=\pm\sigma^x,
\eeq
which shows that the continuum limit around each doubler carries an opposite  chirality. The effective speed of light and mass of the Dirac fermions reads
\beq
\label{eq:parameters_cont_theta}
c=2t_{\rm h}a,\hspace{2ex}m_{\pm}c^2=\frac{\Delta\epsilon}{2}\mp 2t_{\rm h}\sin\left(\frac{\theta}{2}\right).
\eeq
As customary in a Wilson-type discretization~\cite{Wilson1977}, the masses of the fermion doublers are no longer equal. This effect is not specific to the generic flux, as it also occurs in the $\pi$-flux model~\cite{PhysRevX.7.031057},  where it underlies the existence of topological phases of matter. The result  in Eq.~\eqref{eq:parameters_cont_theta} shows that one can achieve different Wilson masses for any flux $\theta\neq 0$, and that we can use the external flux  to tune their relative values. In Fig.~\ref{fig:band_structure} {\bf (a)}, we see that when the flux is set such that $\Delta\epsilon=-4t_{\rm h}\sin(\theta/2)$, only the Dirac fermion at $k_-$ remains massless, which agrees with the predictions in Eq.~\eqref{eq:parameters_cont_theta}. On the other hand,  for $\Delta\epsilon=+4t_{\rm h}\sin(\theta/2)$, it is the massless Dirac fermion at $k_+$ which controls the low-energy properties as depicted in  Fig.~\ref{fig:band_structure} {\bf (b)}, which again agrees with Eq.~\eqref{eq:parameters_cont_theta}.
Note that, in order to use the natural units typical of relativistic quantum field theories, one must set $t_{\rm h}=1/2a$  above.

As one departs from the $\pi$-flux regime, a new term appears in the continuum limit~\eqref{eq:cont_limit} with the new coupling constant
\beq
\label{eq:LV_coupling}
\tilde{c}_\pm=\pm2t_{\rm h}a\cos\left(\frac{\theta}{2}\right).
\eeq
This  coupling  introduces  an explicit  violation of Lorentz symmetry, which actually corresponds to a specific term in the standard model extension~\cite{PhysRevD.58.116002}. Before discussing the origin and consequences of this new term, let us comment on the second line of Eq.~\eqref{eq:cont_limit}, which contains the coupling 
\beq
\frac{g^2}{2}=\frac{V_{\rm v}a}{2}=\frac{V_{\rm v}}{4t_{\rm h}},
\eeq
and is consistent with Eq.~\eqref{eq:parameters} in the  single-flavour limit $N=1$. We can see in Eq.~\eqref{eq:cont_limit} that this interaction term introduces, in addition to  4-Fermi terms for the separate species,  mixing terms due to the back- and Umklapp-scattering that connect the two Fermi points of the original band structure. The appearance of these terms is standard in  other 1D models at half-filling~\cite{Giamarchi:743140}. We note that these interaction-induced couplings between the two doublers play a key role in determining the renormalization-group flows of the microscopic parameters, which can be used to understand the shape of the phase diagram at $\theta=\pi$~\cite{PhysRevB.99.125106}. We thus expect an analogous behaviour for generic fluxes, albeit leading to more complicated flow equations  and  a  richer phase diagram.

Summarizing, we see that besides the new term~\eqref{eq:LV_coupling}, the continuum limit of the synthetic Creutz-Hubbard ladder for arbitrary flux corresponds to a fermion-doubled Gross-Neveu model, albeit with microscopic couplings that depend on the external flux $\theta$. It turns out, however, that this additional term~\eqref{eq:LV_coupling}  changes the physics considerably,  allowing us to explore some sectors  of the SME. In order to unveil this connection, let us find the corresponding  action  $S_{\rm CH}=\int{\rm d}^2x\left(\sum_{\eta}\Pi_\eta\partial_0\Psi_\eta-H_{\rm CH}\right)$, where the canonical momenta for each of the species  are $\Pi_\eta(x)=\ii c\overline{\Psi}_\eta(x)\gamma^0_\eta$, and lead to
\beq
\begin{split}
\label{eq:cont_limit_L}
S_{\rm CH}&=\int\!{\rm d}^2x\left(\sum_\eta\overline{\Psi}_\eta(x)\left(\ii c\Gamma^\mu_\eta\partial_\mu-m_\eta c^2\right){\Psi}_\eta(x)\right)\\
&+\!\int{\rm d}^2x\frac{g^2}{2}\!\left(\sum_\eta\overline{\Psi}_\eta(x){\Psi}_\eta(x)+(-1)^{\frac{x}{a}}\overline{\Psi}_\eta(x){\Psi}_{\overline{\eta}}(x)\right)^{\!\!\!\!2},
\end{split}
\eeq  
where we use Einstein's summation convention for repeated spacetime indices $\mu,\nu,\tau \in\{0,1\}$, and have introduced the modified gamma matrices for each of the doublers
\beq
\label{eq:gamma_Lviolation}
\Gamma^\mu_\eta=\gamma^\mu_\eta+c^{\mu\nu}_\eta g_{\mu\tau}\gamma_{\eta}^\tau.
\eeq
In the context of the SME,  $c^{\mu\nu}$ is a traceless tensor that contains the dominant corrections  to the collisions of unpolarized electrons and positrons due to the  violation of Lorentz invariance~\cite{COLLADAY2001209}. As discussed in the introduction, these terms may arise as an effective microscopic coupling from more fundamental theories at the Planck scale, such as string theory and non-commutative QFTs~\cite{PhysRevD.39.683,PhysRevLett.87.141601}. It is remarkable that these terms arise naturally in the long-wavelength limit of the ladder. In the present context, we get one of these terms  $c^{\mu\nu}_{\eta}$ for each doubler species $\eta\in\{+,-\}$ as a direct consequence of the $\theta\neq\pm\pi$ flux of the 
regularisation~\eqref{eq:H_C}. In fact, we find that
\beq
\label{eq:c_LV}
c^{\mu\nu}_{\pm}=\left\{\begin{matrix}
 \pm \cos(\frac{\theta}{2})& \hspace{2ex}{\rm if}\hspace{1ex} \mu=1,\nu=0 \\
0 & \hspace{-8ex}{\rm else},
\end{matrix}\right.
\eeq
which  becomes non-zero away from the $\pi$-flux regime, such that $\Gamma^0_\eta=\gamma^0_\eta$, and $\Gamma^1_\eta=\gamma^1_\eta+\eta\cos(\frac{\theta}{2})\gamma^0_\eta$. We note that, besides breaking Lorentz invariance, this term also breaks parity and time-reversal symmetry, although $CPT$ is conserved~\cite{COLLADAY2001209}.
As advanced previously, we refer to this continuum limit as a particular type of  Gross-Neveu model extension.

A neat picture of the Lorentz violation caused by this term can be obtained by inspecting the band structure of Eq.~\eqref{eq:bands} for generic fluxes, as  depicted in Figs.~\ref{fig:band_structure}{\bf (a)} and {\bf (b)}. As already discussed in the previous section, when setting $\Delta\epsilon=\pm4t_{\rm h}\sin(\theta/2)$, one obtains a semi-metal with a Fermi point at $k_{\pm}=\pm\pi/2a$. A closer inspection of the bands (see the left- and right-most insets of Figs.~\ref{fig:band_structure}{\bf (a)} and {\bf (b)}) shows that the corresponding Dirac cone has a different propagation speed for right- and left-moving excitations, which reads as follows
\beq
   \label{eq:velocities}
v_{{\rm R}}^{\pm}=2t_{\rm h}a\!\left(\!1\pm\cos\left(\frac{\theta}{2}\right)\!\!\!\right)\!,\hspace{1ex}v_{{\rm L}}^{\pm}=2t_{\rm h}a\!\left(\!1\mp\cos\left(\frac{\theta}{2}\right)\!\!\!\right).
\eeq 
This clearly breaks Lorentz invariance, as  $\Delta\epsilon=\pm4t_{\rm h}\sin(\theta/2)$ implies that the corresponding  mass~\eqref{eq:parameters_cont_theta} vanishes, and the  fermion should thus be travelling at the same speed $c$~\eqref{eq:parameters_cont_theta} in both directions, regardless of the velocity of the frame of an inertial observer. However, for non-zero $c^{\mu\nu}_\eta$, inertial observers moving to the right or left will see the particle travelling at different speeds, such that Lorentz symmetry is broken. 

This velocity difference suggests that there might be non-zero  currents in the groundstate, as particles move faster in one direction than in the other. As shown below, when the ladder has periodic boundary conditions, this flow of particles corresponds to a persistent  chiral  current that circulates in a specific direction set by the sign of the magnetic flux. For open boundary conditions, the persistent chiral current attains a fixed non-zero value at the bulk of the ladder, and gets attenuated as one approaches the left and right boundaries. Moreover, as also discussed below, this chiral current can be used to probe the full phase diagram. 
   
\section{\bf Chiral currents and critical phenomena} \label{sec:Chiral_Current}
     
As discussed  previously, the specific type of Lorentz violation that controls the low-energy properties~\eqref{eq:cont_limit_L} of the model predicts a different propagation speed for right and left-moving particles~\eqref{eq:velocities}. In the underlying lattice discretization, this can yield a net {\it circulating chiral current}  across the ladder
\beq
\label{eq:chiral_current}
J_{\rm c}=\sum_j\left(\ii t_{\rm h}\ee^{\ii\frac{\theta}{2}}c^\dagger_{j+1,{\rm u}}c^{\phantom{\dagger}}_{j,{\rm u}}-\ii t_{\rm h}\ee^{-\ii\frac{\theta}{2}}c^\dagger_{j+1,{\rm d}}c^{\phantom{\dagger}}_{j,{\rm d}}+{\rm H.c.}\right).
\eeq
This current measures the difference of the   right-  and left-wards fermion flows in the upper and lower legs of the ladder $J_{\rm c}=J^{\rightarrow}_{\rm u}-J^{\leftarrow}_{\rm d}$, and thus gives rise to a sense of circulation/chirality (i.e. $\langle J_{\rm c}\rangle>0$  clockwise circulation, and $\langle J_{\rm c}\rangle<0$ anti-clockwise) that connects naturally with the current-carrying edge states of the quantum Hall effect as one increases the number of legs~\cite{PhysRevA.89.023619}. 
     
To avoid possible confusions, let us note that this circulating chiral current is not related to the axial current of  continuum QFTs~\cite{Peskin:1995ev}, which is also  referred to as the chiral current, and is the Noether current associated to axial rotations. Note that for $N=1$~\cite{BERMUDEZ2018149}, the four-Fermi term of the extended Gross-Neveu model~\eqref{eq:H_GNW} can be rewritten as 
\beq
\frac{g^2}{2}\left(\overline{\Psi}(x){\Psi}(x)\right)^2=\frac{g^2}{4}\left(\left(\overline{\Psi}(x){\Psi}(x)\right)^2-\left(\overline{\Psi}(x)\gamma^5{\Psi}(x)\right)^2\right),
\eeq
such that the lattice model~\eqref{eq:H_GNW} including the Lorentz-violating term becomes invariant under continuous axial rotations $\Psi(x)\mapsto\ee^{\ii\alpha\gamma^5}\Psi(x)$, $\forall\alpha\in\mathbb{R}$. In the continuum limit, the Noether current associated to this symmetry is the aforementioned axial current $j_{\rm A}^{\mu}=\overline{\Psi}(x)\gamma^5\gamma^\mu\Psi(x)$, which has a conserved axial charge given by 
\beq
Q_{\rm A}=\int{\rm d}xj_{\rm A}^{0}=\int{\rm d}x\overline{\Psi}(x)\gamma^5\gamma^0\Psi(x).
\eeq
If we calculate the continuum limit of the circulating chiral current~\eqref{eq:chiral_current} using the fermion bilinear expressions of Eq.~\eqref{eq:cont_limit_bilinears}, we find that the leading-order contribution is
\beq
\label{eq:continuum_circulating_current}
J_{\rm c}=2t_{\rm h}\cos\left(\frac{\theta}{2}\right)\int{\rm d}x\overline{\Psi}(x)\Psi(x)+\mathcal{O}(a),
\eeq
which  clearly differs from $Q_{\rm A}$.
We see that away from the $\pi$-flux regime where the above expression~\eqref{eq:continuum_circulating_current} vanishes, the expectation value of the chiral current is proportional to the so-called scalar condensate  $\Sigma_0=\langle \overline{\Psi}(x)\Psi(x)\rangle$, the value of which marks the chiral symmetry breaking by the dynamical mass generation of the standard Gross-Neveu model~\cite{PhysRevD.10.3235}. Therefore, this circulating  current cannot be directly related to the axial current or  to the chiral anomaly due to a background gauge field~\cite{MANTON1985220,AMBJORN1983381}. In spite of this difference, given the transparent interpretation of two chiral windings, namely the anti-clock- and clock-wise circulations, it is customary to refer to this current as the chiral current. 
     
In the following subsection, we shall characterise the phase diagram using this circulating chiral current and its associated susceptibility, which can be combined with the values of a topological invariant   and a strongly-coupled effective theory  to chart the full phase diagram of the model.
     
\subsection{Chiral flows and topological phase transitions}\label{sec:TI_chiral_flow}
     
Using the Hellmann-Feynman theorem, the expectation value of the chiral current can be expressed as the corresponding derivative of the total ground-state energy
\beq
\langle J_{\rm c}\rangle=2\left\langle\frac{\partial H_{\rm CH}}{\partial\theta}\right\rangle=2\frac{\partial E_{\rm gs}}{\partial\theta}.
\eeq
    
This quantity contains useful information about the phase diagram, which can already be seen at the non-interacting level. In this case,
$E_{\rm gs}=\frac{a}{2\pi}\int_{\rm BZ}{\rm d}k\epsilon_-(k)$ is obtained by filling the $N_{\rm s}$ lowest energy states of Eq.~\eqref{eq:bands}, yielding
\beq
\label{eq:chiral_current_free}
\langle J_{\rm c}\rangle=-2t_{\rm h}\cos\left(\frac{\theta}{2}\right)\int_{-\pi}^{\pi}\frac{{\rm d}q}{2\pi}\frac{\frac{\Delta\epsilon}{2}\sin q-2t_{\rm h}\sin(\frac{\theta}{2})\sin^2q}{|\boldsymbol{d}(\frac{q}{a})|},
\eeq
where we have introduced $q=ka$.
Additionally, we shall also be interested in the chiral susceptibility
\beq
\label{eq:chiral_susc}
\chi_{\rm c}=\left\langle\frac{\partial J_{\rm c}}{\partial\theta}\right\rangle=2\left\langle\frac{\partial^2 H_{\rm CH}}{\partial\theta^2}\right\rangle=2\frac{\partial^2E_{\rm gs}}{\partial\theta^2}=\frac{\partial\left\langle J_{\rm c}\right\rangle}{\partial\theta},
\eeq
which can be thus obtained by taking derivatives of the above integral~\eqref{eq:chiral_current_free} with respect to the flux.

\begin{figure}[t]
  \centering
  \includegraphics[width=0.8\linewidth]{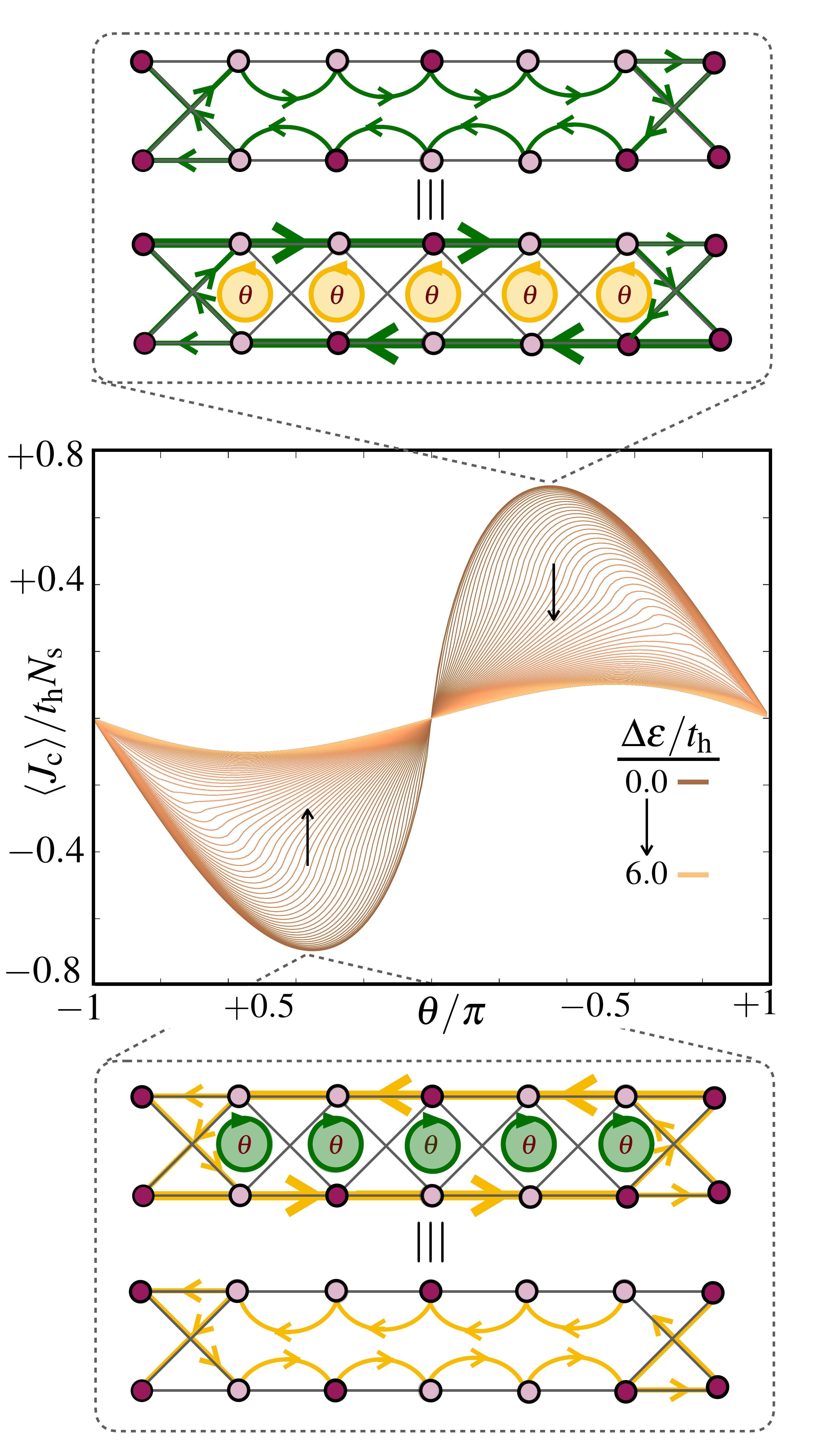}
\caption{\label{fig:chiral_current_theta} {\textbf{Chiral current in the synthetic Creutz ladder:}  the main central panel presents the chiral current as a function of the magnetic flux for different imbalances $\Delta \epsilon$, using a coloring scheme that is specified in the index. In the upper and lower panels, we depict the particle flow clockwise (counter-clockwise) $\langle J_c \rangle >0 $ ($\langle J_c \rangle <0$) for positive (negative) flux $\theta > 0$ ($\theta <0$), and also draw the analog fo the skipping orbits for fermions in a quantum Hall sample.}}
\end{figure}

In the main panel of Fig.~\ref{fig:chiral_current_theta}, we represent the dependence of the chiral current with the magnetic flux for various positive values of the imbalance  $\Delta\epsilon>0$. We observe the characteristic behaviour of a screening current, as also found in  fermionic~\cite{PhysRevB.73.195114} and bosonic~\cite{PhysRevB.64.144515} rectangular ladders: the particles flow clockwise (counter-clockwise) $\langle J_{\rm c}\rangle>0$ ($\langle J_{\rm c}\rangle<0$) for positive (negative) flux $\theta>0$ ($\theta<0$) and, if charged, would thus induce a magnetic field that tends to screen the external one. In the bosonic case, this connects to the Meissner effect of a superfluid, while in the fermionic case it mimics the skipping orbits and edge states in quantum Hall samples (see the upper and lower insets of Fig.~\ref{fig:chiral_current_theta}). This is consistent with the  difference in the velocities of right- and left-moving particles $v_{\rm R}^{+}, v_{\rm L}^{+}$ in Eq.~\eqref{eq:velocities}, which  control the low-energy dynamics in the vicinity of $\Delta\epsilon=4t_{\rm h}\sin(\theta/2)$, which is the  relevant case for the  positive imbalances displayed in the Figure. 

\begin{figure}[t]
  \centering
  \includegraphics[width=0.9\linewidth]{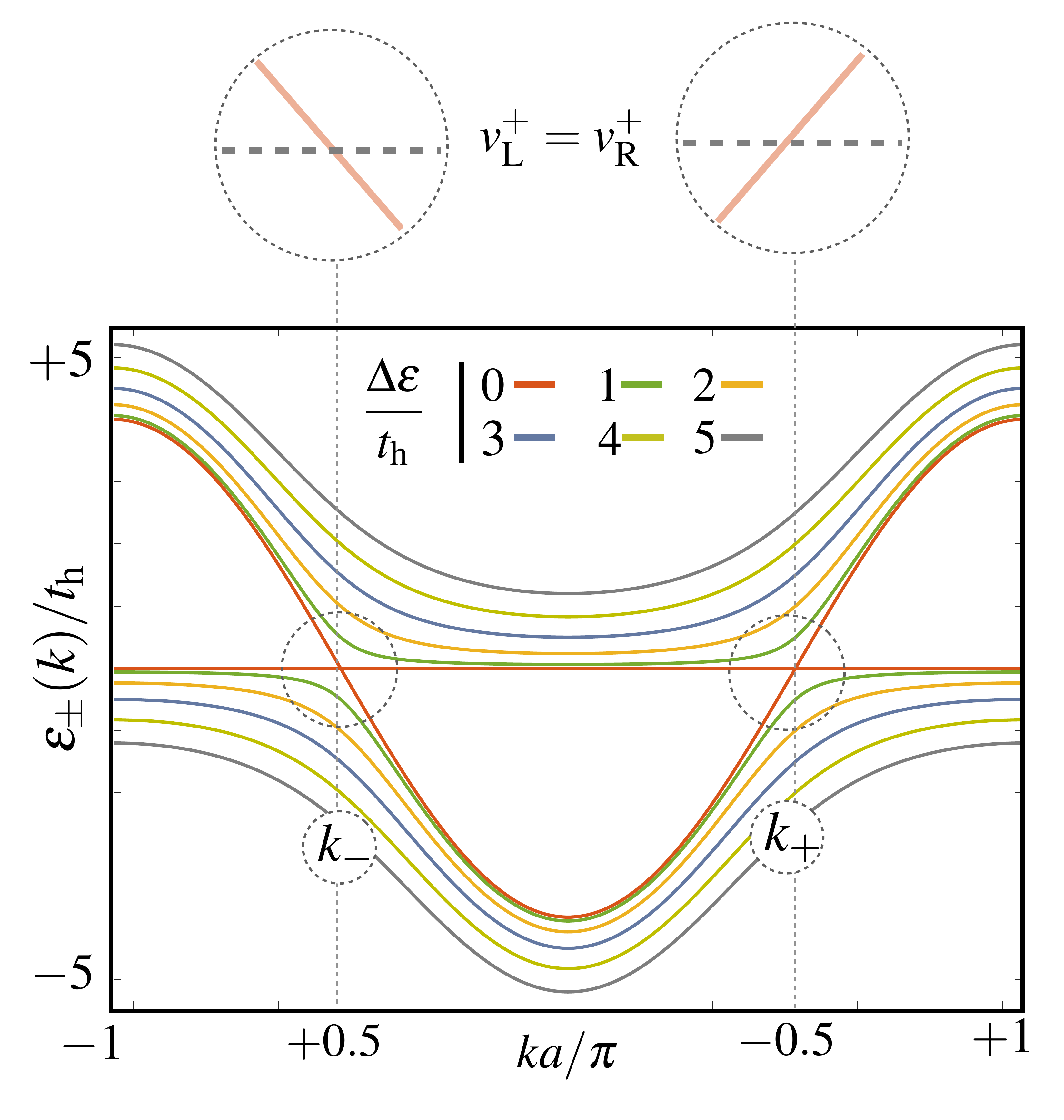}
\caption{\label{fig:bands_theta_0} {\textbf{Bands for the  Creutz ladder for vanishing flux:} Dispersion bands of the imbalanced Creutz ladder for $\theta=0$ for different imbalance $\Delta \epsilon$, with a coloring specified in the inset. For $\theta=0$ and $\Delta \epsilon=0$ (red solid line), the band structure contains a flat and cosine-type band in which the two Fermi points have equal velocities $v^{+}_{\rm R}=v^{-}_{\rm L}$ for right- and left-moving excitations.}}
\end{figure}

We also see in this figure that the chiral flow stops at the time-reversal symmetric points $\theta\in\{-\pi,0,\pi\}$, which is again consistent with both right-and left-moving excitations travelling at the same speed. Note that for a vanishing flux $\theta=0$, the band structure~\eqref{eq:bands} for $\Delta\epsilon=0$ contains a  flat  and a cosine-type band, such that one has to consider the two Fermi points with equal velocities $v_{\rm R}^{+}=v_{\rm L}^{-}$ for the right- and left-moving excitations (see Fig.~\ref{fig:bands_theta_0}). The low-energy theory, in this case, is thus not given by Eq.~\eqref{eq:cont_limit_L} with Eq.~\eqref{eq:c_LV} for $\theta=0$, but rather by the standard Lorentz-invariant  Luttinger liquid for a cosine band, which displays no circulating currents~\cite{Giamarchi:743140}. Accordingly, in both cases $\theta\in\{-\pi,0,\pi\}$, there is no chiral current and we recover the effective Lorentz symmetry.

\begin{figure*}[t]
  \centering
  \includegraphics[width=1\linewidth]{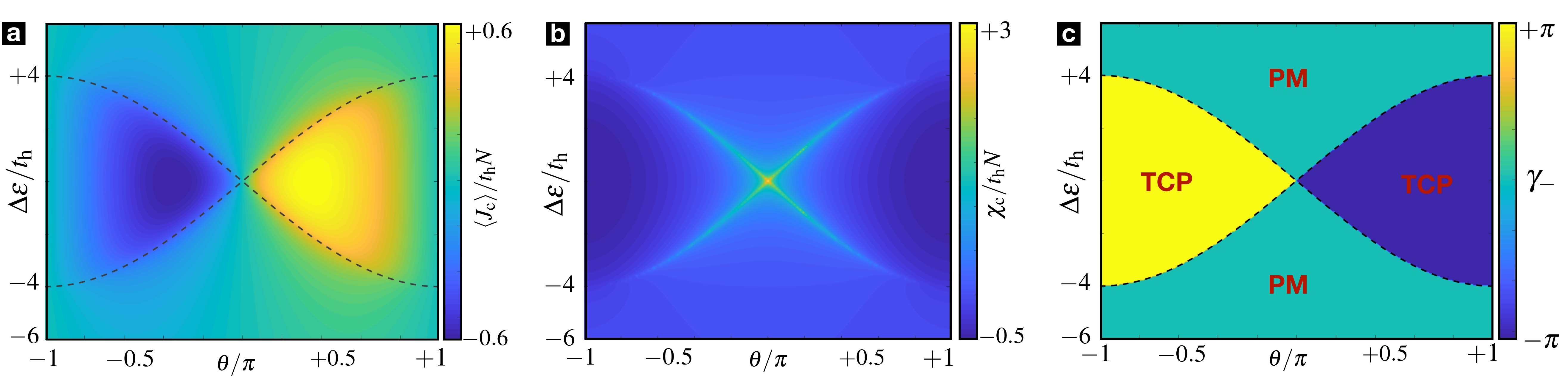}
\caption{\label{fig:phase_diagram_free} {\textbf{Phase diagram of the non-interacting  Creutz ladder:}  {\bf (a)} The persistent circulating chiral current $\langle J_{\rm c}\rangle$ in Eq.~\eqref{eq:chiral_current_free} is represented by a contour plot, and clearly displays a large non-zero value inside  two  lobes that touch at $\theta=0$,  surrounded by  a region with a vanishingly-small particle flow. {\bf (b)} Contour plot for the chiral susceptibility $\chi_{\rm c}$ in Eq.~\eqref{eq:chiral_susc}, the maxima of  which show a divergence with the number of sites $N_{\rm s}$ that delimits the lobes with a large  non-zero chiral current, and marks the onset of a quantum  phase transition. {\bf (c)} Contour plot of Zak's phase in Eq.~\eqref{eq:Zak_phase}
  as a function of the flux and imbalance.
We see that the two lobes with a large chiral flow correspond to the regions where the system is a topological crystalline phase with $\gamma_{\pm}=\pm \pi$, here represented by yellow/blue colors. The phase diagram thus consists of two regions with a topological-crystalline state that supports a large persistent chiral current except for $|\theta|\to \pi$, surrounded by a trivial band insulator with a vanishingly-small chiral current. The dashed black line of {\bf (a)} shows the transition points from the topological to the trivial phases marked by the change of Zak's phase~\eqref{eq:critical_lines_free}, which agree perfectly with the points where the chiral susceptibility diverges.}}
\end{figure*}

For the rectangular Hubbard ladders, the circulating chiral current has a different behaviour. Its dependence with the flux displays a cusp-like behaviour that marks the onset of a Lifshitz transition where the number of Fermi points changes~\cite{PhysRevB.73.195114,PhysRevB.76.195105}. Moreover, as one modifies the filling, the chiral current shows  a linear dependence  prior to the  Lifshitz transition that can be used to extract information about an underlying ladder-version of the Laughlin state in fractional quantum Hall phases~\cite{PhysRevB.92.115446,PhysRevX.7.021033}. Although we do not find this exact phenomenology for  the  cross-linked ladder, the chiral current  still carries information about phase transitions, albeit of a different nature. In order to see these effects, we now study  the dependence of the chiral susceptibility with the magnetic flux. It turns out that the chiral susceptibility shows a clear divergence as $N_{\rm s}\to\infty$ that coincides with the values of the imbalance  $\Delta\epsilon=\pm 4t_{\rm h}\sin(\theta/2)$, where the  masses of the Dirac fermions~\eqref{eq:parameters_cont_theta} vanish. This divergence shall  thus mark the onset of a gap closure and a second-order phase transition instead of the aforementioned Lifshitz one.
     
In Figs.~\ref{fig:phase_diagram_free} {\bf (a)} and {\bf (b)}, we show the contour plots of  the chiral current and the chiral susceptibility, respectively,  as a function of $(\theta/\pi, \Delta\epsilon/t_{\rm h})$. The chiral current shows much larger absolute values within the two lobes (see Fig.~\ref{fig:phase_diagram_free} {\bf (a)}), and the  maxima of the susceptibility serves to delimit the lobes and marks the phase transition (see Fig.~\ref{fig:phase_diagram_free} {\bf (a)}). 

To   understand better the nature of this phase transition, we now discuss the values of the Zak phase~\cite{PhysRevLett.62.2747} for arbitrary magnetic fluxes. This topological invariant, briefly discussed  in the introduction, serves to characterise the topological phases of  the Creutz ladder in the $\pi$-flux  regime~\cite{PhysRevX.7.031057}. For generic fluxes, we find that the additional term $\epsilon_0(k)$  breaking particle-hole symmetry does not contribute to the Berry connection, such that
\beq
\gamma_{\pm}=\mp\int_{\rm BZ}{\rm d}k\frac{{\bf e}_y\cdot(\boldsymbol{d}(k)\times\partial_k\boldsymbol{d}(k))}{2|\boldsymbol{d}(k)|^2},
\eeq
only depends on  the Brillouin-zone vector field $\boldsymbol{d}(k)$  introduced in Eq.~\eqref{eq:non_int_parameters}. We note that the contribution of the Berry connection to this topological invariant is actually concentrated around the two Fermi points $k_{\pm}=\pm\pi/2a$  around which we performed the long-wavelength approximation~\eqref{eq:cont_limit}. This allows to calculate the integrals analytically,  yielding
\beq
\label{eq:Zak_phase}
  \gamma_{\pm}=\mp\frac{\pi}{2}\sum_{\eta=\pm} \!\eta{\rm sgn}\left(\frac{\Delta\epsilon}{2}-\eta2t_{\rm h}\sin\left(\frac{\theta}{2}\right)\!\!\right),
\eeq
which is  the difference of the signs of the two fermion masses  in Eq.~\eqref{eq:parameters_cont_theta}, $   \gamma_{\pm}=\pm\frac{\pi}{2}({\rm sgn}(m_-)-{\rm sgn}(m_+))$. This expression  is analogous to the results  found for higher-dimensional domain-wall fermions~\cite{GOLTERMAN1993219}, and leads to the following critical lines that separate topological from trivial band insulators in the half-filled ladder
\beq
\label{eq:critical_lines_free}
\gamma_-=\left\{   \begin{matrix} 
      0, \hspace{1.5ex} {\rm if}\,\,\,\,\,\,\Delta\epsilon>4t_{\rm h}\left|\sin\left(\frac{\theta}{2}\right)\!\right|,\hspace{3ex} \phantom{,} \\
      \pi,   \hspace{1.5ex} {\rm if}\,\,\,\,\,\,0<\Delta\epsilon<4t_{\rm h}\left|\sin\left(\frac{\theta}{2}\right)\!\right|,\\
     -\pi,   \hspace{1.5ex} {\rm if}\,\,\,\,\,\,0>\Delta\epsilon>-4t_{\rm h}\left|\sin\left(\frac{\theta}{2}\right)\!\right|,\\
            0, \hspace{1.5ex} {\rm if}\,\,\,\,\,\,\Delta\epsilon<-4t_{\rm h}\left|\sin\left(\frac{\theta}{2}\right)\!\right|.\hspace{2ex} \phantom{,} \\
   \end{matrix}\right.
\eeq
Our results show that, in spite of our departure from the $\pi$-flux limit, and the associated breaking of Lorentz-invariance~\eqref{eq:gamma_Lviolation}-\eqref{eq:c_LV}, one can still find a topological phase with a quantised non-zero Zak's phase in certain regions of parameter space. 

This  phase  hosts   edge states localised to the left/right boundaries of the ladder. In the $ \pi$-flux limit~\cite{PhysRevX.7.031057}, where the single-particle Hamiltonian~\eqref{eq:creutz_generic_theta_BZ} belongs to the $\mathsf{AIII}$ class $\{h(k),\sigma^y\}=0$,   the edge states are constrained to be zero-energy modes. In the present case,  the underlying  sub-lattice symmetry is broken, as $\{h(k),\sigma^y\}\neq 0$ for $\theta\neq \pi$. As a consequence, the topological phase is no longer within the  $\mathsf{AIII}$ class, such that   the edge states are no longer restricted to have zero energies. Even if no global symmetry is responsible for the protection of this topological phase, one can find a point-like  inversion symmetry  
 \beq
 c_{i,{\rm u}}\mapsto c_{N_{\rm s}-i,{\rm d}},\hspace{1ex}  c_{i,{\rm d}}\mapsto c_{N_{\rm s}-i,{\rm u}},
 \eeq
under which the Hamiltonian remains invariant $H_{\rm CH}\mapsto H_{\rm CH}$. Accordingly,  groundstates with a  quantized  Zak phase can be understood as  instances of a {\it topological crystalline phase} (TCP)~\cite{fu2011topological,ando2015topological,isobe2015theory}. Note that the above   circulating chiral current~\eqref{eq:chiral_current} is also preserved under this ladder inversion  $J_{\rm c}\mapsto J_{\rm c}$. The groundstate can thus display simultaneously a non-zero topological invariant protected by inversion symmetry, and a non-zero persistent circulating current, departing in this way from other topological crystalline insulators. Note that, in contrast to the two-dimensional quantum Hall effect, this circulating current is not carried by the edge states, which remain localised at the left/right boundaries.  In  Fig.~\ref{fig:phase_diagram_free} {\bf (c)}, we represent  the  contour plot of Zak's phase~\eqref{eq:Zak_phase} as a function of $(\theta, \Delta\epsilon/t_{\rm h})$. By comparing to the contour plot  of the chiral current in Fig.~\ref{fig:phase_diagram_free} {\bf (a)}, we see that the two lobes with a large circulating chiral flow  correspond to  regimes with non-zero  $\gamma_{\pm}=\pm\pi$,  in which the groundstate is a topological phase. As $\theta\to0$, the topological insulator regions give way to a trivial band insulator with a vanishingly-small current. On the other hand, as $\theta\to\pm\pi$ towards the other time-reversal configurations, the circulating chiral current vanishes while the groundstate still supports a non-zero Zak's phase that ultimately yields the $\pi$-flux results of~\cite{PhysRevX.7.031057}.  In the following sections, we  explore how this picture gets modified as we include  Hubbard interactions, starting from strong interactions.
      
\subsection{Dzyaloshinskii-Moriya super-exchange}\label{sec:Strong_Coupling_Sec}

In this section, we discuss the nature of the phase diagram in the strong-coupling limit $V_{\rm v}\gg t_{\rm h}$ (i.e. $g^2\gg 1$ in the language of the Gross-Neveu model~\eqref{eq:parameters}). As mentioned in the introduction,  in the strong-coupling regime,  the half-filled rectangular Hubbard ladder  for spin-full fermions can be described in terms of an effective anti-ferromagnetic Heisenberg ladder, where the origin of the spin-spin interactions can be understood as a super-exchange mechanism~\cite{PhysRev.79.350}. As shown in~\cite{PhysRevX.7.031057}, the strong-coupling limit changes considerably for the cross-linked geometry in the  $\pi$-flux limit. Here, one obtains ferromagnetic Ising interactions with a $\mathbb{Z}_2$ symmetry  instead of the $SU(2)$-invariant Heisenberg couplings. In the Ising case, one finds a symmetry-broken  ferromagnet with magnetization $ M_y\neq 0$ or, equivalently~\cite{BERMUDEZ2018149}, a pseudo-scalar condensate $ \Pi_0\neq 0$ in the language of the Gross-Neveu model. In both cases, there is a spontaneous symmetry breaking (SSB) process where the corresponding $\mathsf{AIII}$ topological insulator gives way to this long-range-ordered phase that breaks the $\mathbb{Z}_2$ symmetry. Since we have shown above  that a current-carrying topological phase also appears for generic fluxes $\theta\neq0$, it is interesting to explore if there is a similar SSB where it disappears in favour of a long-range order.
         
In the strong-coupling regime, due to the nature of the synthetic Hubbard interactions, the low-energy subspace will be spanned by fermion configurations that avoid  double occupancies of two vertically-neighbouring sites of the ladder, as these doublons pay a large energy penalty. The leading-order dynamics comes from second-order processes $\mathcal{O}(t_{\rm h}^2/V_{\rm v})$ where fermions tunnel back and forth  virtually populating these  doublons. At half-filling, all the possible second-order processes can be written in terms  of  orbital spin operators
\beq
\begin{split}
\label{eq:DM_operators}
T_i^x=\half\left(c_{i,{\rm u}}^\dagger c_{i,{\rm d}}^{\phantom{\dagger}}+c_{i,{\rm d}}^\dagger c_{i,{\rm u}}^{\phantom{\dagger}}\right),\\
T_i^y=\textstyle{\frac{\ii}{2}}\left( c_{i,{\rm d}}^\dagger c_{i,{\rm u}}^{\phantom{\dagger}}- c_{i,{\rm u}}^\dagger c_{i,{\rm d}}^{\phantom{\dagger}}\right),\\
   T_i^z=\half\left(c_{i,{\rm u}}^\dagger c_{i,{\rm u}}^{\phantom{\dagger}}-c_{i,{\rm d}}^\dagger c_{i,{\rm d}}^{\phantom{\dagger}}\right).\\
\end{split}
\eeq   
The second-order super-exchange processes lead to the following effective Hamiltonian of coupled orbital spins
\beq
\begin{split}
\label{eq:DM_interaction}
H_{\rm DM}=&\sum_i\bigg(J(1+\xi)T_i^xT_{i+1}^x+J(1-\xi)T_i^yT_{i+1}^y\bigg)\\
+&\sum_i\bigg(\boldsymbol{D}\cdot(\boldsymbol{T}_i\times\boldsymbol{T}_{i+1})+hT_i^z\bigg),
\end{split}
\eeq    
where we have introduced the  couplings
\beq
\label{eq:DM_parameters}
J=\frac{4t_{\rm h}^2}{V_{\rm v}}\cos\theta,\hspace{1.25ex}\xi=\sec\theta,\hspace{1.25ex}\boldsymbol{D}=-\frac{4t_{\rm h}^2}{V_{\rm v}}\sin\theta\textbf{e}_z,\hspace{1.25ex} h=\Delta\epsilon.
\eeq
The first line of Eq.~\eqref{eq:DM_interaction} represents the so-called  XY model~\cite{LIEB1961407,PhysRevA.3.786} with  spin-spin coupling strength $J$ and  anisotropy parameter $\xi$ given in Eq.~\eqref{eq:DM_parameters}. The second-line of Eq.~\eqref{eq:DM_interaction}  contains a so-called {\it  Dzyaloshinskii-Moriya interaction}~\cite{PhysRev.120.91,DZYALOSHINSKY1958241}, which contains an anti-symmetric super-exchange with  the coupling  $\boldsymbol{D}$ given in Eq.~\eqref{eq:DM_parameters}. Additionaly, there is  an external transverse field $h$ also defined in Eq.~\eqref{eq:DM_parameters}. 

Let us note that in the $\pi$-flux regime $\theta=\pi$, the  Dzyaloshinskii-Moriya coupling  and the $xx$ interactions vanish $\boldsymbol{D}=\boldsymbol{0}$, $J(1+\xi)=0$, while the $yy$ interactions  $J(1-\xi)=-8t_{\rm h}^2/V_{\rm v}$ lead to the aforementioned Ising coupling, which favours a ferromagnetic long-range order along the $y$-axis, which coincides with the results of~\cite{PhysRevX.7.031057}. Conversely, for vanishing flux $\theta=0$, it is $\boldsymbol{D}=\boldsymbol{0}$, $J(1-\xi)=0$ which vanish, while the $xx$ interactions $J(1+\xi)=+8t_{\rm h}^2/V_{\rm v}$ favour an anti-ferromagnetic long-range order.   We thus expect that, as one varies the external magnetic flux, the direction and character of the long-range order within the $xy$ plane will be modified. 

Fortunately, the effective model turns out to be exactly solvable via  Jordan-Wigner~\cite{Jordan1928}  and  Bogoliubov~\cite{Bogoljubov1958} transformations. The first one allows us to express the orbital spin model in terms of a   chain of spinless fermions, which in momentum space yields
\beq
\label{eq:JW_DM}
H_{\rm DM}=\sum_{k\in{\rm HBZ}}\boldsymbol{\chi}_k^\dagger
\left(\tilde{\epsilon}_0(k)\mathbb{I}_2+\tilde{\boldsymbol{d}}(k)\cdot\boldsymbol{\sigma}\right)\boldsymbol{\chi}_k^{\phantom{\dagger}}.
\eeq 
Here, we have introduced the Nambu spinor $\boldsymbol{\chi}_k=(f_k^\dagger,f_{-k}^{\phantom{\dagger}})^{\rm t}$, which is expressed in terms of the new creation-annihilation operators of the Jordan-Wigner fermions within the halved Brillouin zone ${\rm HBZ}=[0,\pi/a]$. The quadratic Hamiltonian in Eq.~\eqref{eq:JW_DM} depends on the scalar and vector functions
\beq
\begin{split}
  \tilde{\epsilon}_0(k)&=\frac{4t_{\rm h}^2}{V_{\rm v}}\sin\theta\sin(ka),\\
  \boldsymbol{\tilde{d}}(k)&=\frac{4t_{\rm h}^2}{V_{\rm v}}\sin(ka)\textbf{ e}_y+\left(\Delta\epsilon-\frac{4t_{\rm h}^2}{V_{\rm v}}\cos\theta\cos(ka)\right)\textbf{ e}_z,
\end{split}
\eeq
which play an analogous role to those defined in Eq.~\eqref{eq:non_int_parameters} for the non-interacting limit of the Creutz-Hubbard ladder. Note, however, that the number of spinless Jordan-Wigner fermions is not conserved, as the transformed spin model is not invariant under $U(1)$ symmetry. Applying now a Bogoliubov transformation  yields two energy bands analogous to Eq.~\eqref{eq:bands}
\beq
\label{eq:bands_spins}
\tilde{\epsilon}_{\pm}(k)=\tilde{\epsilon}_0(k)\pm|\boldsymbol{\tilde{d}}(k)|.
\eeq
The groundstate energy, in this case, will be determined by filling all the momenta for which $  \tilde{\epsilon}_{\pm}(k)<0$, which need not be $N_{\rm s}$ fermions as occurs for the original half-filled ladder with $U(1)$ symmetry. Therefore, the presence of a gap-closing phase transition  can be predicted by finding the set of microscopic parameters that allow for  $ \tilde{\epsilon}_{\pm}(k_0)=0$, for some $k_0\in{\rm HBZ}$. One can only find a  solution of this  equation  for a specific momentum $k_0\in\{0,\pi/a\}$ if for  $\Delta\epsilon=\mp(4t_{\rm h}^2/V_{\rm v})\cos\theta$. Accordingly,  we identify two lines of second-order phase transitions at
\beq \label{eq:strong_coupling_critical_line}
   \Delta\epsilon_{\rm c}(\theta)=\pm \frac{4t_{\rm h}^2}{V_{\rm v}}|\cos\theta|.
\eeq
          
\begin{figure*}[t]
  \centering
  \includegraphics[width=1\linewidth]{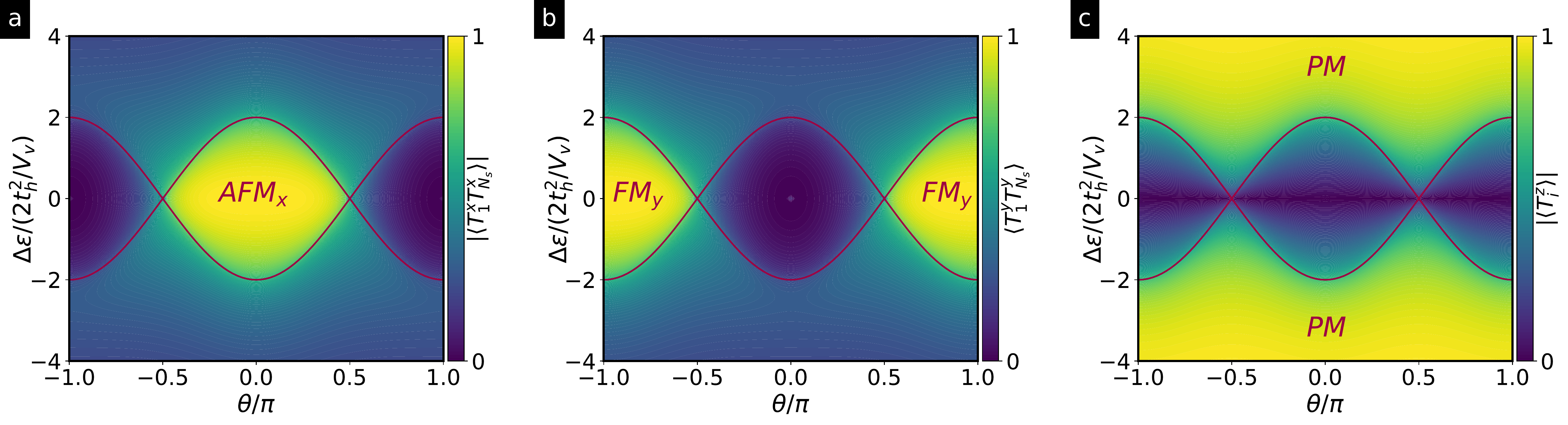}
\caption{\label{fig:phase_diagram_strong} {\textbf{Phase diagram of the  Creutz ladder for strong interactions:}  Phase diagram displaying three different phases: ferromagnetic FM$_y$, anti-ferromagnetic AFM$_x$, and paramagnetic PM phases. The horizontal axis represents the flux per plaquette, whereas the vertical axis corresponds to the ratio of the energy imbalance to the super-exchange strength $J$.  {\bf (a)} The correlator $\langle T^x_1 T^x_{N_{\rm s}} \rangle$  is represented by a contour plot,  quantitatively  distinguishing a central yellow  lobe where this correlator displays AFM$_x$ long-range order. {\bf (b)} We also represent the correlator $\langle T^y_1 T^y_{N_{\rm s}} \rangle$, which instead captures two lateral yellow lobes displaying FM$_y$ long-range order. {\bf (c)} The contour plot of the magnetization along the $z$ direction $|\langle T^z \rangle|$ captures the paramagnetic phases, here represented by the yellow surrounding regions. The red lines correspond to the critical line in Eq.~\eqref{eq:strong_coupling_critical_line}, and perfectly delimit the regions where the above observables characterise the various phases.}}
\end{figure*}
 
Let us now discuss the different phases that are delimited by these critical lines. In addition to the aforementioned SSB ferromagnetic and anti-ferromagnetic phases, the limit $h\gg |J|, |\boldsymbol{D}|$ can be easily understood. In this case, the orbital spins align in the opposite direction of the transverse field~\eqref{eq:DM_interaction}, which can be understood as an orbital paramagnet in which the original fermions are localised to the sites of the lower leg of the ladder. In Fig.~\ref{fig:phase_diagram_strong}, we present some observables that allow to understand how these three different phases are arranged in this strong-coupling limit: {\it (i)} an anti-ferromagnet (AFM$_x)$ aligned along the $x$ axis occurs for $ \Delta\epsilon<\Delta\epsilon_{\rm c}(\theta)$ and $\theta\in(-\pi/2,\pi/2)$, {\it (ii)} a ferromagnet (FM$_y)$ aligned along the $y$ axis occurs for $ \Delta\epsilon<\Delta\epsilon_{\rm c}(\theta)$ and $\theta\in[-\pi,-\pi/2)\cup(\pi/2,\pi]$, and {\it (iii)} a paramagnet (PM) with all spins anti-alligned with the transverse field appears for $ \Delta\epsilon>\Delta\epsilon_{\rm c}(\theta)$ for any flux $\theta\in[-\pi,\pi]$ for sufficiently-large transverse fields.  In Fig.~\ref{fig:phase_diagram_strong} {\bf (a)}, we represent the two-point correlations functions $\langle T_1^xT_{N_{\rm s}}^x\rangle$ in the groundstate of the orbital spin model, which tend to the AFM order parameter in the thermodynamic limit $|\langle T_1^xT_{N_{\rm s}}^x\rangle|\to\langle T_1^x\rangle^2=M_x^2$, which clearly displays a non-zero value  in the inner yellow lobe. In Fig.~\ref{fig:phase_diagram_strong} {\bf (b)}, we represent the two-point correlations functions $\langle T_1^yT_{N_{\rm s}}^y\rangle$ in the groundstate of the orbital spin model, which tend to the FM order parameter in the thermodynamic limit $|\langle T_1^yT_{N_{\rm s}}^y\rangle|\to\langle T_1^y\rangle^2=M_y^2$, which clearly displays a non-zero value  in two yellow lobes that touch the AFM$_x$ phase at $\theta=\pm\pi/2$. 
Finally, in Fig.~\ref{fig:phase_diagram_strong} {\bf (c)}, we represent the  transverse magnetization $|\langle T_i^z\rangle|$, which becomes  larger in the PM regions around the  orbital magnets.

As noted at the end of the previous section, the topological crystalline phases supporting a persistent chiral current at zero interactions should disappear in favor of orbital magnetic phases as one increases the Hubbard interactions. The current strong-coupling analysis shows that these orbital  magnets correspond to an AFM$_x$ or FM$_y$ depending on the magnetic flux. Additionally, the trivial band insulators that appear in the non-interacting limit  should be adiabatically connected to the PM phases, as they both describe a limiting case in which all fermions occupy the sites of the lower leg for a very large energy  imbalance. The question that will be addressed in the following section is how these two limiting cases are connected, and what is the nature of the quantum phase transitions that appear along the way, and how the chiral current/susceptibility can be used to characterise them.
 
\section{\bf Phase diagram:  self-consistent Hartree-Fock method and variational matrix-product states}
\label{section:mf_mps}

To address the question raised at the end of the previous section,  and explore the full phase diagram of the model in the $\{\Delta \epsilon, \theta, V_{\rm v}\}$ volume, we  use two different methods: {\it (i)} a self-consistent mean-field approximation~\cite{bruus_flensberg_2017}, and {\it (ii)} a variational method based on matrix-product states (MPS)~\cite{RevModPhys.77.259}. 

As regards the first one, we start from the Hubbard interactions (\ref{eq:H_CH}) and perform the Hartree-Fock decoupling 
\beq
\begin{split}
  \frac{1}{2}\sum_{\ell}c^{{\dagger}}_{i,\overline{\ell}\phantom{\overline{\ell}}\!\!}c^\dagger_{i,\ell\phantom{\overline{\ell}}\!\!\!}c^{\phantom{\dagger}}_{i,\ell\phantom{\overline{\ell}}\!\!\!}c^{\phantom{\dagger}}_{i,\overline{\ell}}&\approx  \langle {c}^{\dagger}_{i,{\rm u}} {c}^{\phantom{\dagger}}_{i,{\rm u}}  \rangle {c}^{\dagger}_{i,{\rm d}} {c}^{\phantom{\dagger}}_{i,{\rm d}} + \langle {c}^{\dagger}_{i,{\rm d}} {c}^{\phantom{\dagger}}_{i,{\rm d}}  \rangle {c}^{\dagger}_{i,{\rm u}} {c}^{\phantom{\dagger}}_{i,{\rm u}}  \\
&- \langle {c}^{\dagger}_{i,{\rm u}} {c}^{\phantom{\dagger}}_{i,{\rm d}}  \rangle {c}^{\dagger}_{i,{\rm d}} {c}^{\phantom{\dagger}}_{i,{\rm u}} - \langle {c}^{\dagger}_{i,{\rm d}} {c}^{\phantom{\dagger}}_{i,{\rm u}}  \rangle {c}^{\dagger}_{i,{\rm u}} {c}^{\phantom{\dagger}}_{i,{\rm d}},
\end{split}
\eeq
where we have neglected a $c$-number  that will contribute equally to the energies of all possible eigenstates of the mean-field Hamiltonian.
After this Hartree-Fock decoupling, the Hamiltonian~\eqref{eq:H_C}-\eqref{eq:H_CH}  is expressed as the sum of quadratic terms
\beq
\label{eq:MF_Ham}
\begin{split}
{H}^{\rm mf}_{\rm CH} = \sum_{i,\ell} &\left(t_{\rm h}\ee^{-\ii \frac{s_\ell\theta}{2}}c^\dagger_{i+1,\ell}c^{\phantom{\dagger}}_{i,\ell}+t_{\rm d}c^\dagger_{i+1,\ell}c_{i,\overline{\ell}}- \frac{s_\ell\Delta\epsilon}{2} c^\dagger_{i,\ell}c^{\phantom{\dagger}}_{i,\ell}\right.\\
&+ \left.\frac{1}{2}\epsilon_{\ell} \big(\{n_{i,\overline{\ell}}\}\big)c^\dagger_{i,\ell\phantom{\overline{\ell}}\!\!\!}c^{\phantom{\dagger}}_{i,\ell\phantom{\overline{\ell}}\!\!\!}+t_{\ell}\big(\{b_{i,\overline{\ell}}\}\big) {c}^{\dagger}_{i,\ell\phantom{\overline{\ell}}\!\!\!} {c}^{\phantom{\dagger}}_{i,\overline{\ell}} +{\rm H.c.} \right).
\end{split}
\eeq  
Here, we have introduced  a  shift of the on-site energies $\epsilon_\ell$ that depends on the local fermion numbers $n_{i,\overline{\ell}}=\langle c^\dagger_{i,\overline{\ell}
}c^{\phantom{\dagger}}_{i,\overline{\ell}} \rangle$, together with a vertical tunneling  term $t_\ell$ that depends on the bond densities $b_{i,\overline{\ell}}=\langle c^\dagger_{i,\overline{\ell}}c^{\phantom{\dagger}}_{i,\ell\phantom{\overline{\ell}}\!\!\!} \rangle$. The former, which is always a real number, will play an important role when the fermions distribute differently along the two legs of the ladder;  whereas the later, which can be a complex number,   shall  determine how the fermions delocalise over the vertical links of the ladder. These parameters are both proportional to the Hubbard interaction 
\begin{equation}
\label{eq:mf_parameters}
\epsilon_{\ell} \big(\{n_{i,\overline{\ell}}\}\big)=V_{\rm v}\left\langle c^\dagger_{i,\overline{\ell}}c^{\phantom{\dagger}}_{i,\overline{\ell}} \right\rangle,\hspace{2ex}t_{\ell} \big(\{b_{i,\overline{\ell}}\}\big)=V_{\rm v}\left\langle c^\dagger_{i,\overline{\ell}}c^{\phantom{\dagger}}_{i,{\ell\phantom{\overline{\ell}}\!\!\!}} \right\rangle.
\end{equation}
The on-site energy shifts will renormalise the energy imbalance, changing in accordance the extent of the non-interacting phases. In addition, the vertical tunnelings can change the physics considerably, as they modify the effective connectivity of the ladder. In the non-interacting case, inter-leg tunnelings can only occur via cross-link hoppings whereas, in the presence of interactions, these mean-field terms can allow for such hoppings occurring vertically. 

We note that the mean-field approximation requires the observables $\left\{ {n}_{i,\ell},  {b}_{i,\ell}   \right\}^N_{i=1}$ to be determined self-consistently, and we must deal with a number of self-consistency equations that grows linearly with the number of sites. 
The self-consistent loop consists in the following steps: 
{\it (i)} we start by setting an initial configuration of densities $\left\{ {n}_{i,\ell} \right\}$ and bond densities $\{b_{i,\ell}\}$ for different phases. As regards the AFM$_{x}$ ordering, we have chosen a pattern of real numbers with alternating signs for bond densities $\{b_{i,\ell}\}$ and alternation of $0$ and $1$ for $\left\{ {n}_{i,\ell} \right\}$ such that   
$\sum_i \left(n_{i,{\rm u}}+n_{i,{\rm d}}\right)=N_{\rm s}$. Instead, when the target is the topological crystalline phase or the FM$_y$, it is better to start from  a translationally-invariant pattern of complex numbers for $\{b_{i,\ell}\}$.  
Moreover, for the PM phase, we start from a translationally-invariant pattern of real numbers for $\{b_{i,\ell}\}$ and we fixed $n_{i,{\rm u}}=1$ and $n_{i,{\rm d}}=0$ $\forall i$. 
{\it (ii)} Given these initial seeds,  we diagonalise the fermionic tight-binding model defined in Eq. (\ref{eq:MF_Ham}), and get another estimate of the mean-field parameters~\eqref{eq:mf_parameters}. We then iterate this process until  the self-consistent parameters  and the energy converge with an error of $10^{-8}$.

Let us now discuss these different mean-field configurations at the level of the continuum QFT~\eqref{eq:cont_limit_bilinears}. Assuming 
 a translationally-invariant half-filled groundstate, one finds that an occupation imbalance on the legs  $\Delta n_i=n_{i,{\rm u}}-n_{i,{\rm d}}=\Delta n,\,\forall i $ shifts the Wilson masses in Eq.~\eqref{eq:parameters_cont_theta} as follows
\beq
\label{eq:pwilson_mass_mean_field}
m_{\pm}c^2\mapsto \tilde{m}_{\pm}c^2=\frac{\Delta\epsilon-V_{\rm v}\Delta n}{2}\mp 2t_{\rm h}\sin\left(\frac{\theta}{2}\right),
\eeq
This can readily be interpreted as the aforementioned renormalisation of the energy imbalance $\Delta\epsilon\to\Delta\tilde{\epsilon}$. A positive bare energy imbalance $\Delta\epsilon>0$ favors the occupation of fermions on the lower leg $\Delta n<0$, which in turn increases the renormalised energy imbalance $\Delta\tilde{\epsilon}=\Delta\epsilon+V_{\rm v}|\Delta n|$ in Eq.~\eqref{eq:pwilson_mass_mean_field}. Consequently,  the critical lines where one of  the Wilson masses~\eqref{eq:critical_lines_free} gets inverted will be shifted towards smaller values of the bare energy imbalance $\Delta\epsilon$ when the Hubbard interactions are increased, such that the self-consistent topological regions  shrink in favor of the mean-field trivial band insulator.

Motivated by the strong-coupling results of the previous section, another possibility for a translationally-invariant groundstate is that  of a purely imaginary bond density $b_{i,{\rm u}}=\langle c^\dagger_{i,{\rm u}}c^{\phantom{\dagger}}_{i,{\rm d}} \rangle=-b_{i,{\rm d}}=\ii M_y,\,\forall i$. Such arrangement of bond densities corresponds to the orbital ferromagnet FM$_y$ found in the previous section for sufficiently strong couplings and fluxes.  This type of order is induced by the mean-field vertical tunnelings~\eqref{eq:mf_parameters} and, in the continuum QFT~\eqref{eq:cont_limit_bilinears},  leads to a new kind of mass term
\beq
m_\eta c^2\overline{\Psi}_\eta{\Psi}_\eta\to \tilde{m}_\eta c^2\overline{\Psi}_\eta{\Psi}_\eta+\eta V_{\rm v}M_y\overline{\Psi}_\eta\gamma^5_\eta{\Psi}_\eta.
\eeq
These two  self-consistent masses are  renormalised by the imbalance and bond densities, which correspond to the aforementioned scalar and pseudo-scalar condensates of the Gross-Neveu model with a Wilsonian regularization~\cite{BERMUDEZ2018149}. If the interactions are strong enough, the ferromagnet will be formed $M_y\neq 0$, such that the Zak's phase will no longer be quantised~\eqref{eq:critical_lines_free} as a consequence of the two non-zero masses. Moreover, the long-range order is incompatible with the protecting symmetries, and the ferromagnet fully expels the topological crystalline phase via a SSB phase transition. 

Again motivated by the strong-coupling results, we should consider a bond-density wave  that breaks translation invariance. This is given by a purely-real alternating bond density $b_{i,{\rm u}}=\langle c^\dagger_{i,{\rm u}}c^{\phantom{\dagger}}_{i,{\rm d}} \rangle=b_{i,{\rm d}}=(-1)^iM_x,\,\forall i$, corresponding to the orbital anti-ferromagnet AFM$_x$, which was found for sufficiently strong couplings and small magnetic fluxes (see Fig.~\ref{fig:phase_diagram_strong}). In light of this order parameter, we see that in the limit $M_x\to1$, fermions delocalise along the vertical rungs of the ladder with an alternating symmetric/anti-symmetric pattern.  Such an  alternating pattern leads to a different long-wavelength term, as it effectively couples the two species of Dirac fermions by the following Umklapp term  
\beq
\label{eq:umklapp_mass}
m_\eta c^2\overline{\Psi}_\eta{\Psi}_\eta\to \tilde{m}_\eta c^2\overline{\Psi}_\eta{\Psi}_\eta-V_{\rm v}M_x\overline{\Psi}_\eta{\Psi}_{\overline{\eta}}.
\eeq
At zero magnetic flux, where the band structure in Fig.~\ref{fig:bands_theta_0} displays a flat band and a cosine band, one may form a massless Dirac spinor from the right- and left-moving components at $k_\pm=\pm\pi/2a$, such that the  gap-opening Umklapp term  corresponds to yet a different mass term. Although at the mean-field level~\cite{RevModPhys.66.129}, the massless Dirac fermion is unstable towards the alternating bond-density wave for arbitrary-small interactions, the presence of the flat band and the correlations beyond mean-field may change this simple picture. In any case, if this bond-density wave is formed, the topological invariant will no longer be quantised~\eqref{eq:critical_lines_free}, and the associated anti-ferromagnet will fully expel the topological  phase. 

Let us note that this self-consistent mean-field method is equivalent to a large-$N$ limit of the equivalent Gross-Neveu model, where one introduces auxiliary  bosonic fields via a Hubbard-Stratonovich transformation~\cite{coleman_2015}, and sends the number of fermionic species $N\to\infty$. In the $\pi$-flux limit, it suffices to introduce a couple of auxiliary fields, the non-zero expectation values of which   lead to the onset of the so-called scalar and pseudo-scalar condensates discussed in Sec.~\ref{sec:ch_previous_results}. We have checked that our self-consistent mean-field approach provides numerical results that are in complete agreement with the large-$N$ prediction of the $\pi$-flux limit presented in~\cite{BERMUDEZ2018149}. As one modifies the magnetic flux, the large-$N$ approach should be generalised to include an additional auxiliary field, the condensation of which would be connected   to the AFM$_x$ order. Instead of following this approach, which  requires  solving a  new set of coupled non-linear gap equations, the current self-consistent mean-field is sufficiently general so that it can directly encompass any type of ordering, and such an order can be efficiently found by a suitable choice of the initial seed.

In any case, both the mean-field and the large-$N$
methods do not take the inter-particle correlations in full account. For this reason, we benchmark these results with matrix-product state (MPS) methods. In particular, we  use the density matrix renormalization group (DMRG) based on MPS, which is becoming a very useful tool to explore lattice discretizations of quantum field theories~\cite{Carmen_Ba_uls_2020,doi:10.1080/00107514.2016.1151199}. This can be considered as the most stringest test of the validity of the mean-field approach. We consider lattices up to $L=128$ sites and virtual dimension up to $\chi=200$. Our DMRG code is based on MPS and uses a two-site 
optimization scheme. It is built implementing the Abelian symmetries such as
a particle conservation. Indeed, we fix half-filling and open boundary conditions.

\subsection{Weak and intermediate interactions}

\begin{figure}[t!]
  \centering
    \includegraphics[width=0.9\linewidth]{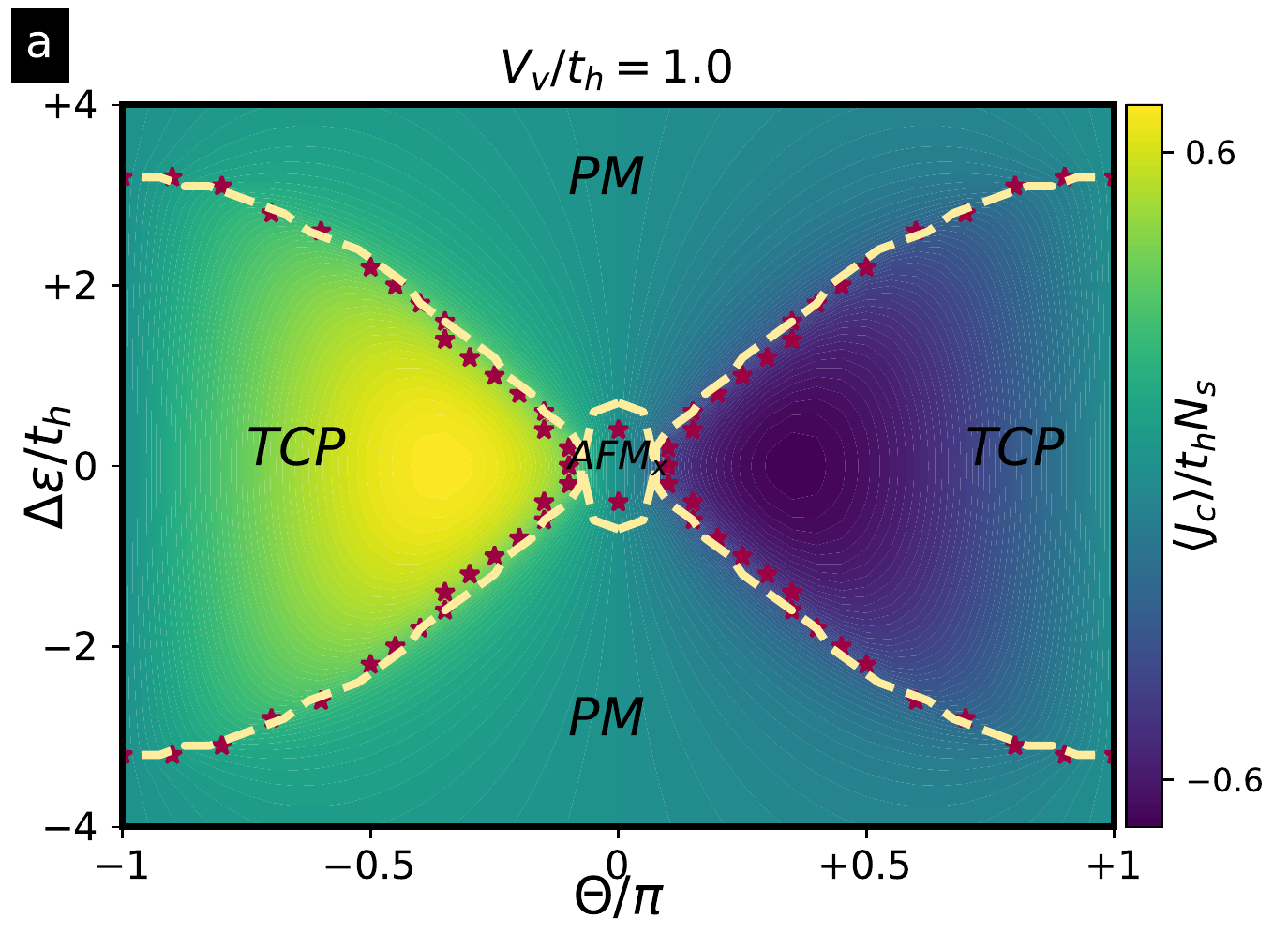} \\
  \includegraphics[width=0.9\linewidth]{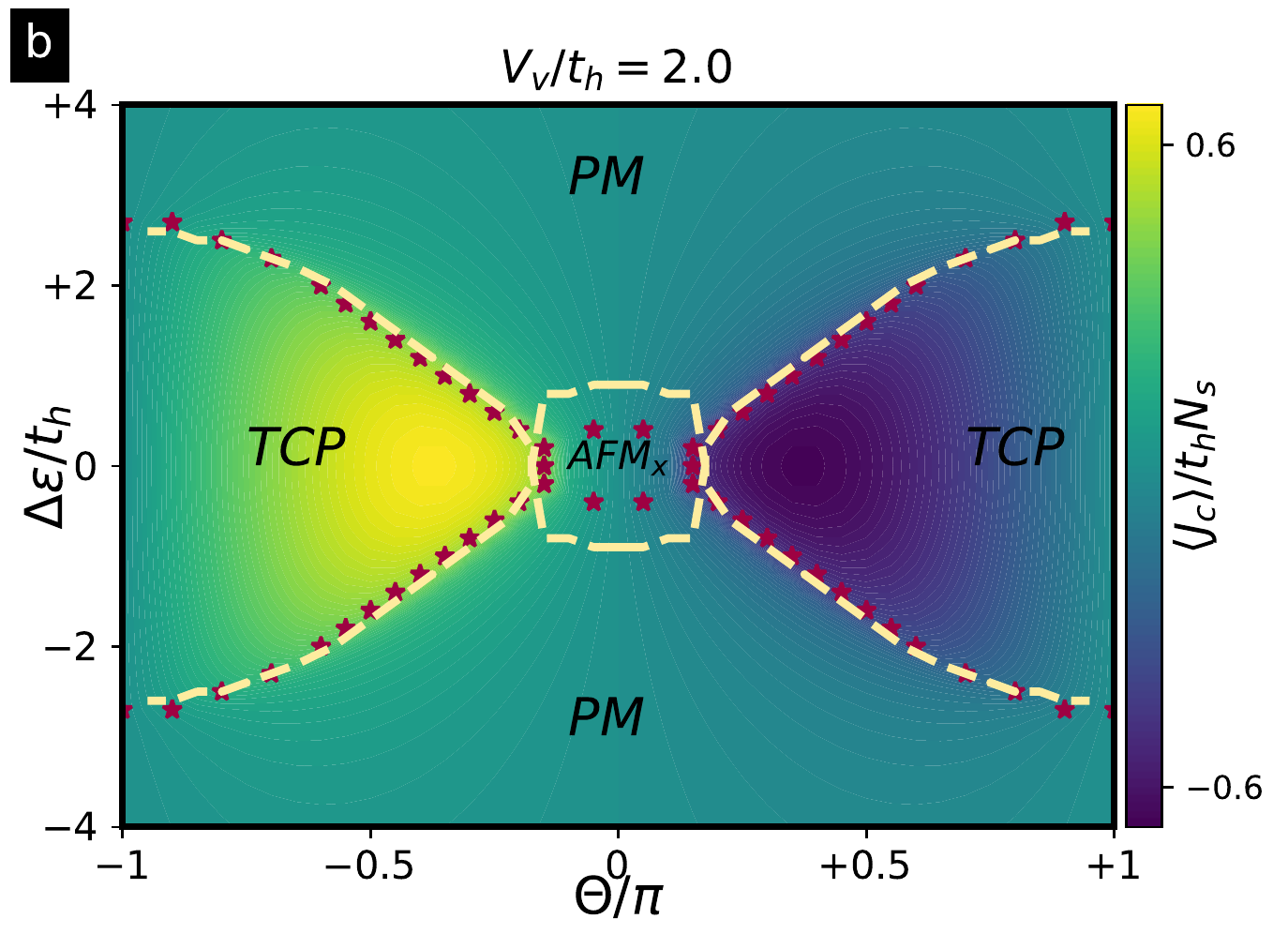}
\caption{\label{fig:phase_diagram_DMRG_weak_int} {\textbf{Phase diagrams  for weak interactions: {\bf (a)} $V_{\rm v}/t_{\rm h}=1$, {\bf (b)} $V_{\rm v}/t_{\rm h}=2$:} The phase diagrams display two regions hosting a topological crystalline phase (TCP), a long-range-ordered anti-ferromagnetic phase (AFM$_x$), and a paramagnetic phase (PM). The horizontal axis
represents the magnetic flux, whereas the vertical axis corresponds to the ratio of the energy imbalance to the tunneling strength. 
The red  stars (yellow dashed lines) show the critical points found from DMRG (self-consistently mean-field) numerics. These  points  are plotted on top of the  contour plot of the  chiral current, obtained by calculating Eq.~\eqref{eq:chiral_current} using DMRG.}}
\end{figure}

\begin{figure*}[t!]
  \centering
    \includegraphics[width=0.8\linewidth]{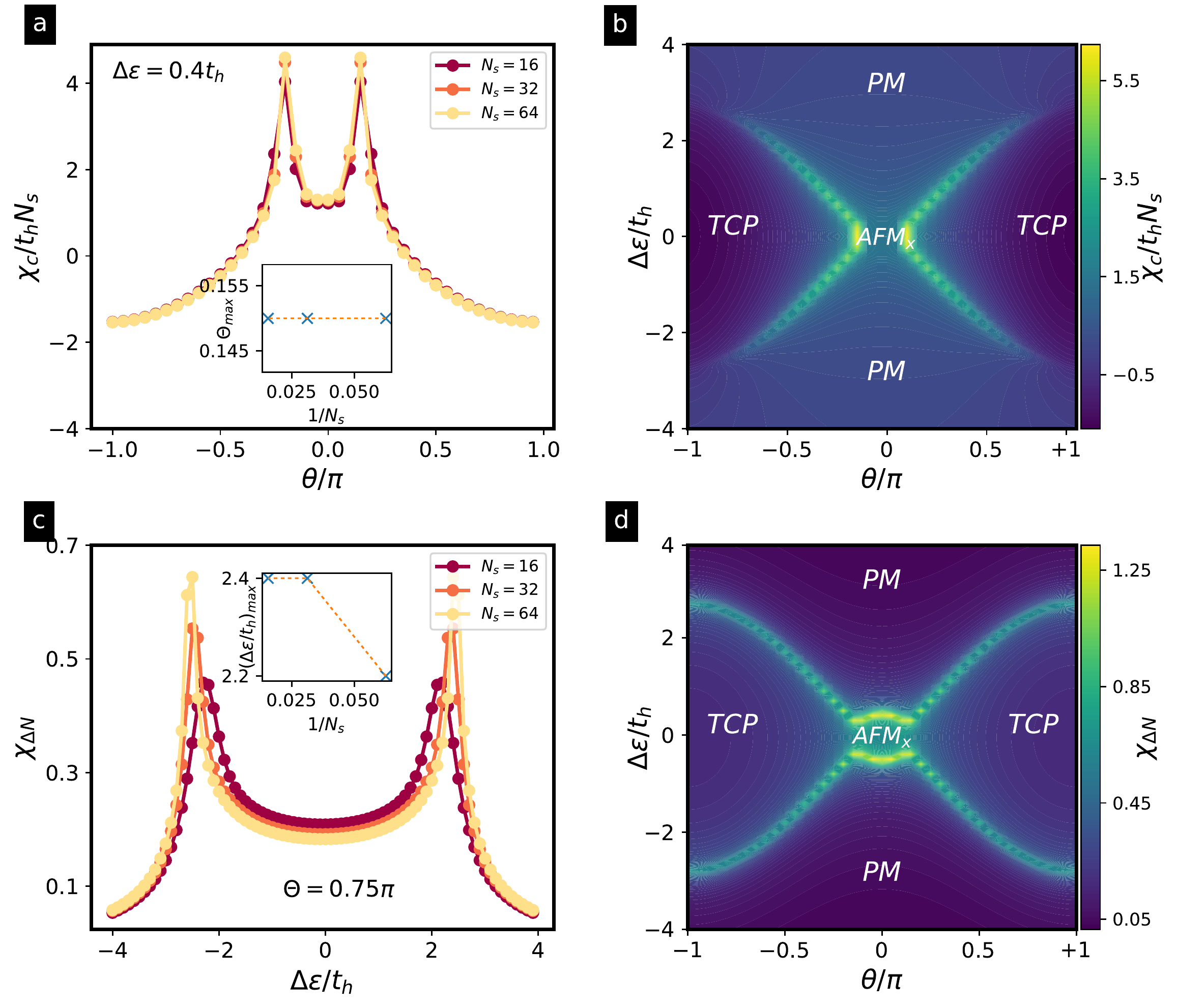}
\caption{\label{fig:Chis_weak_int} {\textbf{Chiral and imbalance susceptibilities:}  {\bf (a)} The  peaks of the chiral susceptibility $\chi_{\rm c}$ with growing system size  $N_{\rm s}\in\{16,32,64\}$ allow to locate the critical point for a cut of phase diagram at $V_{\rm v}=2 t_{\rm h}$  and $\Delta \epsilon=0.4t_{\rm h}$ . {\bf (b)} The contour plot of the chiral susceptibility  for system size $N_{\rm s}=64$, delimiting  the TCP lobes from other phases phases. {\bf (c}) The peaks of the imbalance susceptibility $\chi_{\Delta n}$ with growing system size $N_{\rm s}\in\{16,32,64\}$ allow to locate the critical points also around the $\pi$-flux regime. {\bf (d)} The contour plot of the imbalance susceptibility for system size $N_{\rm s}=64$. }}
\end{figure*}

In this section, we  analyze the effect of correlations in the phase diagram of the Creutz-Hubbard ladder under a  generic magnetic flux, focusing on the regime of  weak interactions. 
 
We start by setting $V_{\rm v}/t_{\rm h}=1.0$ and $V_{\rm v}/t_{\rm h}=2.0$, which correspond to the red and green shaded planes of the schematic phase diagram of Fig~\ref{fig:scheme_phase_diagram}. Our numerical results for the phases of matter that appear in these planes  are presented in Fig.~\ref{fig:phase_diagram_DMRG_weak_int}, where we have used  self-consistent mean-field and DMRG methods. 
The lines represent the critical points where  the topological and SSB phase transitions occur, either obtained with a DMRG method  
based on finite MPS with bond dimension $\chi=200$ (red stars), or by the self-consistent Hartree-Fock method (yellow dashed lines). As can be observed in the figure, for weak interactions, the Hartree-Fock and DMRG critical points separating the TCP and PM regions yield two critical lines that are very similar to each other. The main differences appear in the critical lines separating the AFM$_x$ and  PM phases, as the mean-field method predicts a smaller AFM$_x$. 
Let us now analyze this figure in more detail, and describe  the methodology used to extract these critical points.

In analogy with the non-interacting case~\eqref{eq:H_C}, the groundstate of the weakly-interacting Creutz-Hubbard ladder~\eqref{eq:H_CH}    contains  a  topological crystalline phase (TCP)  as one departs from the $\pi$-flux limit. As discussed qualitatively around Eq.~\eqref{eq:pwilson_mass_mean_field}, one of the expected effects of interactions is to shrink the size of the TCP lobes in favor of the trivial band insulator. Note that the trivial band insulator is adiabatically connected to the orbital PM when the energy imbalance is increased, and we refer to them interchangeably.  This shrinking  can  be readily observed by comparing the vertical extent of the TCP lobes in the non-interacting and interacting cases (compare Figs.~\ref{fig:phase_diagram_free} and~\ref{fig:phase_diagram_DMRG_weak_int}). Moreover,  as shown in Sec.~\ref{eq:entamglement_spec}, the entanglement spectrum on these lobes is two-fold degenerate (see Fig.~\ref{fig:ent_spectrum_entropy} a hallmark of symmetry-protected topological phases~\cite{PhysRevB.81.064439}. This shows that the non-interacting TCPs discussed in Sec.~\ref{sec:TI_chiral_flow}  is adiabatically connected to these correlated TCP lobes. Comparing Figs.~\ref{fig:phase_diagram_DMRG_weak_int} {\bf (a)} to {\bf (b)}, we also see that the shrinking of the TCP in favor of the PM  increases with the strength of the Hubbard interactions, which is a direct consequence of the interaction-induced renormalization of the Dirac-fermion masses~\eqref{eq:pwilson_mass_mean_field} in the continuum QFT.

Let us also recall that, as argued in Sec. \ref{sec:CHL_Continuum_Limit}, the  violation of Lorenz symmetry in the continuum QFT~\eqref{eq:cont_limit_L}-\eqref{eq:gamma_Lviolation} can be associated to a different propagation speed for right- and left- moving
particles (see the insets of Fig.~\ref{fig:band_structure}). This difference can give rise to a net circulating chiral current, serve  to characterise further the phases of the model, and may act as a good indicator to predict the critical lines (see Fig.~\ref{fig:phase_diagram_free} {\bf (a)}-{\bf (b)}). In particular, we recall  that the non-interacting TCP supports large values of this permanent chiral current away from the $\pi$-flux limit (see Fig.~\ref{fig:phase_diagram_free} {\bf (a)}). In the background of  Figs.~\ref{fig:phase_diagram_DMRG_weak_int} {\bf (a)} and {\bf (b)}, we present the expectation value of the  chiral current $\langle J_{\rm c}\rangle$ in Eq.~\eqref{eq:chiral_current} in the presence of interactions, as obtained by the DMRG numerics. We see that, as the TCP lobes shrink in size due to the Hubbard interactions, the area with a large chiral current also decreases. Therefore, the correlated TCPs can still support large  chiral currents around the ladder. 

The critical points delimiting  the TCP lobes  in the phase diagram
are estimated via the  chiral susceptibility
$\chi_{\rm c}=\partial \langle J_{\rm c}\rangle / \partial \theta$~\eqref{eq:chiral_susc}. We recall that, in the non-interacting limit, this susceptibility  presents a divergence at the critical points (see Fig.~\ref{fig:phase_diagram_free} {\bf (b)}). In  Fig.~\ref{fig:Chis_weak_int} {\bf (a)}, we display the chiral susceptibility  for ladders with a different number of sites per leg $N_{\rm s}\in\{16,32,64\}$,  fixing the energy imbalance $\Delta\epsilon=0.4t_{\rm h}$ and interaction strength $V_{\rm v}=2t_{\rm h}$, and varying the external magnetic flux. These numerical results clearly show a peak at two symmetric values of the magnetic flux with respect to the zero-flux axis. The finite size scaling (FSS) of the chiral susceptibility maxima as a function of $N$ is displayed in the inset. 
As one can see in the inset, the peak of the chiral susceptibility  diverges with the size of the ladder, and fitting the maxima of $\theta_{\rm c}$ to $ \theta_c(N)=\theta_c(1+a N^{-1}+b N^{-2})$, we can delimit the TCP lobes and locate the phase transitions. In Fig.~\ref{fig:Chis_weak_int} {\bf (b)}, we show the contour plot of the chiral susceptibility with the corresponding peaks, which clearly shows how the critical lines separate the TCP from other phases. Comparing to the non-interacting case in Fig.~\ref{fig:phase_diagram_free} {\bf (b)}, we clearly see that these critical points are moved towards smaller  imbalances, which underlies  the  shrinking of the TCP lobes.

So far, our numerical benchmark has focused on the TCP phase, and revolved around quantities related to the circulating chiral current. An alternative way of identifying this phase boundary is by studying the occupation leg imbalance 
\beq 
\Delta n = \frac{1}{N_{\rm s}}\sum_j \left( \langle c^{\dagger}_{j,{\rm u}} c^{\phantom{\dagger}}_{j,{\rm u}}\rangle-\langle c^{\dagger}_{j,{\rm d}} c^{\phantom{\dagger}}_{j,\rm d}\rangle \right).
\eeq
The occupation imbalance is induced by a non-zero energy imbalance $\Delta\epsilon$, which motivates the definition of  an imbalance susceptibility $
\chi_{\Delta n} = {\partial  \Delta n }/{\partial \Delta \epsilon}$ that proved to be   a good indicator of  the quantum phase transition in the $\pi$-flux limit~\cite{PhysRevX.7.031057},  and can be easily calculated using our DMRG code. 
In Fig.~\ref{fig:Chis_weak_int} {\bf (c)}, we show the imbalance susceptibility  for ladders  with a different number of sites per leg $N_{\rm s}\in\{16,32,64\}$, fixing the magnetic flux to $\theta=3\pi/4$,  the Hubbard interactions to $V_{\rm v}=2t_{\rm h}$, and varying   the energy imbalance $\Delta\epsilon$. Once again, this susceptibility shows a pair of peaks that are symmetric with respect to the zero-imbalance axis. A FSS shows that this peak actually diverges in the thermodynamic limit, and allows us to locate the position of the critical point. The critical points obtained through this analysis are represented in a contour plot in Fig.~\ref{fig:Chis_weak_int} {\bf (d}). As can be seen from
these results, this susceptibility is more effective than the chiral one in the critical regions around $\pi$-flux, which is to be expected since the chiral current vanishes in these regions. 

\begin{figure}[t!]
  \centering
    \includegraphics[width=0.9\linewidth]{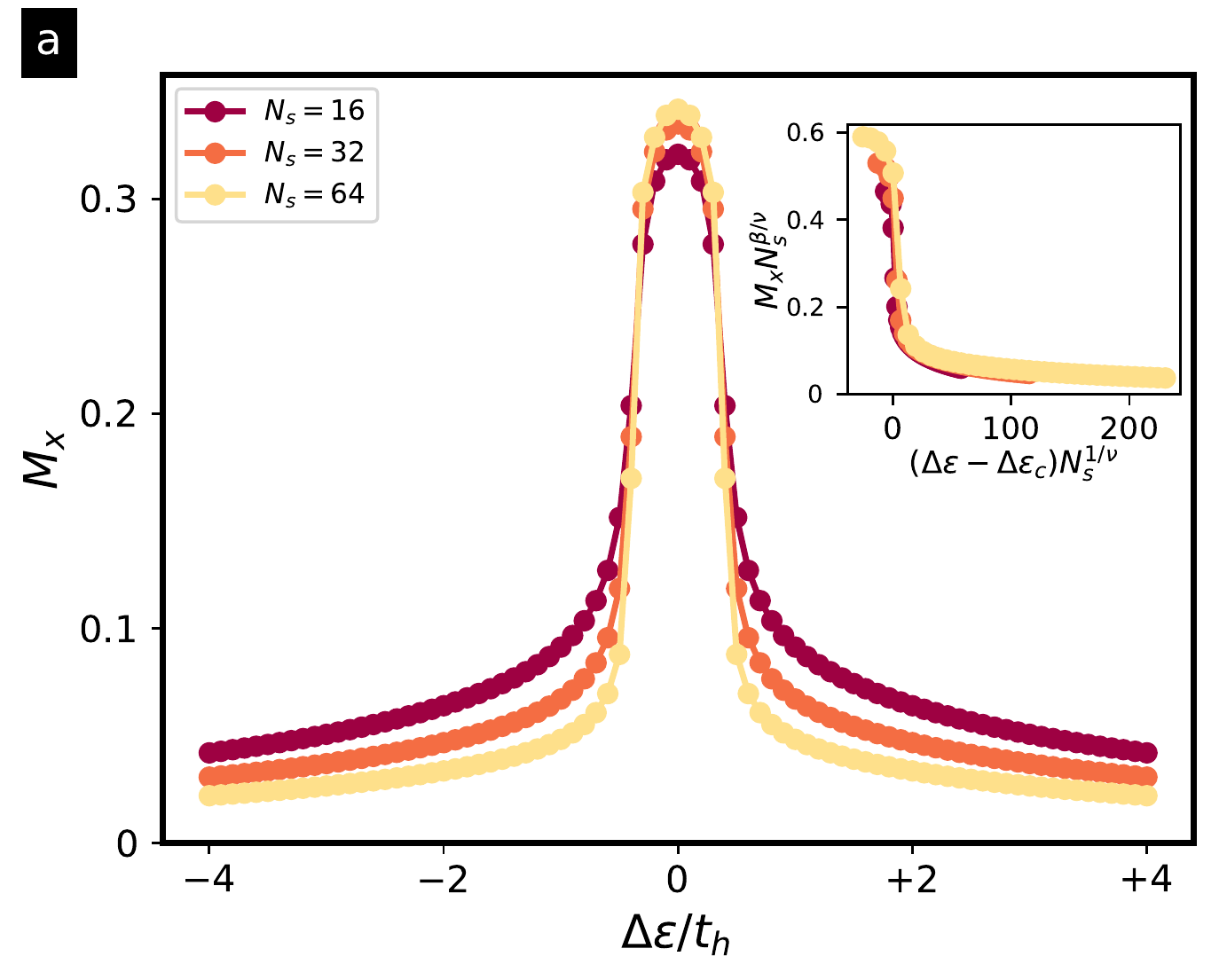} \\
    \includegraphics[width=0.9\linewidth]{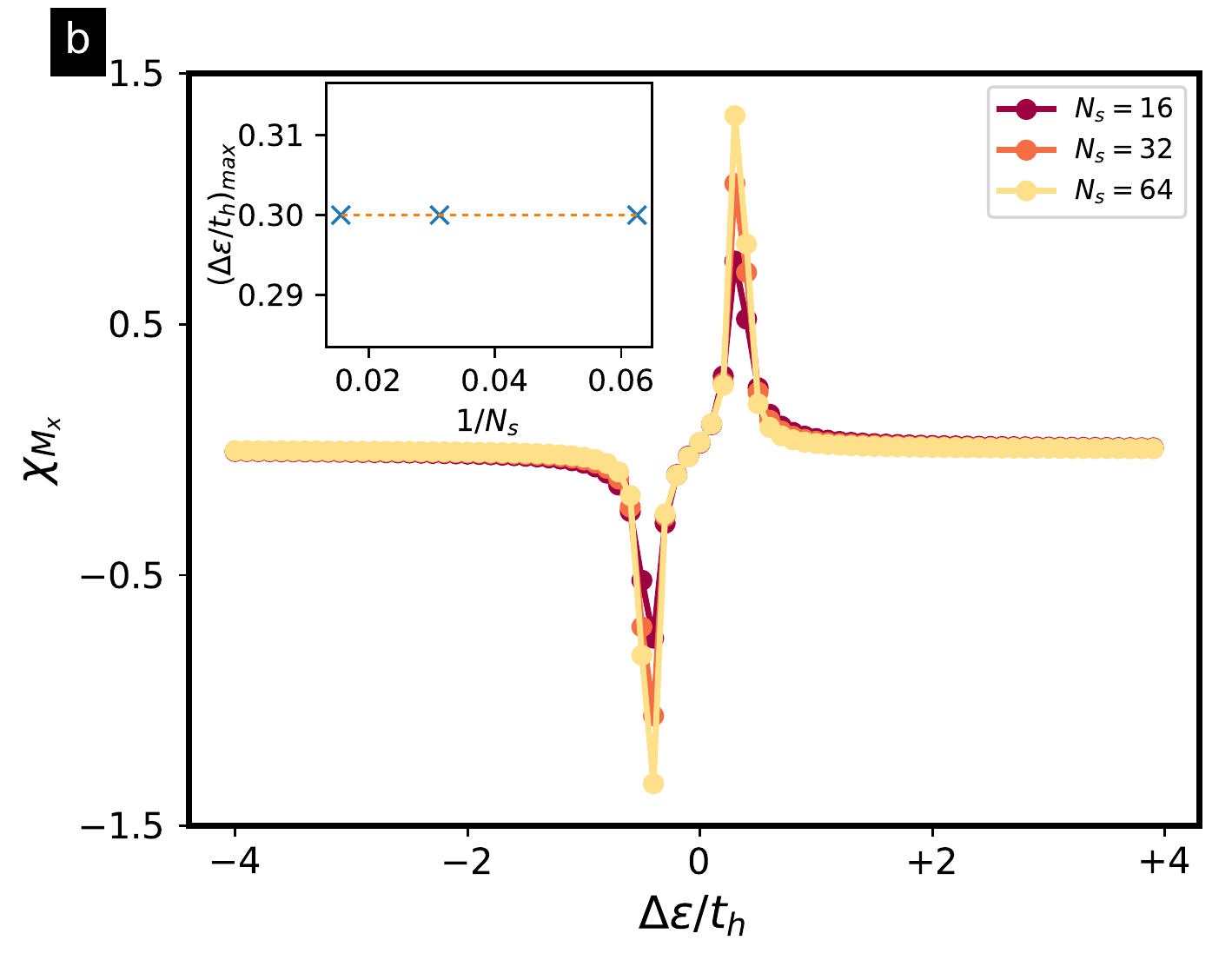}
\caption{\label{fig:Mx_weak_int} {\textbf{Staggered magnetization and  magnetic susceptibility:} {\bf (a)} Staggered magnetization for fixed interaction strength $V_{\rm v}=2 t_{\rm h}$ and flux $\theta=\pi/10$, and different number of sites per length  $N_{\rm s}\in\{16,32,64\}$.  As the length increases, the systems develops a non-zero expectation value only within a symmetric region around zero imbalance. {\bf (b)} The magnetic susceptibility $\chi_{M_x}$ for the same parameters shows a couple of peaks that increase with the system size, and allow us to locate the critical points.}}
\end{figure}

\begin{figure}[t!]
  \centering
    \includegraphics[width=0.9\linewidth]{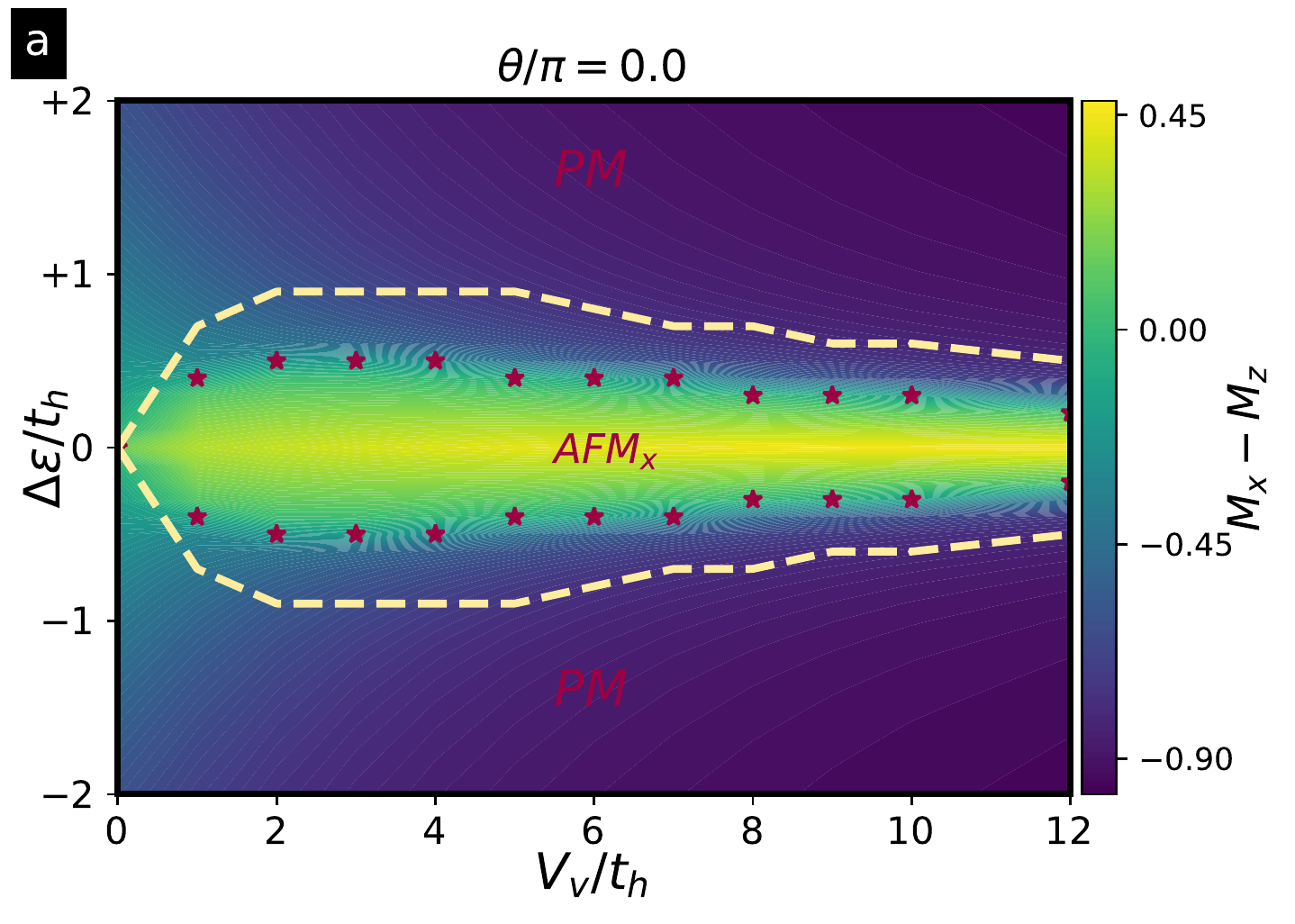} \\
    \includegraphics[width=0.9\linewidth]{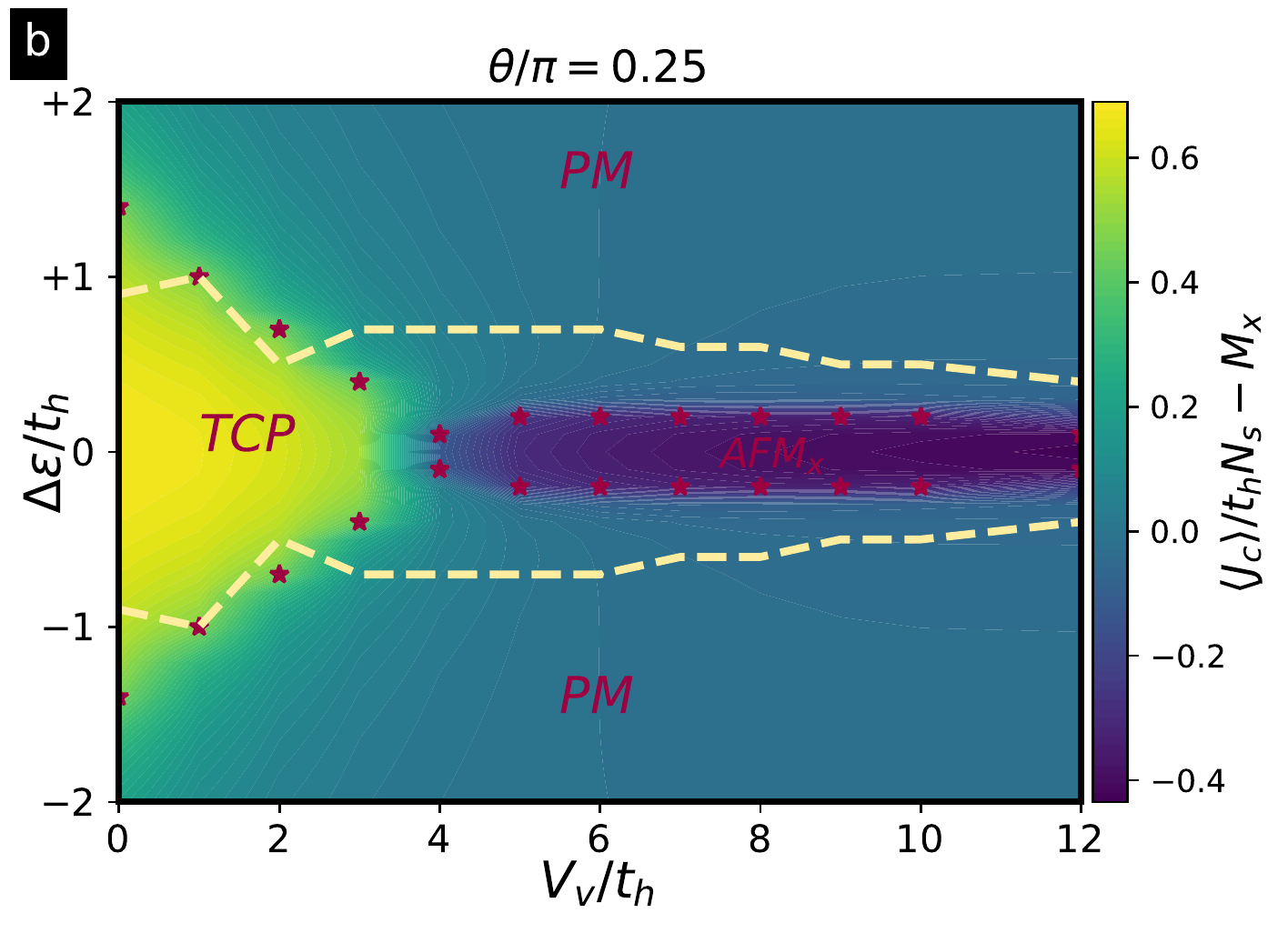} 
\caption{\label{fig:phase_diagram_fluxes} {\textbf{Phase diagram in the plane $(V_v,\Delta \epsilon)$ for fixed flux:} {\bf (a)} The phase diagram for $\theta=0.0$ displays two regions hosting a long-range-ordered anti-ferromagnetic phase (AFM$_x$), and a paramagnetic phase (PM). {\bf (b)} The phase diagram for  $\theta=0.25 \pi$ displays a region hosting a topological crystalline phase (TCP), a long-range-ordered anti-ferromagnetic phase (AFM$_x$), and a paramagnetic phase (PM).
The horizontal axis represents the ratio of interaction to the tunneling strength, whereas the vertical axis corresponds to the ratio of the energy imbalance to the tunneling strength. The red  stars (yellow dashed lines) show the critical points found from DMRG (self-consistently mean-field) numerics. These  points are plotted on top of the  contour plot of the difference between the ferro and anti-ferro order parameters for {\bf (a)} and of the chiral current for {\bf (b)}}}
\end{figure}

The main difference brought up by  the interactions occurs in the region of small magnetic fluxes, where the Umklapp term~\eqref{eq:umklapp_mass} may dominate over the other gap-opening mechanisms. This  term is expected to stabilize a bond-ordered-wave with fermions delocalized along the vertical bonds of the ladder in  alternating symmetric/anti-symmetric  orbitals, resembling an orbital AFM$_x$. Our numerics  show that there are additional critical lines that separate this small-flux region from either the TCP lobes at higher flux, or  the orbital PM  at higher energy imbalances (see   Figs.~\ref{fig:phase_diagram_DMRG_weak_int}). As supported by the numerical results presented in the following paragraphs, this intermediate region 
hosts  AFM$_x$ long-range order due to SSB of a global $\mathbb{Z}_2$  invariance.
To accurately extract  the critical points between the AFM$_x$ and PM, we calculate the staggered magnetization 
\beq
\label{eq:AFM_orbital}
M_x = \frac{1}{N_{\rm s}}\sum_j (-1)^j\left( \langle c^{\dagger}_{j,{\rm u}} c^{\phantom{\dagger}}_{j,{\rm d}}\rangle+\langle c^{\dagger}_{j,{\rm d}} c^{\phantom{\dagger}}_{j,\rm u}\rangle \right),
\eeq
which corresponds to the order parameter of the alternating bond density wave. The advantage of the strong-coupling perspective~\eqref{eq:DM_interaction} is that one identifies the energy imbalance as the transverse magnetic field~\eqref{eq:DM_parameters}, and one can readily define the anti-ferromagnetic susceptibility as $\chi_{M_x}=\partial M_x / \partial \Delta \epsilon$. 
In Fig.~\ref{fig:Mx_weak_int} {\bf (a)}, we present the staggered magnetization $M_x$ for different system sizes. The crossing of the lines in the main panel serves to obtain the critical point of the model for weak fluxes. As proved by the data collapse shown in the inset of Fig.~\ref{fig:Mx_weak_int} {\bf (a)}, the critical exponents correspond to those of the $(1+1)$ Ising universality class.

 

Moreover, we also want to address if the strong-coupling AFM$_x$ phase, which appears for vanishingly-small fluxes, also sets for arbitrarily small values of the interaction strength $V_v$, or if a PM sets in before the AFM$_x$ order takes on. For this aim, we will study the phase diagram sketched in Fig.~\ref{fig:scheme_phase_diagram} fixing the flux $\theta$, and varying the imbalance $\Delta \epsilon$ and the interactions $V_v$, that permit us to explore different planes also the correlation effects at intermediate interactions.  We set  $\theta=0.25\pi$ and $\theta=0$, which correspond to the  green  and brown shaded planes of the schematic phase diagram of Fig.~\ref{fig:scheme_phase_diagram}. Our numerical results for the phases of matter that appear in these planes are presented in Fig.
~\ref{fig:phase_diagram_fluxes}, where we have used the self-consistent 
mean-field (yellow dashed line) and DMRG methods (red stars) to determine the critical lines. In the background, we present a contour plot that serves as a guide to the eye to identify the different phases. As shown in the figure,  the red stars obtained with a DMRG method based on finite MPS with bond dimension $\chi=200$ yield a much better estimate of the critical lines that separate the different phases. 

Let us start discussing  the case of $\theta=0.25\pi$ shown  in Fig.~\ref{fig:phase_diagram_fluxes}{\bf (b)}. Here,  one can distinguish three different phases: TCP, PM, and AFM$_x$. As can be observed in the figure for weak interaction, the Hartree-Fock and DMRG critical points separating the TCP and PM regions yield two critical lines  similar to each other, although differences arise as the interactions are increased. Moreover, the weak-coupling limit can be understood as a ferromagnetic coupling between the two chains in the
spin model description (as shown in the Appendix \ref{sec:appendix}). Along this line the central charge $c$ is equal to $1$. 

We now move to the $\theta=0$ case, which is represented in Fig.~\ref{fig:phase_diagram_fluxes}{\bf (a)}. In this case, one can readily observe that there is no TCP phase, but that the SSB AFM$_x$ sets in at zero imbalance for vanishingly-small interactions. This result is consistent with the qualitative phase diagram of Fig.~\ref{fig:scheme_phase_diagram}, and has very interesting consequences. While magnetism in standard Hubbard ladders only appears for small super-exchange couplings, making its detection with cold atoms rather challenging due to the required temperatures, the AFM$_x$ order in the Creutz-Hubbard ladder for small fluxes appears at much-stronger scales, those set by the bare tunneling.  This can be a very interesting starting point to introduce a non-zero doping and explore the interplay of magnetism and   hole mobility in connection to fermionic mechanisms of superconductivity.

\subsection{Strong Interactions}
\begin{figure}[t]
  \centering
  \includegraphics[width=0.9\linewidth]{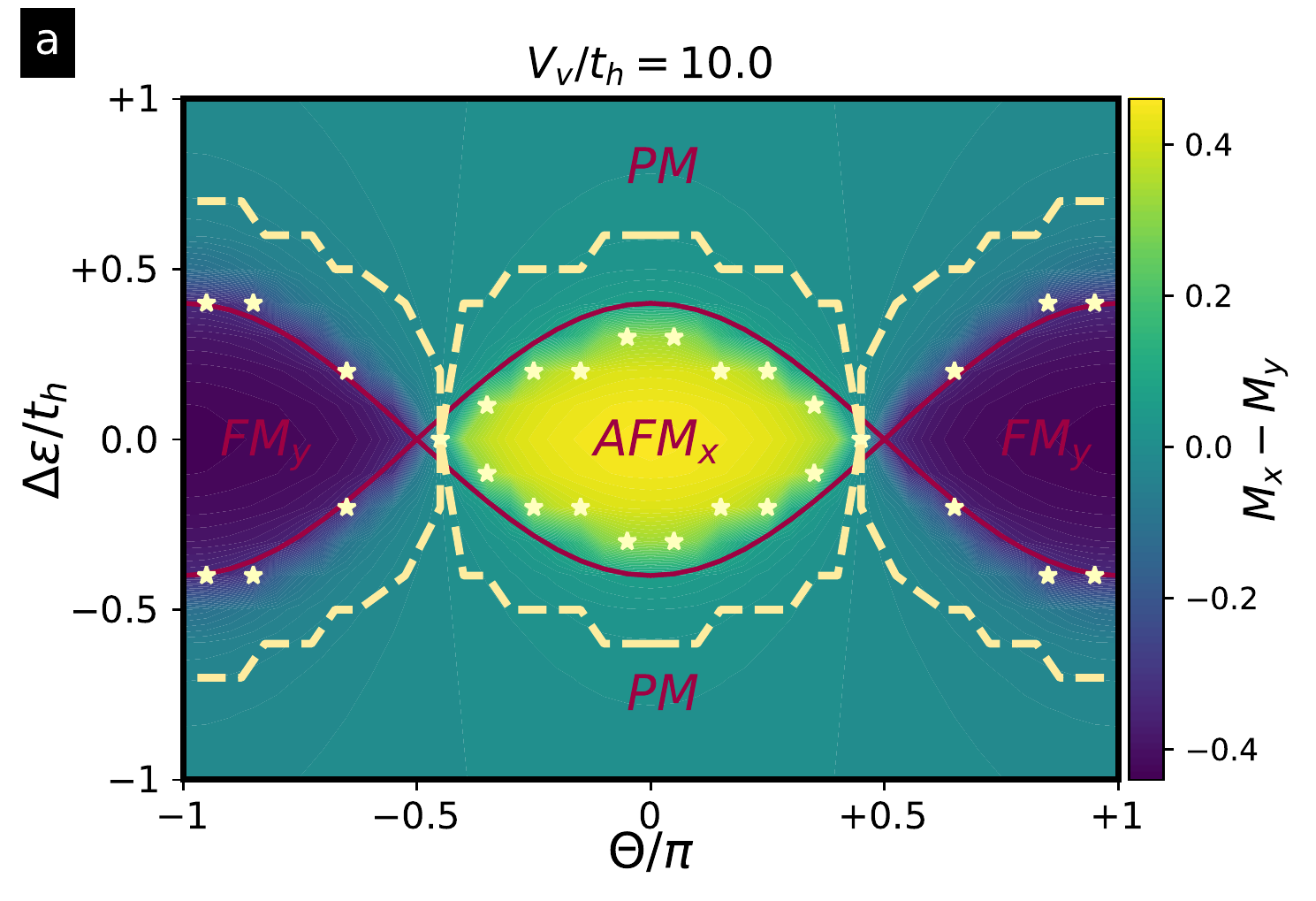} \\
  \includegraphics[width=0.9\linewidth]{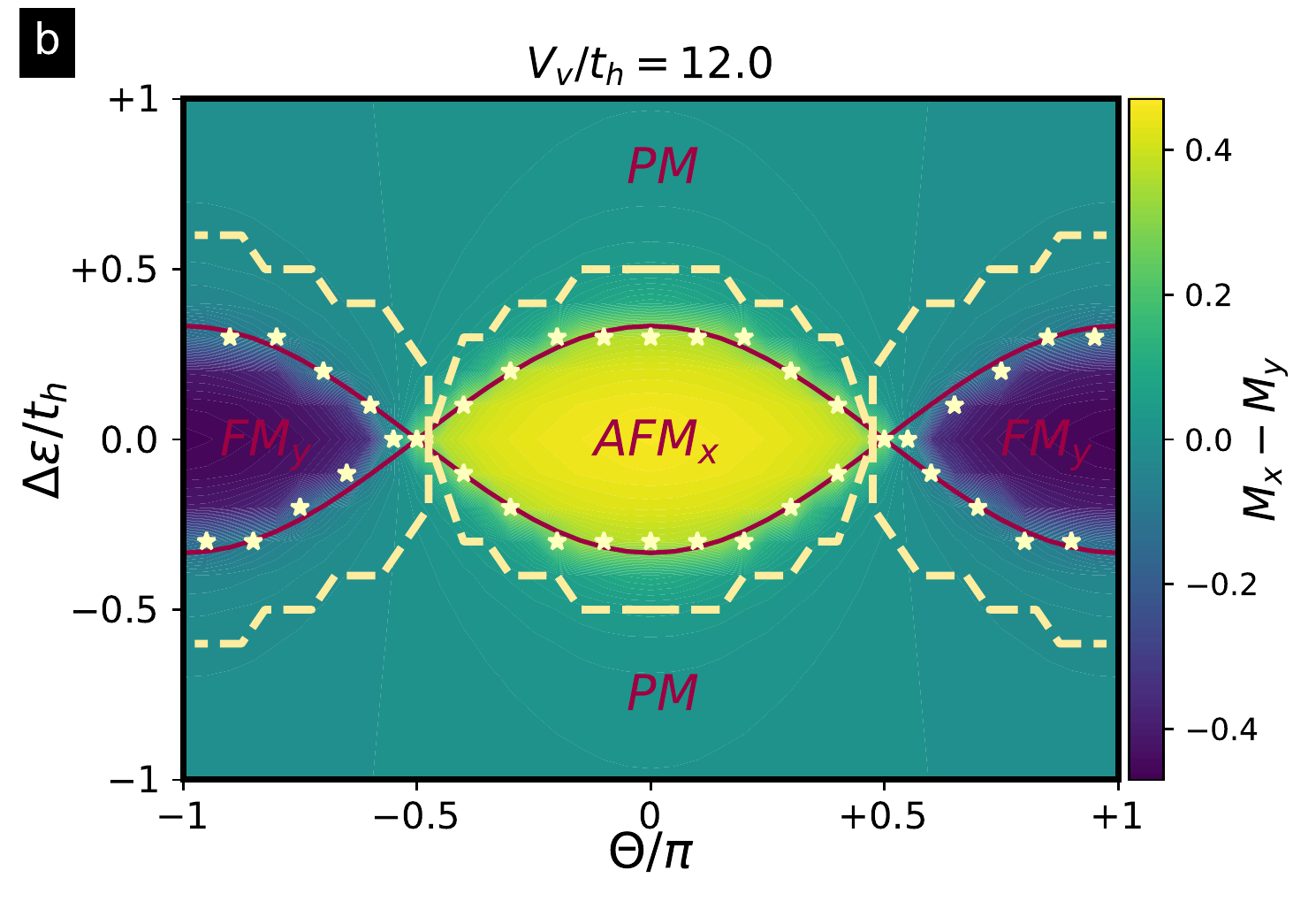}
\caption{\label{fig:phase_diagram_DMRG_strong_int} {\textbf{Phase diagram  for strong interactions:} {\bf (a)} $V_{\rm v}=10t_{\rm h}$ and {\bf (b)} $V_{\rm v}=12t_{\rm h}$. The phase diagram displays a long-range ferromagnetic phase (FM$_y$), an anti-ferromagnetic phase (AFM$_x$), and a paramagnetic phase (PM). The horizontal axis
represents the flux, whereas the vertical axis corresponds to the ratio of the energy imbalance to the tunneling strength. The red line shows the transition points~\eqref{eq:strong_coupling_critical_line} of the effective spin model with D-M interaction in strong coupling limit (\ref{eq:DM_interaction}). The yellow stars show the critical points found by DMRG. Instead, the shaded yellow lines  show the critical points from the self-consistently mean-field. The background contour plot represents the difference between the ferro and anti-ferro order parameters. }}
\end{figure}

In Sec.~\ref{sec:Strong_Coupling_Sec}, we introduced an effective spin
model~\eqref{eq:DM_interaction} with Dzyaloshinskii-Moriya couplings in the limit of very strong interactions. This model is exactly solvable, and allowed us to predict a pair of critical lines~\eqref{eq:strong_coupling_critical_line} separating the orbital ferromagnet FM$_y$,  anti-ferromagnet AFM$_x$, and  paramagnet PM. We remark that these predictions are strictly valid in the limit where $V_{\rm v}/t_{\rm h}\to\infty$, and we should explore when these predictions are valid for  large, yet finite, interactions. In Fig.~\ref{fig:phase_diagram_DMRG_strong_int},  we present the ground-state phase diagrams as a function of $\theta$ and $\Delta \epsilon$ for two values of the interactions $V_{\rm v}=10t_{\rm h}$ and $V_{\rm v}=12t_{\rm h}$, which correspond to the green and magenta shaded planes of the schematic phase diagram of Fig~\ref{fig:scheme_phase_diagram}.  
The red line, in both phase diagrams, represents the analytical critical line~(\ref{eq:strong_coupling_critical_line}), whereas the white stars stand for the DMRG critical points, and the yellow dashed lines represent the mean-field predictions. 
One can see that the agreement of the strong-coupling critical lines with the DMRG data improves as the Hubbard interactions are increased, whereas the mean-field results overestimate the regions with long-range order. Note also that, by comparing Figs.~\ref{fig:phase_diagram_DMRG_strong_int} {\bf (a)} and {\bf (b)}, one can see how the PM region grows
significantly as the interactions are increased, in detriment of the long-range ordered phases, which is a consequence of the scaling of the super-exchange and  Dzyaloshinskii-Moriya couplings~\eqref{eq:DM_parameters} with the inverse of the Hubbard interactions. Let us now discuss in detail how we have characterised these phases and extracted the critical points.

In the background of Fig.~\ref{fig:phase_diagram_DMRG_strong_int}, we present a contour plot for the difference of the orbital ferromagnetic and antiferromagnetic order parameters. Let us note that, in analogy to the anti-ferromagnetic case~\eqref{eq:AFM_orbital}, one can define the orbital ferromagnetic parameter as
\beq
\label{eq:FM_orbital}
M_y = \frac{1}{N_{\rm s}}\sum_j \ii \left( \langle c^{\dagger}_{j,{\rm u}} c^{\phantom{\dagger}}_{j,{\rm d}}\rangle-\langle c^{\dagger}_{j,{\rm d}} c^{\phantom{\dagger}}_{j,\rm u}\rangle \right),
\eeq
which attains non-zero values for a purely-imaginary arrangement of  bond densities. Note that such an arrangement is reminiscent of a set of alternating fermionic currents that flow vertically  between the legs of the  ladder.  However, in contrast to the circulating chiral current~\eqref{eq:chiral_current}, the corresponding vertical bare tunnelings are absent from the original Hamiltonian~\eqref{eq:H_C}, and we would thus lack a continuity equation for such fermionic currents.


\begin{figure*}[t]
\centering 
\includegraphics[width=0.8\linewidth]{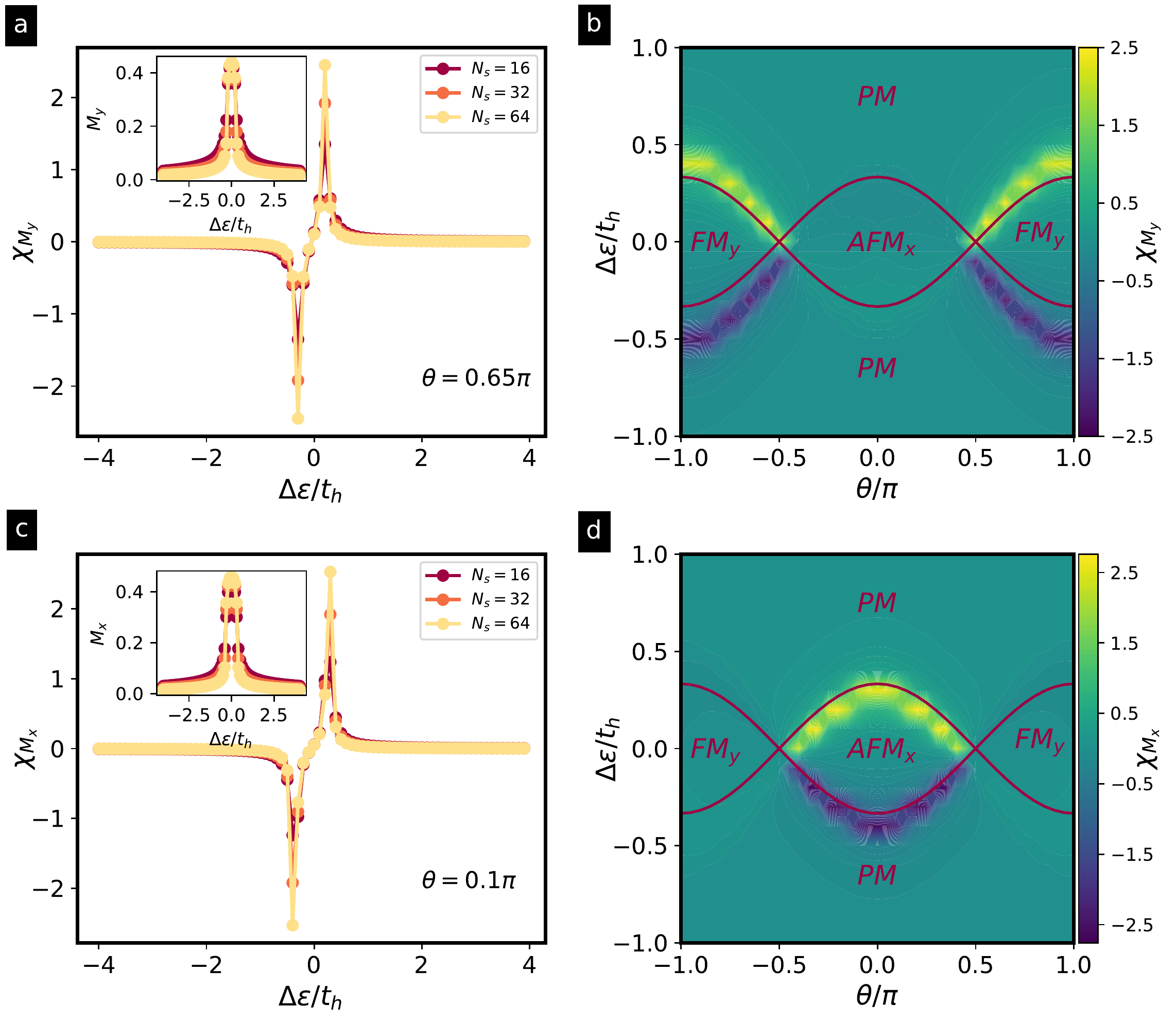}
\caption{\label{fig:chi_mags_SI} {\textbf{Magnetic subsceptibilities  for strong interactions:} {\bf (a)} the divergence of $\chi_{M_y}$ with growing system size indicate the phase transition between FM$_y$ and PM phases for a cut of phase diagram at $V_{\rm v}=10 t_{\rm h}$ and $\theta=0.65 \pi$. 
{\bf (b)} The contour plot of $\chi_{M_y}$ for system size $N_s=64$ distinguishes  two FM$_y$ lobes. The red line shows the analytical critical  points~\eqref{eq:strong_coupling_critical_line}  of the effective spin model~(\ref{eq:DM_interaction}), strictly valid for infinitely-large Hubbard interactions.
{\bf (c)} the divergence of $\chi_{M_x}$ with growing system size indicates the phase transition between AFM$_x$ and PM phases for  $V_{\rm v}=10 t_{\rm h}$ and $\theta=0.1 \pi$.
{\bf (d)} The contour plot of $\chi_{M_x}$ for system size $N_{\rm s}=64$ delimits the  AFM$_x$ phase. The red line shows the analytical critical  points~\eqref{eq:strong_coupling_critical_line}.}}
\end{figure*}

In order to avoid numerical problems due to the incomplete symmetry breaking of the magnetic order parameters $M_x,M_y$, we determine instead the  corresponding structure factors   
\beq
\begin{split}
\label{eq:structure_factors}
S_{T_y T_y}(k) = \frac{1}{N_{\rm s}^2} \sum_{i,j} \ee^{\ii ka (i-j)} \langle T^{y}_i T^{y}_j \rangle,\\
S_{T_x T_x}(k) = \frac{1}{N_{\rm s}^2} \sum_{i,j} \ee^{\ii ka (i-j)} | \langle T^{x}_i T^{x}_j \rangle |,
\end{split}
\eeq   
where the orbital spin operators are defined in Eq.~\eqref{eq:DM_operators}.
The  zero-momentum component of these structure factors yield the  desired  magnetizations in the thermodynamic limit $M_y=\left( S_{T_y T_y} (0) \right)^{1/2}$, $M_x=\left( S_{T_x T_x} (0) \right)^{1/2}$. In the  insets of Figs.~\ref{fig:chi_mags_SI} {\bf (a)} and {\bf (c)}, we show the two $M_y,M_x$ order parameters, which attain a non-zero value for small energy imbalance, and  for  large $\theta=0.65\pi$ and small $\theta=0.1\pi$ magnetic fluxes, respectively. To locate the corresponding critical points delimiting  the phase boundaries, we calculate   the respective susceptibilities  $\chi_{M_y}=\partial M_y / \partial \Delta \epsilon$, and $\chi_{M_x}=\partial M_x / \partial \Delta \epsilon$, which again show peaks whose FSS should determine the critical lines accurately. In the main panels of Fig.~\ref{fig:chi_mags_SI} {\bf (a)} and {\bf (c)}, we can see how the susceptibilities peak at symmetric values of the zero-imbalance axis, and how these peaks grow with the lattice size. The numerical results shown in Figs.~\ref{fig:chi_mags_SI} {\bf (b)} and {\bf (d)} show that the susceptibility peaks agree considerably well with the analytical predictions already for these finite interactions and  ladder lengths.

\begin{figure*}[t]
\centering
\includegraphics[width=0.8\linewidth]{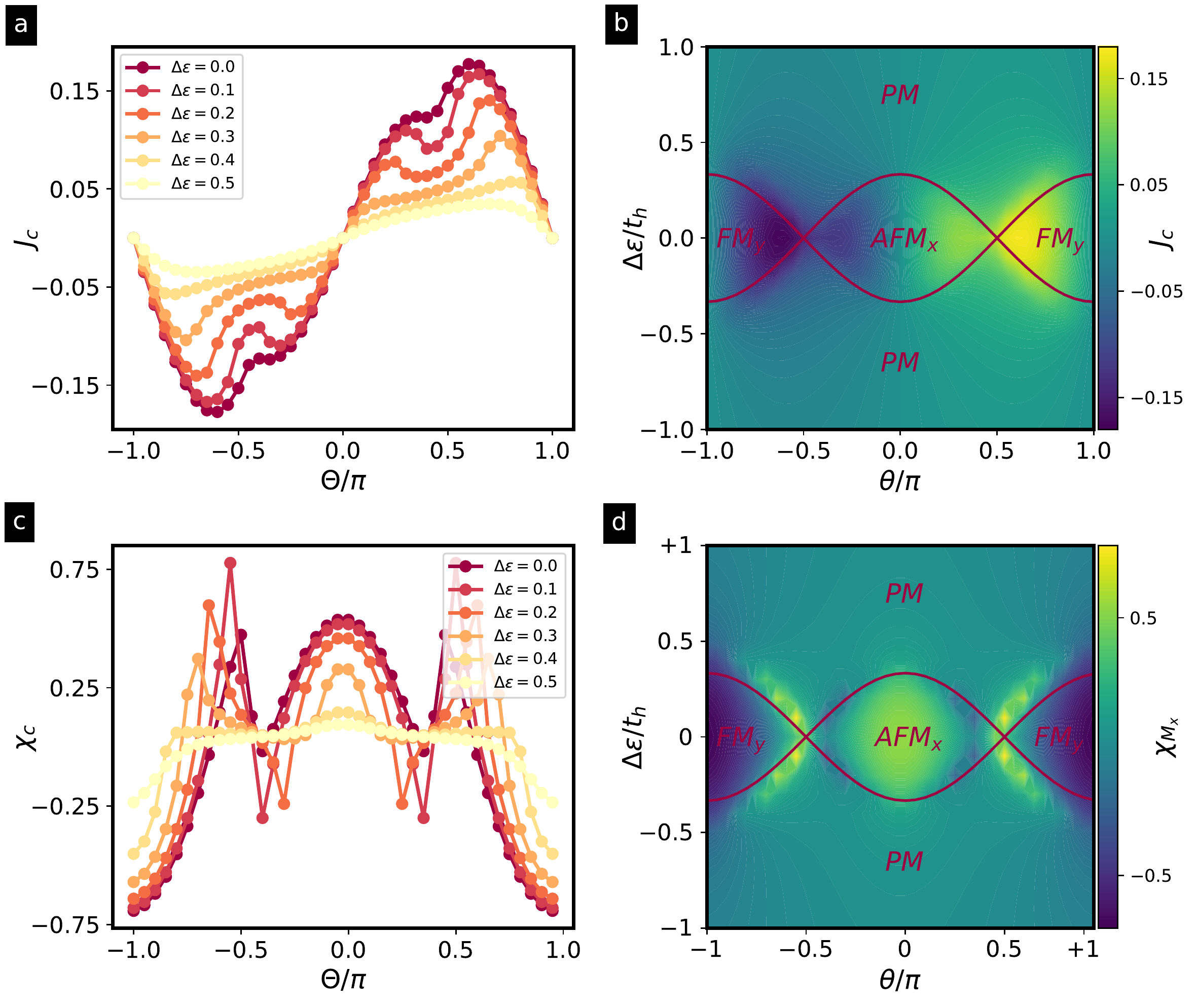}
\caption{\label{fig:chiral_sus_strong_int} {\textbf{Chiral flows for  strong interactions:}
{\bf (a)} Chiral current as a function of the magnetic flux for different energy  imbalances $\Delta \epsilon$ and $V_v=12 t_h$. {\bf (b)} The contour  plot of the the chiral current  qualitatively distinguishes the two phases. The red line shows the transition points of the effective spin model with D-M interaction in strong coupling limit (\ref{eq:DM_interaction}). {\bf (c)} Divergence of chiral susceptibility
for different imbalance $\Delta \epsilon$ that indicates the phase transition between FM$_y$-AFM$_x$ phases. {\bf (d)} The maximum of $\chi_c$ that delimit the FM$_y$ and AFM$_x$ lobes.}}
\end{figure*}

In Fig.~\ref{fig:chiral_sus_strong_int}, we show that the correlated FM$_y$ and AFM$_x$ can still support large chiral currents around the ladder.
In fact, the ferromagnetic and anti-ferromagnetic long-range order phase supports persistent chiral current for small imbalance $\Delta \epsilon$, and it disappears in the orbital magnetic phase. This suggests that the correct indicator for the phase transition FM$_y$-AFM$_x$ is the chiral susceptibility. Moreover this phenomenon is related to the Chiral Mott insulator phase introduced for quasi-one-dimensional systems \cite{dhar2012bose,dhar2013chiral,tokuno2014ground} and for two-dimensional systems \cite{zaletel2014chiral,aji2007spin,chua2011exact,messio2012kagome,yan2011spin,huerga2014chiral}.
In Fig.~\ref{fig:chiral_sus_strong_int} {\bf (a)}, we show the numerical results of the chiral current in function of the flux $\theta$ for different imbalance interactions. 
In Fig.~\ref{fig:chiral_sus_strong_int} {\bf (b)}, we present the contour plot
of the chiral current. It shows an inversion around the critical line predicted by the eq.~\eqref{eq:strong_coupling_critical_line}. To locate the critical points delimiting the phase boundaries, we calculate the chiral susceptibility $\chi_c$. In Fig.~\ref{fig:chiral_sus_strong_int} {\bf (c)} we can see how the chiral susceptibility symmetric peaks  at symmetry values of the zero-$\theta$ axis. Moreover, the numeric results shown in  Fig.~\ref{fig:chiral_sus_strong_int} {\bf (d)} show that the susceptibility peaks agree well with the critical line predicted by the eq.~\eqref{eq:strong_coupling_critical_line} for the FM$_y$-AFM$_x$ phase transition. 


\subsection{Entanglement spectroscopy}
\label{eq:entamglement_spec}

Thus far, we have used a conventional condensed-matter approach to characterise the phase diagram of the model, which is based on exploiting currents, susceptibilities, and correlation functions to identify phases with long-range order or symmetry-protected topological phases. In this section,  we follow an alternative approach based on the ground-state entanglement to understand the phase diagram of the model.

Among the various existing measures of entanglement on a bipartite scenario \cite{plenio2014introduction}, we compute the entanglement spectrum and the entanglement entropy that enjoy a privileged status within the realm of quantum many-body system \cite{amico2008entanglement}. As regards the first one, we define a bipartition of the system, and write the ground-state as $|\psi_{g.s.} \rangle=\sum_n \lambda_n |\psi_n\rangle_L |\psi_n\rangle_R $, where $L$ and $R$ are two subsystem, and $\lbrace \lambda_n\rbrace$ are the corresponding Schmidt coefficients. The entanglement spectrum is defined as the set of all the Schmidt coefficients
in logarithmic scale $\epsilon_n=-2 \log(\lambda_n)$, and it can be directly extracted  from the MPS calculations. As originally pointed 
out in \cite{li2008entanglement}, in the context of the characterization of the Haldane phase of Heisenberg-type magnets, the degeneracy of the entanglement spectrum identifies the symmetry protecting topological phase.

\begin{figure*}[t]
\centering
\includegraphics[width=1\linewidth]{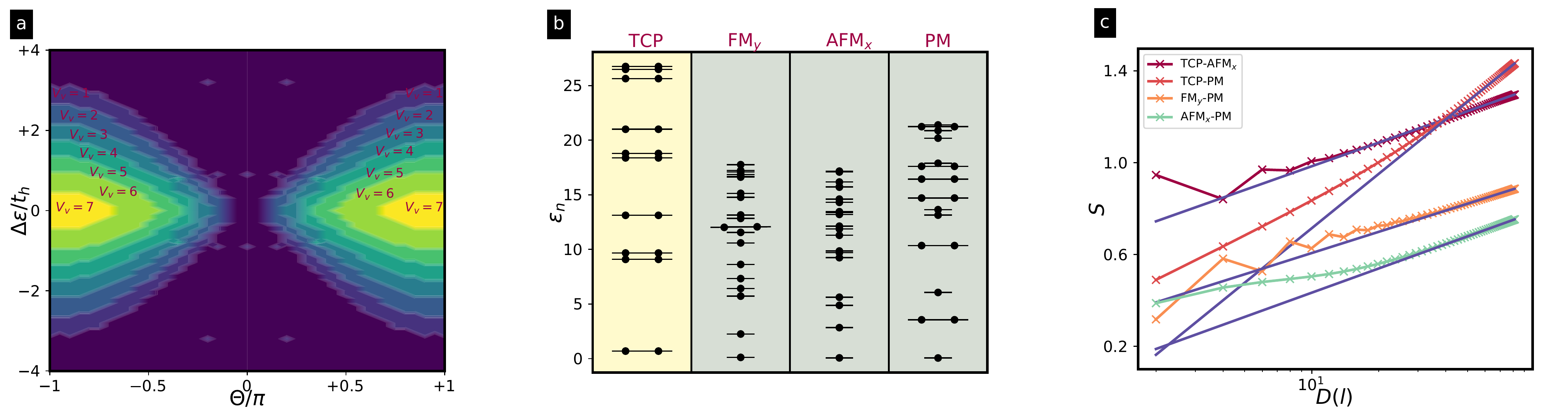}
\caption{\label{fig:ent_spectrum_entropy} {\textbf{Entanglement spectrum and entanglement entropy:} {\bf (a)} Degeneracies of the entanglement spectrum in the $(\theta,\Delta \epsilon)$ plane for different interactions. We show that the region in which the entanglement spectrum shows an exact two-fold degeneracy, and thus corresponds to the topological crystalline phase, gets reduced by increasing the interactions. {\bf (b)} Low-lying eigenvalues of the entanglement spectrum for the different phases. For a ladder of length $N_{\rm s}=128$, and for a half-chain bipartition of the chain. Onlye the TCP phase dysplays an exact two-fold degeneracy. {\bf (c)} Entanglement entropy for a $l$-sites block reduced density matrix $\rho_l$ obtained from a ladder of length $N_s = 128$. The blue, red, and yellow crosses correspond to the data for the critical points. The solid lines correspond to the fittings with the conformal field theory predictions~\eqref{eq:cft_entropy}, where $c$ is the central charge. The fitted central charges are $c = 1.02$ (for the TCP-PM transition), $c = 0.503$ (for the TCP-AFM$_x$ transition), $c = 0.49$ (for the FM$_y$-PM transition), and $C=0.45$ (for the AFM$_x$-PM transition), respectively.}}  
\end{figure*}

It has been established that the entanglement spectrum is degenerate for 
symmetry-protected topological phases \cite{pollmann2010entanglement,fidkowski2010entanglement,pollmann2012symmetry}. In particular, this degeneracy is robust against symmetric perturbations as long as the many-body gap of the system  remains open. In Fig. \ref{fig:ent_spectrum_entropy} {\bf (a)}, we present the entanglement spectrum degeneracy for the Creutz-Hubbard model in function of $\left( \theta / \pi, \Delta \epsilon / t_h \right)$ for different values of interactions strength $V_{\rm v}$. The dark blue region represents the region where the AFM$_x$ and PM are sited and the entanglement spectrum is trivial. Instead, the two lobs represent the TCP phase in which the entanglement spectrum is fully degenerate. The effect of the interactions shrinks the two TCP lobes. In fact, as shown in the Fig. \ref{fig:ent_spectrum_entropy} {\bf (a)}, the spectrum degenerate region is reduced.       
As shown in Fig. \ref{fig:ent_spectrum_entropy} {\bf (b)} (yellow column),  the entanglement spectrum in the TCP phase is clearly  doubly degenerate, whereas in the PM, FM$_y$ and AFM$_x$ phases such a degeneracy of all eigenvalues  is lost (green columns). This supports our claim about the topological nature of the wide region of the phase diagram labeled as TCP, and demonstrates that the topological crystalline  phase survives to considerable strong interactions. 
In particular, in the left panel of Fig. \ref{fig:ent_spectrum_entropy} {\bf (b)}, we present the entanglement spectrum for weak interactions for different values of the imbalance $\Delta \epsilon$ and fluxes $\theta$. We consider a bipartition in the middle of chain, for $\Delta \epsilon=0$ and $\theta=0.5 \pi$. 
Increasing $\Delta \epsilon$ and decreasing $\theta$, we enter in the AFM$_x$ or in PM phase, and the entanglement spectrum is trivial and almost completely
degenerate, as shown in the green columns. 

As regards the block entanglement entropy, it is defined as 
\beq
S(l)=-{\rm Tr} \left( \rho_l \ln \rho_l \right),
\eeq     
where $\rho_l={\rm Tr}_{N_{\rm s}-l} \left( |\epsilon_{g.s.} \rangle \langle \epsilon_{g.s.}|  \right)$ is the reduced density matrix of the left block with $l$ sites
for bipartition of each leg of the ladder of $N_{\rm s}$ sites. Remarkably enough, not only the 
block entanglement serve as a probe of criticality due to its divergence at phase transition, but its scaling with system size also reveals the central
charge $c$ of the conformal field theory underlying the critical behavior \cite{vidal2003entanglement,calabrese2004entanglement,calabrese2009entanglement}. For a critical system with open boundary conditions, the block
entanglement entropy scales as
\beq 
\label{eq:cft_entropy}
S(l)= \frac{c}{6} \ln \left( \frac{2 N_{\rm s}}{\pi} \sin \frac{\pi l}{N_{\rm s}} \right)+C,
\eeq   
where we introduce a non universal constant $C$.
Since such entanglement entropy can be easily recovered from our MPS results, calculating the central charge of the different critical lines of our phase diagram can serve as an additional confirmation of our previous derivations. To explore the scaling region of the quantum phase transition
between topological and trivial band insulators,  we represent in Fig. \ref{fig:ent_spectrum_entropy} {\bf (c)} the
critical entanglement entropy as a function of the so-called chord length
$D(l,N_{\rm s})=(2N_{\rm s}/ \pi) \sin (\pi l / N_{\rm s})$, along the critical lines in 5 representative cases. For the TCP-PM transition the central charge value 
agrees with $c=1$. This result can be understood well in the weakly 
interesting limit. In this regime, the Creutz-Hubbard model for generic flux can be understood as two coupled Ising chains, as shown in the Appendix \ref{sec:appendix}.
In the limit of $\pi$-flux, the model is mapped to a couple of Ising chains
in transverse magnetic field. Accordingly, the corresponding CFT should
have central charge of $c=1/2+1/2=1$ such that we would expect the scaling
$S(l)=(1/6) \ln [ (2N_{\rm s}/\pi) \sin(\pi l/N_{\rm s})+a]$. In Fig. \ref{fig:ent_spectrum_entropy} {\bf (c)}, we also show that prediction are confirmed also for a various fluxes. 

For the strongly interacting regime, we showed in previous sections that the FM$_{y}$-PM and AFM$_x$-PM transitions can be accounted for by an XY model with D-M interaction. In the limit
of $\theta=\pi$, or $\theta=0$, the model becomes a single Ising model in a 
transverse field. Accordingly, the corresponding CFT should have a central charge of $c=1/2$, and $S(l)=\frac{1}{12} \ln [(\frac{2N_{\rm s}}{\pi}) \sin (\frac{\pi l }{N_{\rm s}}) ]+C$. In contrast to the previous case, the critical phenomena is governed by the CFT of single Majorana fermion with central charge $c=1/2$.

Finally, in the intermediate interacting regime we argue that the relevant physics to understand the TI-FM phase transition is by approximating a complicated non standard model. In Fig. \ref{fig:ent_spectrum_entropy} {\bf (c)}, we show the entanglement entropy between the $N$-physical sites in the original basis  and therefore measure $S(l)=\frac{1}{12} \ln [(\frac{2N_{\rm s}}{\pi}) \sin (\frac{\pi l }{N_{\rm s}}) ]+C$ for a system of length $N_{\rm s}$ with a central charge of $c=1/2$. 
   
\section{\bf Cold-atom Raman lattice scheme}
\label{sec:cold_atoms}

 As outlined in the introduction, despite the fact that laser-cooled  gases of neutral atoms   move at velocities well below the speed of light~\cite{RevModPhys.80.885}, these systems yield an ideal platform  for the quantum simulation~\cite{Feynman_1982} of relativistic QFTs~\cite{1911.00003}. Following the perspectives of lattice field theories, one need not look for cold-atom quantum simulators  directly described by  continuum QFTs~\eqref{eq:cont_limit_L}, but instead look for a lattice model with critical points around which  a long-wavelength description  coincides with the target relativistic QFT. In this section, we describe a scheme based on cold atoms in optical Raman lattices for  the quantum simulation of the  Creutz-Hubbard ladder  under an arbitrary   flux~\eqref{eq:H_C}-\eqref{eq:H_CH}. As emphasized previously, such a quantum simulator would provide the first experimental realization of Lorentz-violating terms~\eqref{eq:gamma_Lviolation} studied in the context of the SME~\cite{PhysRevD.58.116002,PhysRevD.65.056006,COLLADAY2001209}, and combined with four-Fermi interactions of the Gross-Neveu type.

We note that the fermionic Creutz ladder has been recently realized in experiments with two-orbital ultracold-atoms in shaken optical lattices~\cite{PhysRevLett.121.150403,Kang_2020}.  In principle,  interaction effects could  be explored by including more hyperfine states. Note, however, that such interacting models would likely correspond to a spin-full version of the Creutz ladder with short-range Hubbard-type interactions (see Fig.~\ref{fig:scheme_ladders} {\bf (e)}), which is different from the target model~\eqref{eq:H_C}-\eqref{eq:H_CH}. To build the desired quantum simulator,  the   scheme proposed in Ref.~\cite{PhysRevX.7.031057}  exploits the idea of synthetic dimensions~\cite{PhysRevLett.108.133001} by mapping the ladder legs to a couple of hyperfine atomic states from the groundstate manifold~\cite{PhysRevLett.102.135702,Mazza_2012}. The attractive feature  of this scheme is that the interactions along the vertical direction~\eqref{eq:H_CH} in Fig.~\ref{fig:scheme_ladders} {\bf (g)}, the so-called synthetic dimension,  are mapped onto  contact Hubbard interactions, which are readily implemented by the $s$-wave scattering of two ultracold atoms trapped in the same site of the optical lattice~\cite{PhysRevLett.81.3108,PhysRevLett.89.220407}.  

\begin{figure*}[t]
  \centering
  \includegraphics[width=0.8\linewidth]{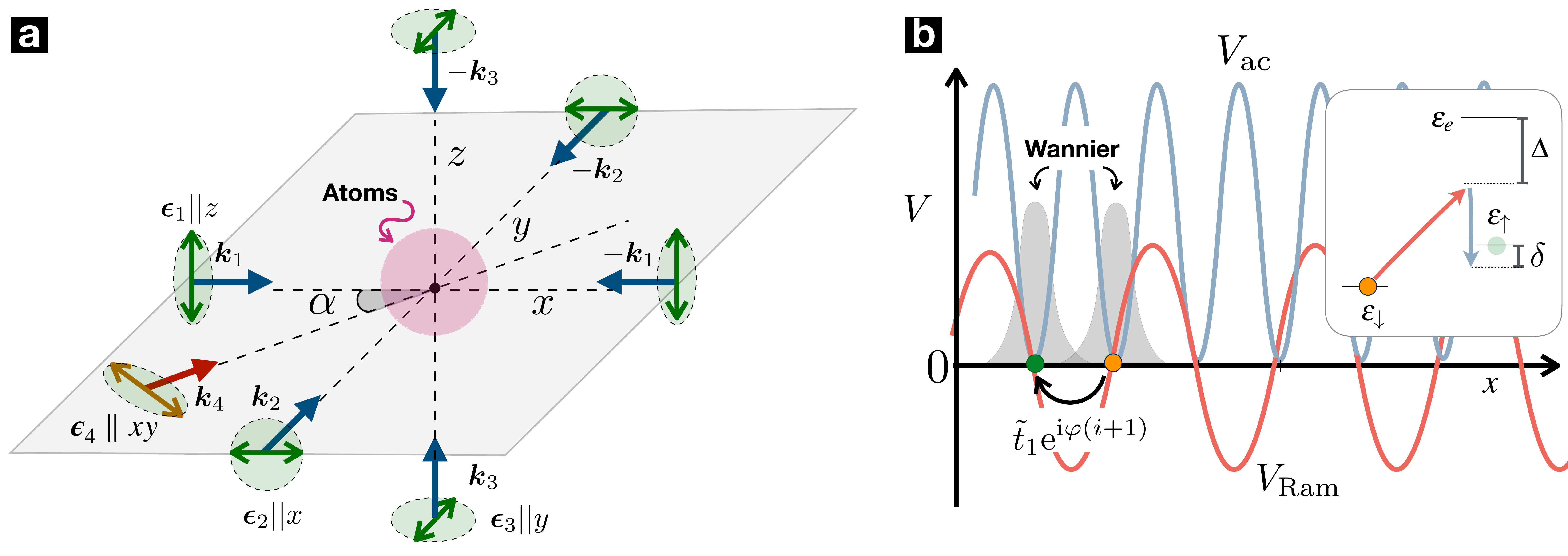}
\caption{\label{fig:scheme_raman_lattice} {\textbf{Scheme of the optical Raman potential for the quantum simulation of the Creutz-Hubbard ladder:}  {\bf (a)} A cloud of atoms, represented by a pink shaded sphere, is subjected to three pairs of counter-propagating laser beams with mutually-orthogonal polarisations, here depicted with blue arrows, which create a cubic optical lattice. A fourth laser beam, represented by a red arrow,  propagates in the $xy$ plane with an angle $\alpha$ with respect to the $x$ axis, sharing the polarization with the standing wave along that axis $\epsilon_4||\epsilon_1$. The interference of the standing- and traveling-wave  drives a two-photon Raman transition between two hyperfine states when the beatnote is tuned close to the resonance of the latter $\Delta\omega=\omega_1-\omega_4=\epsilon_\uparrow-\epsilon_\downarrow-\delta=\omega_0-\delta$ by virtually populating a highly-off-resonant excited state $\delta\ll\omega_0\ll\Delta$ (see the inset of {\bf (b)}). As shown in {\bf (b)}, due to this interference, this Raman term $V_{\rm Ram}$ has a doubled period with respect to the static optical lattice $V_{\rm ac}$. The atoms trapped at the minima of this optical lattice do not see any intensity of the Raman beams, and there is no local transitions. On the other hand, the overlap between neighboring Wannier functions   of oppossite spin mediated by this Raman potential can lead to a spin-flipping tunneling of strength  $\tilde{t}_1$ with a site-dependent phase $\varphi$, which is controlled by the tilting angle $\alpha$.  }}
\end{figure*}

Synthetic dimensions have been successfully exploited for the quantum simulation of rectangular ladders under a  background magnetic field. The  associated magnetic flux across the square plaquettes of the ladder   can be  simulated by a local Raman transition between the two hyperfine states~\cite{PhysRevLett.112.043001,Mancini:2015uqa,stuhl_lu_aycock_genkina_spielman_2015}, which induces a Peierls-substituted vertical tunneling that connects the upper and lower legs of the ladder (see Fig.~\ref{fig:scheme_ladders} {\bf (c)}). For the Creutz ladder, the situation is complicated further due to the cross-link nature of the inter-leg tunelings (see Fig.~\ref{fig:scheme_ladders} {\bf (g)}). As proposed in~\cite{PhysRevX.7.031057}, these cross-linked terms could be implemented by a Raman-assisted tunneling scheme~\cite{Jaksch_2003} that activates a spin-flipping tunneling against the energy penalty of a linear lattice tilt.  Note that, as the cross-link tunnelings take place in both directions, i.e. $({\rm u},i)\to({\rm d},i+1)$ and $({\rm d},i)\to({\rm u} ,i+1)$, the scheme does not require the use of state-dependent lattices as occurs for the simulation of the Hofstadter model~\cite{Jaksch_2003}, minimising in this way the expected heating due to residual spontaneous emission in the fermionic Raman couplings. The implementation of the horizontal tunnelings (see Fig.~\ref{fig:scheme_ladders} {\bf (g)})  requires a different  mechanism to assist the spin-conserving tunneling while imprinting an effective Peierls' phase. In the scheme of Ref.~\cite{PhysRevX.7.031057},
this is provided by an additional shallower lattice potential with a doubled wavelength, the  intensity of which is periodically modulated. This allows for a Floquet-assisted tunneling mechanism with the desired Peierls' phase, mimicking the role of the external magnetic flux. 

We note that the combination of Floquet-assisted tunneling with Hubbard interactions, as would be required for this full quantum simulator of the Creutz-Hubbard ladder~\cite{PhysRevX.7.031057}, may present limitations due to a heating mechanism where the atoms get excited by resonantly  absorbing quanta from the periodically-driven radiation~\cite{PhysRevLett.119.200402,PhysRevX.10.011030,RevModPhys.89.011004}. Although there has been  promising  progress in the  minimization of   this heating in specific optical-lattice implementations~\cite{PhysRevA.98.023623,PhysRevLett.121.233603,Gorg2019,PhysRevX.11.011057}, it would be desirable to come up with a different scheme that implements the Peierls phases in the cross-linked geometry. The goal of this section is to present such a scheme.

 We show that the recent   progress in synthetic spin-orbit coupling with cold atoms in the so-called optical Raman  potentials~\cite{liu_zhang_materials_liu} can be exploited for the full quantum simulator of the Creutz-Hubbard ladder~\eqref{eq:H_C}-\eqref{eq:H_CH}. In these schemes, one combines   periodic ac-Stark shifts from a standing wave  with Raman potentials that stem from the   interference of lasers in  standing-  and traveling-wave configurations.  As discussed in more detail below, when the beatnote of the standing- and traveling-wave laser beams is tuned close to the spin-flip transition frequency, the Raman cross terms  do not drive local spin flips~\cite{PhysRevLett.110.076401,PhysRevLett.112.086401,PhysRevLett.113.059901,ziegler2020correlated,ziegler2021largen}, but  instead assist a spin-flipping tunneling that can be used to simulate spin-orbit coupling~\cite{PhysRevLett.121.150401,Songeaao4748,Wu83,liang2021realization}. Although these Raman schemes may also be limited  by   residual photon scattering from the off-resonant excited state, the associated heating is not as severe as that of fermions trapped in  state-dependent lattices since, if the available laser intensity is not the limiting factor, this heating can  be minimised by working with larger Raman detunings $\Delta$ with respect to the excited state (see inset of Fig.~\ref{fig:scheme_raman_lattice} {\bf (b)}). In any case, one may also consider using lanthanide~\cite{Gerbier_2010,PhysRevX.6.031022,PhysRevA.94.061604,PhysRevLett.117.220401} or alkali-earth~\cite{Kolkowitz2017,Bromley2018,liang2021realization} atoms in order to minimise it further.

Let us now describe in detail how these periodic optical potentials can be  exploited for the specific quantum simulation of the Creutz-Hubbard ladder~\eqref{eq:H_C}-\eqref{eq:H_CH}, and how to tune the value of the synthetic magnetic flux $\theta$. In contrast to the scheme that realizes Rashba- or Dirac-type spin-orbit coupling~\cite{liu_zhang_materials_liu}, we consider a configuration where the traveling wave does not propagate orthogonally to the $x$-axis standing wave, but instead at an angle $\alpha$ (see Fig.~\ref{fig:scheme_raman_lattice} {\bf (a)}). In addition, we consider  two additional standing waves with orthogonal polarisations along the remaining spatial directions, which will lead to periodic optical potentials along the $y$ and $z$ axes. In the regime where the lasers are far detuned with respect to an excited state, this creates a periodic optical potential for the atoms in two hyperfine states $\sigma\in\{{{\uparrow}},{{\downarrow}}\}$, namely
\beq
\label{eq:optical_potential}
V\!=\!\sum_{\alpha}\!\frac{V_{0,\alpha}}{2}\!\cos^2\!(\boldsymbol{k}_{\alpha}\cdot\boldsymbol{x})\mathbb{I}_2+\frac{\tilde{V}_{0}}{2}\!\cos (\boldsymbol{k}_{1}\cdot\boldsymbol{x})\ee^{\ii(\boldsymbol{k}_{4}\cdot\boldsymbol{x}-\Delta\omega t)}\!\sigma^++{\rm H.c.},
\eeq
where $\sigma^+=\ket{{\uparrow}}\bra{{\downarrow}}$. Here, $V_{0,\alpha}$ are the amplitudes of the ac-Stark shifts induced by  pairs of counter-propagating laser beams  along each of the axes $\boldsymbol{k}_\alpha=k_\alpha{\bf e}_\alpha$ with $\alpha\in\{1,2,3\}\equiv\{x,y,z\}$. These pairs of laser beams have mutually orthogonal polarizations, such that the interference patterns along each axis in the first term of Eq.~\eqref{eq:optical_potential}   can be tuned independently, and  thus
corresponds to a standard cubic  optical lattice. Additionally, $\tilde{V}_{0}$ is the amplitude of the two-photon Raman transition, whereby the spin state  gets flipped by virtually populating an excited state. As shown in the inset of (see Fig.~\ref{fig:scheme_raman_lattice} {\bf (b)}), this transition involves absorbing a photon from the travelling wave, which  propagates along   $\boldsymbol{k}_4=k_4\cos\alpha{\bf e}_1+k_4\sin\alpha{\bf e}_2$, and subsequently re-emitting it in the standing wave. This leads to a beatnote  of frequency  $\Delta\omega=\omega_4-\omega_1$ that can drive the spin flips when tuned close to the resonance  $\Delta\omega\approx\omega_0$. A crucial aspect of this type of schemes, which can already be appreciated at this stage, is that the period of the spin-flipping potential is doubled with respect to that of the static optical lattice (see Fig.~\ref{fig:scheme_raman_lattice} {\bf (b)}). As such, the atoms residing at the minima of the latter do not see any Raman intensity that would drive local transitions, but it is only in the spin-flip tunneling where this Raman term contributes.

As noted at the beginning of this section, these systems can be described by the following non-relativistic QFT
\beq
\begin{split}
\label{eq:atom_ft}
H&=\!\int\!\!{\rm d}^3x\!\!\sum_{\sigma,\sigma'}{\Phi}^{\dagger}_{\sigma}(\boldsymbol{x})\left(-\frac{\boldsymbol{\nabla}^2}{2m}+\bra{\sigma}V(\boldsymbol{x})\ket{\sigma'}\right){\Phi}^{\phantom{\dagger}}_{\sigma'}(\boldsymbol{x})\\
&+\!\!\int\!\!\!{\rm d}^3x\!\!\!\int\!\!\!{\rm d}^3x'\!\!\sum_{\sigma,\sigma'}\!\!{\Phi}^{{\dagger}}_{\sigma}(\boldsymbol{x}){\Phi}^{{\dagger}}_{\sigma'}(\boldsymbol{x}')\frac{2\pi a_{s}}{m}\delta(\boldsymbol{x}-\boldsymbol{x}'){\Phi}^{\phantom{\dagger}}_{\sigma'}(\boldsymbol{x}'){\Phi}^{\phantom{\dagger}}_{\sigma}(\boldsymbol{x})\,,
\end{split}
\eeq  
where $m$ is the mass of the atoms, and $a_s$ the $s$-wave scattering length for the low-energy two-body collisions. We have introduced the fermionic field operators ${\Phi}^{{\dagger}}_{\sigma}(\boldsymbol{x}), {\Phi}^{\phantom{\dagger}}_{\sigma}(\boldsymbol{x})$ that create-annihilate an atom at position $\boldsymbol{x}$ with  internal state $\sigma$. Note that these fields differ from the relativistic ones~\eqref{eq:cont_limit}, which will be recovered in the long-wavelength limit of the lattice model. To obtain such a lattice description, we consider the regime of deep optical lattices $V_{0,\alpha}\gg E_{\rm R,\alpha}=k_\alpha^2/2m$, where the atoms are tightly confined to the minima of the periodic blue-detuned optical lattice (see Fig.~\ref{fig:scheme_raman_lattice} {\bf (b)}). The lattice sites thus correspond to $\boldsymbol{x}_{\boldsymbol{i}}^0=\sum_\alpha\frac{\lambda_{\alpha}}{2}(i_\alpha+\half){\bf e}_\alpha$, where $\lambda_\alpha=2\pi/k_\alpha$ are the wavelengths of the standing-wave laser beams, and $i_\alpha\in\mathbb{Z}_{N_{\rm s}}$. We now express the fields  in the  Wannier basis
\beq
{\Phi}^{\phantom{\dagger}}_{\sigma}(\boldsymbol{x})=\sum_{\boldsymbol{ i}}w(\boldsymbol{x}-\boldsymbol{x}^0_{\boldsymbol{i}})f_{\boldsymbol{i},\sigma}^{\phantom{\dagger}},\hspace{1ex} {\Phi}^{{\dagger}}_{\sigma}(\boldsymbol{x})=\sum_{\boldsymbol{ i}}w(\boldsymbol{x}-\boldsymbol{x}^0_{\boldsymbol{i}})f_{\boldsymbol{i},\sigma}^{{\dagger}},
\eeq
where $w(\boldsymbol{x}-\boldsymbol{x}^0_{\boldsymbol{i}})$ are the Wannier functions localised around each minima $\boldsymbol{x}^0_{\boldsymbol{i}}$ of the optical potential, and $f_{\boldsymbol{i},\sigma}^{{\dagger}}, f_{\boldsymbol{i},\sigma}^{\phantom{\dagger}}$ is a set of dimensionless lattice operators that describe the creation-annihilation of fermions in the lowest band of the optical lattice. Substituting these expressions in the QFT~\eqref{eq:atom_ft}, one obtains a lattice model with parameters that depend on the  overlaps of these Wannier functions. These parameters decay exponentially fast with the distance, and are typically restricted to on-site and nearest-neighbor terms~\cite{PhysRevLett.81.3108,PhysRevLett.89.220407}. 

This lattice model is composed of spin-conserving terms 
\beq
\label{eq:spin_conserving_terms}
H_{\rm sc}\!=\!\sum_{\boldsymbol{i}}\!\!\left(\!\!\sum_{\sigma,\alpha}\!\!\! \left(\!\!-t_\alpha f_{\boldsymbol{i},\sigma}^{{\dagger}}f_{\boldsymbol{i}+{\bf e}_\alpha,\sigma}^{\phantom{\dagger}}+{\rm H.c.}\!\right)\!\!+\!\!\!\sum_{\sigma\neq\sigma'}\!\!\!\frac{U_{\uparrow\downarrow}}{2}\!n_{\boldsymbol{i},\sigma}^{\phantom{\dagger}}n_{\boldsymbol{i},\sigma'}^{\phantom{\dagger}}\!\!\!\right)\!\!,
\eeq 
where $n_{\boldsymbol{i},\sigma}^{\phantom{\dagger}}=f_{\boldsymbol{i},\sigma}^{{\dagger}}f_{\boldsymbol{i},\sigma}^{\phantom{\dagger}}$, and  we have introduced the tunnelings 
\beq
\label{eq:tunneling_strengths}
t_\alpha=\frac{4}{\sqrt{\pi}}E_{\rm R}\left(\frac{V_{0,\alpha}}{E_{\rm R}}\right)^{\frac{3}{4}}\ee^{-2\sqrt{\frac{V_{0,\alpha}}{E_{\rm R}}}},
\eeq
assuming that the standing waves have the same wavelength $\lambda_\alpha=2\pi/k, \,\forall\alpha$. Likeweise, the Hubbard interaction reads
\beq
\label{eq:Hubbard_cold_atoms}
U_{\uparrow\downarrow}=\sqrt{\frac{8}{\pi}}ka_sE_{\rm R}\left(\frac{V_{0,1}V_{0,2}V_{0,3}}{E_{\rm R}^3}\right)^{\frac{1}{4}}.
\eeq
By setting $V_{0,2},V_{0,3}\gg V_{0,1}\gg E_{\rm R}$, the tunnelings along the $y$ and $z$ axes~\eqref{eq:tunneling_strengths} are inhibited $t_2,t_3\ll t_1$, and we obtain an effective Fermi-Hubbard model in one spatial dimension. Note that, in addition to these spin-conserving  terms,  there will also be local terms corresponding to on-site energies that can be approximated by an overall quadratic trapping potential.

In addition to these contributions, we have the spin-fliping Raman potential~\eqref{eq:optical_potential} that stems from the interference of the standing and traveling waves along the $x$ axis, and will lead to spin-flipping terms in the lattice model. In this case, the local terms are not simply an on-site potential, as discussed in the previous paragraph but, instead,  describe  processes where an atom  tightly confined to a minimimum of the optical potential experiences a Raman transition that flips its internal state $\ket{\uparrow}\leftrightarrow\ket{\downarrow}$. This type of terms can indeed be exploited to implement the synthetic dimensions discussed in the introduction ~\cite{PhysRevLett.112.043001,Mancini:2015uqa,stuhl_lu_aycock_genkina_spielman_2015}. Note that these terms correspond to a vertical tunneling within the synthetic ladder, and would thus spoil the   cross-linked nature of the Creutz-ladder of Fig.~\ref{fig:scheme_ladders} {\bf (g)}. Fortunately, the corresponding Wannier  integrals  vanish since the Raman potential is odd with respect to each lattice site  while the Wannier functions are even (see Fig.~\ref{fig:scheme_raman_lattice} {\bf (b)}),
\beq
\int\!\!{\rm d}^3x w(\boldsymbol{x}-\boldsymbol{x}^0_{\boldsymbol{i}})\frac{\tilde{V}_{0}}{2}\!\cos (\boldsymbol{k}_{1}\cdot\boldsymbol{x})\ee^{\ii\boldsymbol{k}_{4}\cdot\boldsymbol{x}}w(\boldsymbol{x}-\boldsymbol{x}^0_{\boldsymbol{i}})=0.
\eeq
The leading spin-flipping terms are thus described by tunnelings along the $x$ axis that stem from the following overlaps
\begin{widetext}
\beq
\int\!\!{\rm d}^3x w(\boldsymbol{x}-\boldsymbol{x}^0_{\boldsymbol{i}})\frac{\tilde{V}_{0}}{2}\!\cos (\boldsymbol{k}_{1}\cdot\boldsymbol{x})\ee^{\ii\boldsymbol{k}_{4}\cdot\boldsymbol{x}}w(\boldsymbol{x}-\boldsymbol{x}^0_{\boldsymbol{j}})\approx
\frac{\tilde{V}_0}{2}\ee^{-\fourth m\nu|\boldsymbol{x}^0_{\boldsymbol{i}}-\boldsymbol{x}^0_{\boldsymbol{j}}|^2}\cos\left(\half\boldsymbol{k_1}\cdot(\boldsymbol{x}^0_{\boldsymbol{i}}+\boldsymbol{x}^0_{\boldsymbol{j}})\right)\ee^{\frac{\ii}{2}\boldsymbol{k_4}\cdot(\boldsymbol{x}^0_{\boldsymbol{i}}+\boldsymbol{x}^0_{\boldsymbol{j}})},
\eeq
\end{widetext}
where we have performed a Gaussian approximation around the minima of each lattice site, where the lattice potential can be approximated by a harmonic oscillator of frequency  $\nu=2\sqrt{V_{0,1}E_{\rm R}}$. As occurs for the spin-conserving terms, these tunnelings decay rapidly with the distance, such that we can restrict to nearest-neighbors. In this case, the overlaps display a crucial alternation of the sign, $\cos\big(\half\boldsymbol{k_1}\cdot(\boldsymbol{x}^0_{\boldsymbol{i}}+\boldsymbol{x}^0_{\boldsymbol{i}+{\bf e}_1})\big)={\rm exp}\{\ii\pi(i_1+1)\}$, which has been exploited to simulates synthetic spin-orbit coupling~\cite{PhysRevLett.121.150401,Songeaao4748,Wu83}. The novelty of our scheme is that  the non-orthogonal propagation of the traveling wave shown in  Fig.~\ref{fig:scheme_raman_lattice} {\bf (a)} allows us to pick up an arbitrary phase factor that shall also play a crucial role, namely  ${\rm exp}\{\frac{\ii}{2}\boldsymbol{k_4}\cdot(\boldsymbol{x}^0_{\boldsymbol{i}}+\boldsymbol{x}^0_{\boldsymbol{i}+{\bf e}_1})\}={\rm exp}\{\ii\eta(i_1+1)\}$, where 
\beq
\label{eq:eta_param}
\eta=\frac{k_4}{k_1}\cos\alpha.
\eeq
 Note that there is also a dependence of this phase factor on the particular lattice site along the $y$ axis, but the negligible tunnelings along this direction allow us to gauge it away without changing the dynamics.

Let us now tune the Raman beatnote  slightly off-resonance with respect to the internal transition $\Delta\omega=\omega_0-\delta$, where the detuning is constrained to $\delta\ll\omega_0$. Considering the discussion in the previous paragraph,
and working in a rotating frame, we find that the Raman potential leads to the following spin-flipping tunnelings and energy imbalance
\beq
\begin{split}
\label{eq:spin_flipping_terms}
H_{\rm sf}\!=&\!\sum_{\boldsymbol{i}}\tilde{t}_1 \!\!\left(\ee^{\ii\varphi(i_1+1)}f_{\boldsymbol{i},\uparrow}^{{\dagger}}f_{\boldsymbol{i}+{\bf e}_1,\downarrow}^{\phantom{\dagger}}+\ee^{-\ii\varphi(i_1+1)}f_{\boldsymbol{i},\downarrow}^{{\dagger}}f_{\boldsymbol{i}+{\bf e}_1,\uparrow}^{\phantom{\dagger}}\right)\!\!+{\rm H.c.}\\
&+\!\sum_{\boldsymbol{i}}\frac{\delta}{2}\!\!\left(f_{\boldsymbol{i},\uparrow}^{{\dagger}}f_{\boldsymbol{i},\uparrow}^{\phantom{\dagger}}-f_{\boldsymbol{i},\downarrow}^{{\dagger}}f_{\boldsymbol{i},\downarrow}^{\phantom{\dagger}}\right),
\end{split}
\eeq
where we have introduced the following parameters
\beq
\label{eq:spin_flip_tunneling_strength}
\tilde{t}_1=\frac{\tilde{V}_0}{2}\ee^{-\frac{\pi^2}{4}\sqrt{\frac{V_{0,1}}{E_{\rm R}}}},\hspace{1ex}\varphi=\pi(1+\eta).
\eeq

The final step to define our quantum simulator  is to map  the fermionic operators of the Creutz-Hubbard ladder~\eqref{eq:H_C}-\eqref{eq:H_CH} via a synthetic dimension  to the cold-atom operators
\beq
\label{eq:gauge_transformation}
c_{i,{\rm u}}=\ee^{-\ii\frac{\varphi}{2}(i+1)}f_{i,\uparrow},\hspace{2ex} c_{i,{\rm d}}=\ee^{+\ii\frac{\varphi}{2}i}f_{i,\downarrow},
\eeq
such that the cold-atom lattice model in Eqs.~\eqref{eq:spin_conserving_terms}-\eqref{eq:spin_flipping_terms} maps directly onto  the Creutz-Hubbard ladder~\eqref{eq:H_C}-\eqref{eq:H_CH} with the following identification of microscopic parameters
\beq
t_{\rm h}=t_1,\hspace{1ex}t_{\rm d}=-\tilde{t}_1,\hspace{1ex}\Delta\epsilon=\delta,\hspace{1ex}\theta=\varphi,\hspace{1ex}V_{\rm v}=U_{\uparrow\downarrow}.
\eeq
Note that  the various  observables discussed in our study of the Creutz-Hubbard ladder may be modified by virtue of the  $U(2)$ gauge transformation~\eqref{eq:gauge_transformation}. One should thus identify the correct cold-atom operators that contain the relevant information about the groundstate phases of the Creutz-Hubbard ladder.

It follows from our discussion that all of these parameters can be  tuned independently. The horizontal tunneling is controlled by the intensity of the standing-wave lasers via Eq.~\eqref{eq:tunneling_strengths}. If, on the other hand,  one modifies the intensity of the travelling-wave laser, only the spin-flip tunneling will change~\eqref{eq:spin_flip_tunneling_strength}, which translates into a modification of the cross-link tunneling of the Creutz-Hubbard ladder
 (see Fig.~\ref{fig:scheme_ladders} {\bf (g)}). In order to modify the synthetic magnetic flux independently, Eqs.~\eqref{eq:eta_param}-\eqref{eq:spin_flip_tunneling_strength} show that one must either modify the angle $\alpha$  of propagation  of the travelling wave. Similarly, the energy imbalance is controlled by the beatnote detuning, which can be readily modified by changing the frequency of traveling-wave laser. Finally, the Hubbard interactions  can be tuned by exploiting a  Feshbach resonance in the scattering length~\eqref{eq:Hubbard_cold_atoms}.

\section{\bf Conclusions and outlook}
In this work, we have described the connection between Lorentz-violating
four-Fermi QFTs and models of correlated topological crystalline phases. 
By focusing on the  Creutz-Hubbard ladder for arbitarry magnetic flux, we have presented a through analysis based on  analytical and numerical tools, showing that, as a consequence of the violation of Lorentz invariance, persistent chiral currents can coexist with strong correlations in the   topological crystalline phase. We have discussed in detail the phase diagram of the model and the nature of the phase transitions, showing that the chiral current and its susceptibility provide relevant information, which  can be complemented with other observables and entanglement-related quantities. We have also discussed a experimental scheme based on ultra-cold Fermi gases in tilted Raman lattices for the quantum simulation of these phenomena. We refer the reader again to Sec.~\ref{sec:summary}, where all these results have been summarised.     We would like to  conclude by noting that our work constitutes an example of the useful
dialogue and exchange of ideas between the high-energy physics, condensed
matter, quantum information, and quantum optics, hopefully stimulating
further cross-disciplinary efforts in the future. 
From the above results, we are convinced that the synthetic Creutz-Hubbard ladder with generic flux can become a work horse in the theoretical and experimental study of correlated topological phases of matter, connecting  
high-energy physics and condensed matter, and proving a clear path to for its implementation via ultra-cold atoms. This would provide the first experimental realization of Lorentz-violating QFTs within the Standard Model extension.

As an outlook, it would be very interesting to exploit  these cross-disciplinary approaches to understand the physics of the 
Creutz-Hubbard ladder at generic flux, and thus  the  Gross-Neveu model extension,
 as one moves away from  half-filling and thus explores finite fermion densities in the continuum QFT. It will also be interesting to explore real-time dynamical effects in this model, in particular in connection with anomalies, the circulating chiral current, and the underlying violation of Lorentz symmetry. This would require to upgrade the flux $\theta$ to a dynamical variable with its own dynamics described by a $U(1)$ lattice gauge theory, following a similar gauging as in the case of a $\mathbb{Z}_2$ gauge group~\cite{PhysRevX.10.041007}. As discussed for other lattice models in~\cite{PhysRevD.99.014503, PhysRevB.100.115152}, $\mathbb{Z}_N$ discrete gauge groups could also be explored as a proxy for the $U(1)$ ladder gauge theory. It will be very interesting to study the interplay of this gauge theory  with the topological crystalline insulator and its associated edge states. We note that both the finite-density and real-time phenomena provide very challenging problems where the advantage of the quantum simulator could be exploited.  Finally, it would also be interesting to explore how other terms within the SME can be incorporated in a lattice model, and how this model could be realised in experiments of ultra-cold atoms.        
     
\section*{Acknowledgements}

ICFO group acknowledges support from: ERC AdG NOQIA; Agencia Estatal de Investigación (R\&D project CEX2019-000910-S, funded by MCIN/ AEI/10.13039/501100011033, Plan National FIDEUA PID2019-106901GB-I00, 
FPI, QUANTERA MAQS PCI2019-111828-2, Proyectos de I+D+I Retos Colaboración
RTC2019-007196-7);  Fundació Cellex; Fundació Mir-Puig; Generalitat de Catalunya through the CERCA program, AGAUR Grant No. 2017 SGR 134, QuantumCAT \ U16-011424, co-funded by ERDF Operational Program of Catalonia 2014-2020; 
EU Horizon 2020 FET-OPEN OPTOLogic (Grant No 899794); National Science Centre, Poland (Symfonia Grant No. 2016/20/W/ST4/00314); Marie Sk\l odowska-Curie grant STREDCH No 101029393; “La Caixa” Junior Leaders fellowships (ID100010434) and EU Horizon 2020 under Marie Sk\l odowska-Curie grant agreement No 847648 (LCF/BQ/PI19/11690013, LCF/BQ/PI20/11760031,  LCF/BQ/PR20/11770012, LCF/BQ/PR21/11840013). A.B. acknowledges support from the Ram\'on y Cajal program RYC-2016-20066,  CAM/FEDER Project S2018/TCS- 4342 (QUITEMADCM),  and PGC2018-099169-B-I00 (MCIU/AEI/FEDER, UE).

\appendix
\section{Weak interactions and coupled Ising chains}
\label{sec:appendix}
In this appendix, we discuss more in detail the week interacting limit $t_h \gg V_v$ of the Creutz-Hubbard ladder for generic flux, and in particular we will show how in this limit we can map the Creutz ladder as spin ladder. 

We start from  the free part of the Creutz-Hubbard Hamiltonian $H_c$ defined in the eq.(\ref{eq:H_C}). Introducing the spinor notation and performing the 
transformation $c_j= \left(- i \sigma_z \right)^j \tilde{c}_j$, we obtain that the hopping along the two legs becomes now identical at the price of having site-dependent diagonal hopping terms that connect neighboring sites in opposite legs of the ladder:
\beq
\begin{split}
H_C &= \sum_j \tilde{c}^{\dagger}_j \Big[-t_{\rm h} I - t_{\rm d} \cos \left(\frac{\theta}{2} (2 j+1) \right) \sigma_x  \\ &+ t_{\rm d} \sin \left(\frac{\theta}{2} (2 j+1) \right) \sigma_y \Big] c_{j+1} + \frac{\Delta \epsilon}{4} \tilde{c}^{\dagger}_j \sigma_z \tilde{c}_j + {\rm H.c.}.
\end{split}
\eeq
Introducing the following rung operators:
\beq
\begin{split}
\label{eq:transformation}
r_{j,1} &=\frac{i^j}{\sqrt{2}} \left[ i \tilde{c}_{j,u}+(-1)^j \tilde{c}^{\dagger}_{j,d} \right] ,\\
r_{j,2} &=\frac{i^j}{\sqrt{2}} \left[ \tilde{c}_{j,u}+(-1)^j \tilde{c}^{\dagger}_{j,d} \right]  ,\\
\end{split}
\eeq   
Under this canonical transformation, the Hamiltonian is transformed onto
\beq
\begin{split}
H_c &= \sum_{j n} \big[ -t_{\rm h} \left( r^{\dagger}_{j+1,n} r_{j,n} +H.c. \right) \\
&+ t_{\rm d} \left( (-1)^{j+1} i e^{i \frac{\theta}{2}(2j+1)} r^{\dagger}_{j+1,n} r^{\dagger}_{j,n} + H.c.  \right) \\
&+ \frac{\Delta \epsilon}{4} \left( 2 r^{\dagger}_{j,n} r_{j,n} -1 \right) \big] 
\end{split}
\eeq
In this particle-hole rung basis Eq. \ref{eq:transformation}, we identify two independent subsystems which no longer display particle number
conservation, but instead have parity conservation.
A Jordan-Wigner transformation \cite{jordan1993paulische}, namely,
\beq 
r^{\dagger}_{j,n} = \Pi_{i<j} (-\sigma^z_{i,n}) \sigma^+_{j,n} = (r_{j,n})^{\dagger}
\eeq
reveals the Ising nature of the two subsystems, and leads to
a Hamiltonian that can be understood as a two-leg quantum
spin ladder
\beq
\begin{split}
H_c &= \sum_{j n} \big[ -t_{\rm h} \left( \sigma^{+}_{j+1,n} \sigma^{-}_{j,n} +H.c. \right) \\
&+ t_{\rm d} \left( (-1)^{j+1} i e^{i \frac{\theta}{2}(2j+1)} \sigma^{+}_{j+1,n} \sigma^{+}_{j,n} + H.c.  \right) + \frac{\Delta \epsilon}{4} \sigma^z_{j,n} \big]. 
\end{split}
\eeq
We now consider the model with Hubbard-like interactions for the first time. In particular, we take a look at the regime of small interactions $V_v <t$ and study how these interactions alter the results of the non-interacting model. Applying the series of transformations presented above we obtain that the interaction term reads
\beq
\begin{split}
\hat{H}_{\rm Hub} =\frac{V_{\rm v}}{2} \sum_j \Big( \hat{r}^{\dagger}_{j,1} \hat{r}_{j,1} + \hat{r}^{\dagger}_{j,2} \hat{r}_{j,2} + \ii \hat{r}^{\dagger}_{j,1} \hat{r}_{j,2} & \\
- \ii \hat{r}^{\dagger}_{j,2} \hat{r}_{j,1}
-2 \hat{r}^{\dagger}_{j,1} \hat{r}_{j,1}\hat{r}^{\dagger}_{j,2} \hat{r}_{j,2} \Big) .
\end{split}
\eeq   
For half filling we have $\langle \hat{r}^{\dagger}_{j,1} \hat{r}_{j,2} -\hat{r}^{\dagger}_{j,2} \hat{r}_{j,1} \rangle=0$, simplifying the Hamiltonian to
\beq 
\hat{H}_{\rm Hub}=\frac{V_{\rm v}}{2}\sum_j \left( \hat{r}^{\dagger}_{j,1} \hat{r}_{j,1}+\hat{r}^{\dagger}_{j,2} \hat{r}_{j,2}-\hat{r}^{\dagger}_{j,1} \hat{r}_{j,1}\hat{r}^{\dagger}_{j,2} \hat{r}_{j,2} \right).
\eeq
The Jordan-Wigner transformation translates this expression to a ferromagnetic coupling between the two spin models:
\beq 
\hat{H}_{ \rm Hub}= \frac{V_{\rm v}}{4} \sum_j \left( 1-\sigma^z_{j,1} \sigma^z_{j,2} \right). 
\eeq
In $\pi$-flux regime, the imbalanced Creutz-Hubbard model can be understood as two coupled quantum Ising chains. For weak interactions $V_{\rm v}<t_{\rm h}$, we can treat the mutual effect of chains on each other through an MF decoupling which renormalizes the original transverse magnetization term.

\bibliographystyle{apsrev4-1}
\bibliography{chiral_current_characterization_creutz_hubbard_ladder}

\end{document}